\documentclass[12pt,cite,epsfig]{article}
\usepackage{times,epsfig}
\usepackage{caption}
 \usepackage{amsmath}	   
\newcommand{\epo}{\;.}

\newcommand{\nc}{\newcommand}
\nc{\bb}{\bibitem}
\nc{\be}{\begin{equation}}
\nc{\ee}{\end{equation}}
\nc{\pa}{\partial}
\nc{\parsym} {\stackrel{\leftrightarrow}{\pa}}
\nc{\ra}{\rightarrow}
\nc{\la}{\leftarrow}
\nc{\etp}{{\eta^\prime}}
\nc{\omg}{\omega}
\nc{\gam}{\gamma}
\nc{\hvar}{\epsilon}
\nc{\epsProd} {$\epsilon \epsilon^\prime$}

\nc{\indentB}{\indent \indent}
\nc{\piZ}{\pi^0}
\nc{\piP}{\pi^+}
\nc{\piM}{\pi^-}
\nc{\piC}{\pi^\prime}
\nc{\tlambda}{\widetilde{\lambda}}
\nc{\Leftright}{\leftrightarrow}

\nc{\second}{ {\prime \prime} }

\nc{\ggam}{\gamma \gamma}

\nc{\bea}{\begin{eqnarray}}
\nc{\eea}{\end{eqnarray}}
\nc{\beas}{\begin{eqnarray*}}
\nc{\eeas}{\end{eqnarray*}}
\nc{\non}{\nonumber}

\def\hhhv{\rule[-3.mm]{0.mm}{9.mm}}
\def\hhhvw{\rule[-2.mm]{0.mm}{7.mm}}

\begin{document}
\begin{titlepage}
\vbox{~~~ \\

\title{The $\eta/\etp \ra \pi^+ \pi^- \gam $ Decays within BHLS$_2$ and the Muon HVP
%
   }
 \author{
M.~Benayoun$^{a \dag}$, L.~DelBuono$^a$, F.~Jegerlehner$^b$ \\
\small{$^a$ LPNHE des Universit\'es Paris VI et Paris VII, IN2P3/CNRS, F--75252 Paris, France }\\
\small{$^b$ Humboldt--Universit\"at zu Berlin, Institut f\"ur Physik, Newtonstrasse 15, D--12489 Berlin,
Germany }\\
}
\date{\today}
\maketitle
\begin{abstract}
The departure of the latest FNAL experimental average  for the muon anomalous magnetic moment 
$a_\mu=(g_\mu-2)/2$  measurements having increased  from  $4.2 \sigma$  \cite{FNAL:2021} 
to  $5.0 \sigma$ \cite{FNAL:2023}, with respect to the White Paper (WP)  
consensus\cite{WhitePaper_2020}, it may indicate a hint for new physics. 
As the most delicate piece of  $a_\mu$ is its leading order
HVP part $a_\mu^{HVP-LO}$,  methods to ascertain its theoretical value 
are crucial to interpret appropriately this departure with the measurement.
We, therefore, propose to  examine closely the dipion spectra from the 
$\eta/\etp \ra \pi^+ \pi^- \gam$ decays  in the Hidden Local Symmetry (HLS)
context using  its BHLS$_2$ broken variant. We thus have at our disposal
a framework where the close relationship of the dipion spectra from the $\eta/\etp$ and
$\tau$ decays and of the $e^+e^- \to \pi^+\pi^-$ annihilation can be 
simultaneously considered.
A special focus is put to the high statistic dipion spectra from the $\eta$ 
decay collected by the KLOE/KLOE2 Collaboration and $\etp$ decay collected
by the BESIII Collaboration, and it is shown that the BHLS$_2$ framework provides a fair account of their dipion spectra. More precisely, it is first proven that a single  Omnès representation real polynomial  is requested, common to both the $\eta$ and $\etp$ dipion spectra. Moreover, it is shown that fits involving  the $\eta/\etp/\tau$ dipion spectra, and excluding the $e^+e^- \to \pi^+\pi^-$ annihilation data, allow for a  prediction of the pion vector form factor data $F_\pi(s)$ which fairly agree with the usual dipion spectra collected in the $e^+e^- \to \pi^+\pi^-$ annihilation channel.  Even if more precise $\eta/\etp/\tau$ dipion spectra would help to be fully conclusive, this confirms the Dispersive Approach results for $a_\mu^{HVP-LO}$ and points towards a common non experiment-dependent origin to this tension with the now well accepted LQCD result.

\vspace{1cm}

\it{$\dag$ Maurice Benayoun has passed on September 15th, 2023.}

\end{abstract}
}
\end{titlepage}
\tableofcontents

\section{Preamble~: Various Aspects of the Dispersive Approach to the Muon HVP}
\label{preambule}

\indentB The hadronic vacuum polarization (HVP) $a_\mu \equiv (g_\mu - 2)/2$ 
plays a central role in
precision physics, in particular, in the Standard Model prediction of
the Muon Anomalous Magnetic Moment, but as importantly, for a
precise calculation of the running electromagnetic fine structure
constant $\alpha_{em}(s)$ and of the electroweak mixing parameter $\sin^2
\theta_W(s)$. Thereby, accurate predictions suffer from the
non-perturbative contributions from the low--lying hadron
physics uneasy to address precisely from first principles. 

Recently \cite{FNAL:2023}, the Muon $g-2$ FNAL experiment has 
re-estimated the previous average value of their run
1 data sample  \cite{FNAL:2023}  and the
latest  BNL measurement \cite{BNL2}  by also considering their run 2 and 3 data
samples; this turns out to increase the statistics by a factor of $\simeq 4$.   
Moreover, the Muon $g-2$ FNAL Collaboration achieved  an improvement by 
about a factor of 2 of their systematics uncertainty.  The derived updated  average~:   
$$a_\mu^{exp.} = 116 592 059(22) \times 10^{-11} (0.19 {\rm ppm}) $$
increases the deviation   from  the White Paper (WP) Standard Model  consensus
\cite{WhitePaper_2020}, from  $4.2~ \sigma$  \cite{FNAL:2021} to  $5.0 ~\sigma$ 
\cite{FNAL:2023}. The  difference 
$\delta_a =a_\mu^{exp.}-a_\mu^{th.}$ is now
$\delta_a= 24.4 \pm 4.5$ in units of $10^{-10}$, dominated by the uncertainty
agreed upon by the WP theory consensus \cite{WhitePaper_2020}. This departure
from theoretical expectations deserves, of course,  to be explored as,
indeed, the overall pattern reflected by the various model/theoretical
approaches is unclear, even contradictory. 

\vspace{0.5cm}

The WP Standard Model  consensus for $a_\mu^{th.}$   resorts to a
data-driven dispersion relation  (DR) approach, where the experimental
low-energy hadron production cross-sections provide the
non-perturbative input to calculate the HVP effects. Fortunately, the
problem can be restricted to a precise knowledge of the process
$e^+e^- \to \gamma^* \to \mbox{hadrons}$, and for what concerns the
muon $g-2$, the $e^+e^- \to \pi^+\pi^-$ channel provides the dominant
contribution to the model uncertainty. 

Regarding its non-perturbative hadronic content, the standard dispersive-based (DR) evaluation of the HVP consists of  deriving the
 contribution  of {\it each}  $e^+e^- \to \gamma^* \to \mbox{hadrons}$
 annihilation channel by combining the different spectra collected
by the different experiments in the   hadronic channel considered
employing  algorithms of different levels of sophistication.
The full hadronic HVP value is then defined, for what concerns its 
non--perturbative content, by the sum of these different contributions.
The WP Standard Model consensus \cite{WhitePaper_2020}
is based on a combination  of two such evaluations \cite{Davier:2010nc,Teubner3}.

Although the main challenge  is then,  seemingly,  the simple
$\pi\pi$-production process, the experimental challenge is highly
complex depending on a precise understanding of the detectors and,
on the theory side, the radiative corrections required to disentangle
hadronic effects from electromagnetic contamination. 
Unfortunately,
the data samples provided by the different experiments do not exhibit
a satisfactory consistency  -- and even some can be in strong contradiction 
\cite{CMD-3:2023alj} with the others. 
Using the  $\tau \to \pi^-\pi^0 \nu_\tau$ 
decay information, first proposed by \cite{alemany:1997tn}, has been considered 
to discriminate  among the $\pi^+ \pi^-$  spectra, but did not lead
to  convincing enough conclusions.

\vspace{0.5cm}

It is widely considered that all low-energy hadronic processes derive 
from QCD even though, in the non-perturbative low-energy regime, tools 
to make valid predictions of real-time hadronic cross sections are
missing. Nevertheless, as   hadron physics is accepted to derive from QCD,
it follows that {\it the various specific hadronic decay processes 
are highly correlated to each other}. It is thus motivated to address 
these correlations, especially in order to constrain the
non-perturbative sector of the $e^+e^- \to \gamma^* \to \mbox{hadrons}$
annihilations.

Although we lack methods to predict a process like $e^+e^-
\to \pi^+\pi^-$, we know that QCD implies well-defined symmetry
patterns like approximate chiral symmetry, and gives rise to Chiral
Perturbation Theory (ChPT), a systematic expansion about the chiral
symmetry point. It allows one to work out reliable predictions from
first principles for the  low energy tail of the QCD hadron
spectrum (up to about the $\eta$ meson mass).  

 With this in mind, an attempt to consider the $e^+e^- \to \pi^+\pi^-$ 
annihilation  not only in relation with the $\tau^\pm \to \pi^\pm \pi^0 \nu_\tau$ 
decay {\it but  also   with other related spectra} is important;
it motivates  a unified modeling\footnote{Considering individual channels in
isolation, as usually done, does not help much to uncover
inconsistencies between different experimental data sets sometimes involving
different final states.} by a
version of the Resonance Lagrangian approach (RLA) \cite{Ecker1,Ecker2} -- we adopted the
Hidden Local Symmetry (HLS) version\footnote{See also \cite{Meissner:1987ge} for an equivalent version.}  \cite{HLSOrigin,FKTUY} -- 
needed to extend the Chiral perturbation theory toward higher energy
 to cover the $\rho$, $\omega$ and  $\phi$ energy 
 range\footnote{A precise evaluation of the photon HVP implies 
 a precise account of  
 the energy range $\sqrt{s}\equiv [2 m_\pi, 1.05$ GeV],
 the largest contribution of the non--perturbative region  which extends
 up to $\simeq 2$ GeV as experimentally observed \cite{KEDR2016,BES2009}.}.
 To practically  succeed in such a program
the original HLS model -- see for instance \cite{HLSRef} for a review --
has been  supplied  with appropriate symmetry-breaking mechanisms with
various levels of sophistication to derive the  earlier versions of 
the BHLS model in \cite{taupaper,ExtMod3,ExtMod4}, or the 
more refined 
BHLS$_2$ version \cite{ExtMod7}, updated in \cite{ExtMod8}.  
 
One thus achieved
 a simultaneous consistent fit of the $e^+e^- \to \pi^+\pi^-$ data
from CMD-2~\cite{CMD2-1998-2}, SND~\cite{SND-1998}, KLOE~\cite{KLOE10,KLOE12,KLOEComb},  , BaBar~\cite{BaBar,BaBar2}, BESIII~\cite{BES-III,BESIII-cor} and CLEO-c~\cite{CESR} and the $\tau \to
\pi^-\pi^0 \nu_\tau$ decay spectral functions collected by
ALEPH~\cite{Aleph}, CLEO~\cite{Cleo} and
Belle~\cite{Belle} (see~\cite{taupaper,ExtMod3,ExtMod7,ExtMod8}). 
This  updated  BHLS$_2$ fairly recovers the known properties of the $[\pi^0,\eta,\etp]$ 
system thanks to its kinetic breaking mechanism \cite{ExtMod8}.
   
Besides keeping the neutral vector current 
conserved, this breaking mechanism also generates  a violation of the 
charged vector current conservation and a departure of $F_\pi^\tau(s=0) =1 $ 
by a few per mil. Such an option finds support in the own Belle fit results 
reported  in  Table VII of their \cite{Belle};  additional
$\tau$ spectra are needed to conclude -- see the discussion in Section 3  of  \cite{ExtMod8} --
as such a breaking mechanism might affect  $\tau $ based predictions for the muon HVP.

Beside the $\pi^+\pi^-$ annihilation channel and the $\tau \to
\pi^-\pi^0 \nu_\tau$ decay spectra,  BHLS$_2$ \cite{ExtMod7,ExtMod8}, 
also addressed successfully the 
$\pi^+\pi^-\pi^0, (\pi^0/\eta) \gam$ and  $K \overline{K}$  
final states in the fully correlated way represented by a single Lagrangian.
A few additional radiative partial
width decays are also considered, noticeably those for 
$\pi^0/\eta·/\etp \ra \gam \gam$, and some more $VP\gam$ radiative decays.

\vspace{0.5cm}

In view of the significant inconsistencies of the data samples
collected by some experiments, the global fit approach has two
advantages: first, more data are expected to reduce the uncertainties
of the HVP evaluations and, second, provides consistency checks of each
$e^+e^- \to \gamma^* \to \mbox{hadrons}$ data set versus the other
samples collected in the same annihilation channel {\it  or in another one}.

In the present work, we go a step further by also involving the
$\eta/\etp \to \pi^+\pi^- \gamma$ decay modes in order to obtain additional $\pi\pi$
dipion spectra from experiments with systematics quite different from those encountered in
 $e^+e^-$ annihilations. As will be seen below, these decays  allow for a new test of the
  self--consistency of the DR based estimates of $a_\mu$~: Indeed,  the  $\eta/\eta'$
decay spectra can provide a DR evaluation for $a_\mu(\pi^+\pi^-, \sqrt{s} < 1~{\rm GeV})$
which can be fruitfully compared with those directly derived from directly integrating 
the $e^+e^- \ra \pi^+\pi^-$ annihilation data. One may expect that the  $\eta/\etp$ dipion 
spectra  benefit from systematics largely independent of those in the $e^+e^-$ annihilation.

\vspace{0.5cm}

Besides the DR approach which gave rise to several evaluations of the muon HVP $a_\mu$
listed in the White Paper  \cite{WhitePaper_2020}, the challenging Lattice QCD  (LQCD)
approach has been used by several groups and produced  results with relatively 
poor precision at the time of the White Paper. They were not used to
define the so-called WP Standard Model consensus reported in \cite{WhitePaper_2020}
which, based on some DR estimates, provided the leading order (HVP-LO) consensus
 $a_\mu^{\rm  LO}[{\rm th.}] = 693.1(4.0)\times 10^{-10}$. 
Using the LQCD approach, the BMW Collaboration which first got
\cite{BMW_amu,WhitePaper_2020} 
$a_\mu^{\rm  LO}=(711.1 \pm 7.5 \pm 17.4) \times 10^{-10}$  later on
improved their calculation and got
$a_\mu^{\rm LO}=(707.5 \pm 5.5) \times 10^{-10}$ \cite{BMW_amu_final}
at clear variance with the WP
consensus just reminded. 
This evaluation finds support from the new evaluations by other LQCD groups~:
  $a_\mu^{\rm LO}=(720.0 \pm 12.4_{\rm stat} \pm 9.9_{\rm syst}) \times 10^{-10}$ 
 (Mainz/CLS 19)~\cite{Gerardin:2019rua,Ce:2022kxy} and 
 $a_\mu^{\rm LO}=(715.4 \pm 16.3_{\rm stat}\pm9.2_{\rm syst}) \times 10^{-10}$ 
 (RBC/UKQCD18)~\cite{RBC_UKQCD_amu}.   
  
The lattice calculation of $a_\mu^{\rm LO}$ 
thus brings the SM prediction of $a_\mu$ into an acceptable agreement with 
the experiment but generates a  significant 
disagreement between the LQCD results and the different data-driven dispersive results;  this looks now well established. It adds to the former puzzle from data versus predictions a puzzle between Lattice QCD  and the DR approaches, which deserves clarification.


\section{Introduction}
\label{Introduction}

\indentB  In this article, we focus on the traditional way of estimating the contribution 
of the non--perturbative energy region to the photon HVP which relies  on dispersive methods 
using as basic ingredients the $e^+e^-$  annihilation cross sections to all the possible 
exclusive hadronic final states collected up to $\sqrt{s} \simeq 2$ GeV.
 
The different successive broken variants of the HLS model, especially BHLS$_2$
 \cite{ExtMod7,ExtMod8}, provides a  well adapted framework
to address   the 
most relevant  $e^+e^-$ annihilations to hadronic channels in the  crucial part of
 the low energy region ($\sqrt{s} \leq 1.05$ GeV), namely 
 the  $e^+ e^-$ annihilations to the 
$ \pi^+ \pi^-, K \overline{K}/\pi^+ \pi^- \pi^0 /\pi^0 \gamma /\eta \gamma$
final states; these already provide more than 80\% of the muon HVP, when 
integrated  up to the $\phi$ meson mass. 

A BHLS$_2$ based computer code was used for this analysis which 
 considered the large number of available data samples (several dozens), 
more than 1400 data points and thus,
practically, the whole set of the available data samples has been exhausted.  
They have been 
listed, analyzed, and  discussed in full detail previously, especially in the 
recent  articles \cite{ExtMod7,ExtMod8}, where a large number of previous references 
 can be found\footnote{\label{CMD3-pi}
 The CMD-3 Collaboration has recently published a high-statistics measurement
 of the $e^+e^- \ra \pi^+ \pi^-$ cross section \cite{CMD-3:2023alj} which deserves a
 specific analysis beyond the scope of the present work which is focused on a quite 
 different topic; nevertheless, the information provided by the CMD-3 Collaboration
 in their article regarding the consistency of their spectrum with the previously collected
 data samples may indicate that, as it is,  their measurement is not consistent with any subset
 of the relevant existing data samples and thus it should hardly accommodate a global framework
 like HLS; so, it should not impact the conclusions of the present work. }.
This computer code takes faithfully into account the whole uncertainty
information  provided together with these data samples and, therefore,
 yielding satisfactory global fit probabilities turns out to  
have simultaneously  a satisfactory model, a satisfactory handling data of the 
samples collected in several
physics channels and, also, a satisfactory dealing with their 
reported uncertainty information.

In this perspective, given data samples exhibiting contradictory aspects
compared to most of the others may  lead to either discard  them or, 
when meaningful,  motivate several solutions that avoid  mixing up 
contradictory  spectra; this has led us in our previous studies  
\cite{ExtMod7,ExtMod8} to provide
different HVP evaluations based on some of the reported dipion KLOE samples, 
-- namely  \cite{KLOE10,KLOE12,KLOEComb} --
on the one hand and separately on their Babar analog \cite{BaBar,BaBar2}
on the other hand.

Regarding the various dipion spectra, the  studies found  strong contradictions 
between the so-called KLOE8 data sample \cite{KLOE08}, or the recently published SND spectrum 
\cite{SND20}, and the bulk of the other considered data samples,  have been 
discarded. 

Comparing our own evaluations with those based on Dispersion Relations 
collected  in \cite{WhitePaper_2020}, one does not observe any loss in precision
with any of the various reported values of the muon $g-2$; however,  
differences between central values can be observed, clearly
related to the contradictory properties of some data samples, especially KLOE 
\cite{KLOE10,KLOE12}
 versus Babar \cite{BaBar,BaBar2} reported since a long time 
 \cite{ExtMod3,ExtMod4,BM_roma2013}.
 
As noted above, the contribution of the listed HLS channels 
to the HVP is large; however, it is also worth mentioning that 
their contribution to the HVP uncertainty is almost negligible
compared to those of the rest of the non--perturbative region.
Moreover, as the HLS approach implies tight connections between the various 
annihilation channels, it allows performing stringent consistency
checks on the different data samples involving the same physics 
channels or, also, the other channels addressed by the HLS Lagrangian.
It is worthwhile pointing out  this important property, specific to global models 
like  BHLS$_2$ and also  stressing that, by far, most of the
available data samples fulfill this drastic constraint.

On the other hand, as indicated in the previous Section, the updated version BHLS$_2$ variant 
\cite{ExtMod8} of the broken HLS model
\cite{ExtMod7} allows  to fairly address the physics of  the [$\pi^0,~\eta,~\etp$] system within 
the HLS corpus. Indeed, beside the $e^+ e^- \ra (\pi^0/\eta) \gam$ annihilations, the PS decays 
to $\gamma \gamma$ and the $VP\gamma$ couplings,  the pseudoscalar meson (PS)  mixing properties in 
the octet-singlet \cite{Kaiser_2000,leutwb,leutw}
and quark flavor \cite{feldmann_1,feldmann_2,feldmann_3}
basis parametrizations have been analyzed, leading to
a satisfactory comparison with expectations. 
 
Among the other processes involving the  properties of the 
[$\pi^0,~\eta,~\etp$] system,  the $\etp \ra \pi^+ \pi^- \gam$ decay 
spectrum deserves special attention. The measurements of this decay
process started long ago -- as early as 1975 \cite{Grigorian_BOX} -- and  
several  experiments have collected samples of limited statistics
\cite{JADE_BOX,CELLO_BOX_OK,TASSO_BOX_OK,PLUTO_BOX_OK,TPC_2gamma_BOX,
ARGUS_BOX_OK,LeptonF_BOX,Crystal_Barrel_BOX_OK}
motivated by a reported 20 MeV mass shift  of the $\rho$ peak 
compared to its observed value  in the $e^+e^- \ra \pi^+ \pi^-$ annihilation.  
 
\vspace{0.5cm}

 This effect was soon attributed to an interference between the $\etp \ra \rho \gamma$
($\rho \ra \pi^+ \pi^-$)  resonant amplitude and the Wess-Zumino-Witten (WZW)
 anomalous
 $\etp \pi^+ \pi^- \gamma$ contact  term \cite{WZ,Witten}; this
 so-called box anomaly was expected to occur alongside the triangle anomaly
 responsible of the two--photon decays of the $\pi^0,~\eta$ and $\etp$ mesons.
 A basic HLS approach including this 
 anomalous interaction  term beside the dominant  $\etp \rho^0 \gamma$ coupling 
 \cite{box} confirmed this guess.

However, the dipion $\etp$ spectrum from BESIII Collaboration \cite{BESIII_BOX_OK} 
published much later,  thanks to its large statistics (970,000 events), modified 
the picture~: It led to conclude that supplementing the ($\rho^0,\omg$) resonance
 contributions by only a contact term is insufficient to reach a satisfactory
  description  of the dipion spectrum.

On the other hand, the reported dipion spectrum observed in the parent
$\eta \ra \pi^+ \pi^- \gam$ decay
has undergone much less measurements. Beside  former 
spectra\footnote{The numerical content of these spectra can only
be derived from the paper figures.} from Layter {\it et al.}
\cite{Layter_BOX} and Gormley  {\it et al.} \cite{Gormley_BOX_OK},
 WASA-at-COSY reported for a 14,000 event spectrum \cite{WASA_BOX_OK} 
whereas the KLOE/KLOE2 Collaboration collected a 205,000 
event spectrum \cite{KLOE_BOX_OK}.

As the dipion spectra reported from
the recent  measurements of the $\eta/\etp \ra \pi^+ \pi^- \gam$  decays
carry high statistics,
it thus becomes relevant to re-examine if (and how) they fit within the recently defined 
BHLS$_2$ framework of the HLS model, especially thanks to its kinetic breaking
(See Appendix \ref{HLS_PS})
which has already allowed for a satisfactory description
of the [$\pi^0,~\eta,~\etp$] system properties \cite{ExtMod8}.
Moreover, even if the physics of the $\eta/\etp$ mesons is interesting {\it per se}, 
a better understanding of their properties is important, given
their important  role in the  Light-by-Light (LbL) contribution  to 
 the muon anomalous magnetic moment.
 
\vspace{0.5cm}

The layout of the paper is as follows. Section \ref{Kroll-cnd} aims at reminding 
the Kroll
conditions \cite{Kroll:2005}  which reduce the number of free parameters of 
the kinetic breaking
mechanism from 3 to 1; it also reminds and corrects Lagrangian pieces relevant 
for the present study.  Section \ref{eta-etp-amplitudes} is intended to
identify the Lagrangian pieces contributing to the considered 
$\eta$ and $\etp$ radiative decays and displays  the involved diagrams;
 the BHLS$_2$  amplitudes for these
are constructed in Section \ref{BoxEta} for the $\eta \ra \pi^+ \pi^- \gam$
 decay and in Section \ref{BoxEtp} for the $\etp \ra \pi^+ \pi^- \gam$ one.
The relation between the anomalous HLS amplitudes and their Wess--Zumino--Witten
(WZW) \cite{WZ,Witten} analogs is given in Section \ref{WZW}. The derivation
of the dipion mass spectrum in the $\eta/\etp$ radiative decays is done in
Section \ref{MassSpectra-1} and the role of the additional polynomial factor in the  $\eta/\etp$ radiative decays is thoroughly examined  in  Section \ref{FSI}. 
The polynomial factors accompanying the pion form factor parameterize an approximation to higher inelastic effects (beyond the two-pion channel) and are associated with the production 
vertex\footnote{The polynomials addressed here are process dependent and also appear in the Omn\`es representation of the pion form-factor related $\pi\pi$ production in $e^+e^-$-annihilation~\cite{Leutwyler:2002hm}\cite{Colangelo_BOX_th}\cite{Colangelo:2006cd}.}.

Section \ref{HLS_eta_etp_Decays} is the central part of 
the present study; Subsection \ref{exp-spectra} presents exhaustively the 
available $\eta/\etp \ra \pi^+ \pi^- \gam$  data samples; for this purpose
it is  important to note that all the available spectra carry an arbitrary
absolute normalization and that, accounting for the 
$\eta/\etp \ra \pi^+ \pi^- \gam$ partial widths 
implies using also an external piece of (PDG \cite{RPP2022}) information. A 
detailed study
of the additional polynomial degrees is the subject of Subsection \ref{fits-1} which 
reports on the fits performed separately with the $\eta$ and $\etp$ spectra 
 to find the appropriate degrees of the requested polynomials.
 This allows to perform the fits of the dipion spectra
reported in Subsection \ref{fits-2} where it is proved that a unique 
polynomial can  satisfactorily account for both  the $\eta$ and
$\etp$ dipion spectra simultaneously. 

Subsection \ref{fits-3} is devoted to comparing our polynomial  results
to  those reported in the literature.
The role of intermediate  $\rho^\pm$ exchanges is emphasized  in Subsection 
\ref{TR2}. The global BHLS$_2$ fits  performed to simultaneously 
describe the dipion spectrum lineshapes  examined in the previous 
Subsections and the PDG information for the partial widths 
$\Gamma(\eta/\etp \ra \pi^+ \pi^- \gam)$  is worked out
in Subsection \ref{absolute_norm}.
A brief numerical analysis of some parameter values returned by the fits of the $\eta/\etp$ dipion spectra is the subject of Section \ref{BriefPar}.
Finally in Section \ref{MAMM} one examines the issues relative to the connection between the $\eta/\etp \ra \pi^+ \pi^- \gam$ decays (and also the $\tau$ decay data) and the hadronic contribution to the muon anomalous magnetic moment $a_\mu$. Section
\ref{conclusion} summarizes the conclusions reached by the  present study.

In order to ease the  paper reading, the main pieces of information
regarding the HLS model  are briefly reminded in Appendix \ref{HLS_org}, whereas
its symmetry-breaking mechanisms are briefly summarized in Appendices \ref{HLS-BKY} to
\ref{HLS_PS}. An Erratum to the previous broken version of the BHLS$_2$ version
is the subject of Appendix \ref{Erratum}. To ease the reading of the present work,   
one has also found it appropriate to give the most relevant parts of the non--anomalous 
and anomalous BHLS$_2$  pieces under the Kroll Conditions -- reminded just  
below -- in Appendices \ref{AAP-VVP} and \ref{PPP}.

\section{The Kroll Conditions and VPP Lagrangian Pieces}
\label{Kroll-cnd}
\indent \indent In the FKS approach  \cite{feldmann_1,feldmann_2,feldmann_3}
to the [$\pi^0,~\eta,~\etp$] system, it has been found appropriate 
to impose the  Kroll conditions  \cite{Kroll:2005}  to  axial current 
matrix elements. Applied  to the  BHLS$_2$ axial currents, these conditions~: 
\be
\begin{array}{llll}
\displaystyle  <0|J^a_\mu| \eta_b(p)> =ip_\mu f_a \delta_{ab}~~, 
&\displaystyle  |\eta_b(p)> =| b \overline{b}(p)>~~,
&\displaystyle J^a_\mu=\overline{a} \gamma_\mu \gamma_5 a~~, 
&\displaystyle \{a,b =u,d,s\}~~.
 \end{array} 
 \label{kroll-1}
\ee
lead to two non--trivial relations \cite{ExtMod8} -- referred to below as $A_\pm$ 
solutions --  among the $\lambda_i$ parameters of the generalized 't~Hooft term 
\cite{tHooft,leutw} (see Appendix \ref{Erratum}); 
one gets~:
\be
\left \{
\begin{array}{lll} 
 \displaystyle { \rm Solutions} ~~A_\pm~ \Longleftrightarrow   
 &\displaystyle   \lambda_0=  \sqrt{2} \lambda_8 = \pm  \sqrt{\frac{3}{2}}
 \lambda_3 
\end{array}
\right \} .
\label{kroll-7b}
\ee
which reduces the actual parameter freedom of the kinetic breaking from 
three to only one. 

One thus should note that the Kroll Conditions  tightly couple the breaking 
in the BHLS$_2$ Lagrangian of the original U(3)  symmetry
to  SU(3)$\times$U(1) and a particular  Isospin breaking piece
(via $\lambda_3 \ne 0$); it also lead to $F_\pi^\tau(s=0)=1-\lambda_3^2/2$.

\vspace{0.5cm}

The $\pm 1 $ factor in Equations (\ref{kroll-7b})  is propagated below as $d_\pm$; so,
$A_+$ corresponds to $d_+$ and $A_-$  to $d_-$. The non--anomalous pieces 
${\cal L}_{\eta^\prime \pi^\pm}$ and ${\cal L}_{\eta \pi^\pm}$ of the BHLS$_2$ 
Lagrangian acquire  simplified expressions compared to \cite{ExtMod8}~:
 \be
\left \{
\begin{array}{lll}
\displaystyle  {\cal L}_{\pi^0 \pi^\pm}= & \displaystyle 
\frac{iag}{2} (1+\Sigma_V)  (1-\frac{\lambda_0^2}{3})\left[~
\rho^- \cdot \pi^+ \parsym \pi^0   - \rho^+ \cdot  \pi^- \parsym \pi^0 
\right ]\\[0.5cm]
\displaystyle  {\cal L}_{\eta \pi^\pm}= & \displaystyle 
- \frac{iag}{2} 
\left[ 1+ \Sigma_V \right]  
\left [\epsilon - \frac{A_\pm}{2} \sin{\delta_P}
\right ]
   \left[~
\rho^- \cdot \pi^+ \parsym \eta   - \rho^+ \cdot \pi^- \parsym \eta
\right ]\\[0.5cm]
\displaystyle  {\cal L}_{\eta^\prime \pi^\pm}= & \displaystyle 
-\frac{iag}{2} 
\left[ 1+ \Sigma_V \right]  
\left [\epsilon^\prime + \frac{A_\pm}{2} \cos{\delta_P}\right ]
\left[~
\rho^- \cdot \pi^+ \parsym \eta^\prime   - \rho^+ \cdot \pi^- \parsym \eta^\prime
\right ]
\end{array}
\right .
\label{kroll-7d}
\ee
where~:
 \be
 A_\pm = \Delta_A+d_\pm \lambda_0^2~,
\label{kroll-7e}
\ee
exhibiting the BKY $\Delta_A$ and $\delta_P$ is defined by~:
  \be
  \begin{array}{lll}
\displaystyle    \cos{\delta_P}= ~~~\frac{1}{\sqrt{3}}
\left [ \sin{\theta_P} + \sqrt{2}\cos{\theta_P}
\right ]~~~,~~~&\displaystyle   
\displaystyle    \sin{\delta_P}=- \frac{1}{\sqrt{3}}
\left [ \cos{\theta_P} - \sqrt{2}\sin{\theta_P}
\right ]
\end{array}
 \label{def1}
 \ee
 in terms of $\theta_P$, the third mixing angle \cite{WZWChPT} which is one
 among the BHLS$_2$  fit parameters. It has been shown in \cite{ExtMod8}
 that the BKY parameter  $\Sigma_V$ can be dropped out without any loss in generality.
 
 One should note that, if $ {\cal L}_{\pi^0 \pi^\pm}$
 is leading order, both  ${\cal L}_{\eta \pi^\pm}$ and ${\cal L}_{\etp \pi^\pm}$
 are manifestly ${\cal O}(\delta)$, {\it i.e.} first order in breakings. Finally,
 it is worthwhile to remind that terms of order ${\cal O}(\delta^2)$ or
  higher  in amplitudes are  discarded.
 
\section{The $\eta/\etp \ra \pi^- \pi^+ \gam$ Decays in the BHLS$_2$ Framework}
\label{eta-etp-amplitudes}
\indentB The amplitudes for the  $\eta/\etp \ra \pi^- \pi^+ \gam$ decays
{\it a priori} involve the $APPP$, $VPPP$ and $AVP$ 
 sectors of the full BHLS$_2$ Lagrangian \cite{ExtMod7,ExtMod8}. 
 The  interaction terms 
 involved are displayed in Appendices \ref{AAP-VVP} and \ref{PPP} in terms
 of the {\it physical} pseudoscalar fields and {\it ideal} vector fields
 which should be replaced by their physical partners following  the method 
 developped in \cite{ExtMod7}. The $V -\gam$ transition couplings can be
found  in \cite{ExtMod7}, Appendix \ref{HLS_unbrk} and the  relevant non--anomalous
VPP couplings have been displayed, for convenience, in Section \ref{Kroll-cnd} just above.
\begin{figure}[!ht]
\begin{center}
{\includegraphics[height=5.5cm]{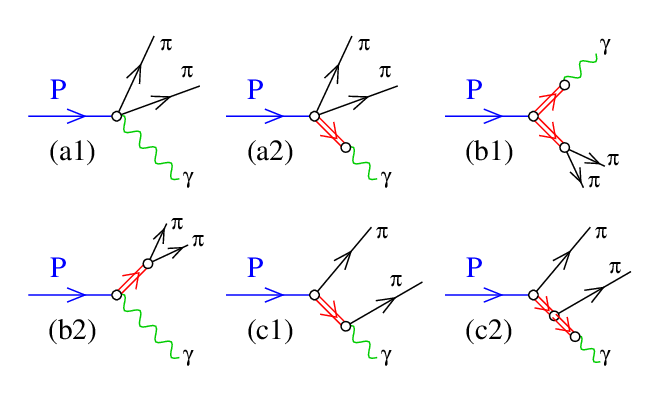}}
\end{center}
\begin{center}
\caption{\protect\label{Fig:vmd_tree} The classes of tree diagrams. $P$ stands for either of 
$\eta$ and $\eta^\prime$; in diagrams  $a$ and $b$, the double  lines stand
for the neutral vector mesons (subject to mixing), in diagrams  $c$, the
intermediate vector meson is $\rho^\pm$ whereas the external one is neutral. 
The pions are charged. The vanishing of the $AVP$ couplings (see text)
implies that  diagrams  (b2) and (c1) do not contribute to the decay amplitudes.}
\end{center}
\end{figure}
The classes of diagrams {\it a priori} involved in the $\eta/\etp$ decays
 to $\pi^- \pi^+ \gam$ are displayed in Figure~\ref{Fig:vmd_tree}.
 Namely, diagram (a1) illustrates the $APPP$ interaction, whereas diagram (a2)
 sketches  the  $VPPP$ contributions with  $V -\gam$ transitions 
 ($V=\rho^0,~\omg,~\phi$) provided by the non--anomalous BHLS$_2$ Lagrangian
 (\cite{ExtMod7}, Appendix A).  These two kinds of diagrams are generally
 named box anomaly terms.

 Diagram (b1) sketches the diagram class involving $VVP$
 couplings; these diagrams provide the major contribution to the
 $\eta/\etp$ dipion spectra.  
 As one assumes   $c_3=c_4$ thanks to former works \cite{ExtMod3},
  all contributions involving $AVP$ couplings, as those 
  depicted  in Figures (b2) and (c1), identically vanish.
 Finally, the (c2) diagram class  illustrates the
   diagrams reflecting the 2 possible choices for the $\pi^\pm \pi^\mp$ pair,
  each involving an intermediate $\rho^\pm$ exchange.
 
 In the following, for the $\eta$ and $\etp$ decays, 
 the non-resonant (a1) and (a2) contributions  are gathered
 into  the $T^{NR}$ partial amplitude, whereas the (b1) and (c2) resonant contributions are
 given by resp. the $T^{R1}$ and $T^{R2}$  terms.

\section{The $\eta \to \pi^+ \pi^- \gamma$ Amplitude within BHLS$_2$}
\label{BoxEta}
\indent \indent As three kinds of diagrams contribute,  the full $T(\eta)$, amplitude 
 for the $\eta \to \pi^+ \pi^- \gamma$ decay is written~:
 \be
 \displaystyle 
T(\eta) = T^{NR}(\eta) + T^{R1}(\eta) + T^{R2}(\eta)
\label{BoxEta-1}
\ee
and they include the common tensor object~:
 \be
 \displaystyle 
 F=
\epsilon^{\mu \nu \alpha \beta} \varepsilon_\mu (\gamma , q) q_\nu p^-_\alpha
p^+_\beta 
\label{BoxTensor}
\ee
typical of the anomalous Lagrangian piece expressions; $F$ exhibits the obvious 
momentum notations.
This factor is understood in the $T(\eta/\etp)$  amplitude expressions 
here and below to lighten writing; it is restored in the final expressions 
involving the  differential decay widths.

As already stated, the first term in the
expansion (\ref{BoxEta-1}) gathers the non-resonant ($APPP/VPPP$) contributions 
whereas the second and third terms  collect the resonant contributions 
of different structures generated via the VVP Lagrangian 
and commented  in the Section just above.

\vspace{0.4cm}
The $T^\eta_{NR}$  term  can be
written ($A_\pm=\Delta_A+d_\pm \lambda_0^2$)~:
\be
\begin{array}{ll}
T^{NR}(\eta)=  \displaystyle - \frac{ie}{4\pi^2 f_\pi^3} 
  \left[ 1-\frac{3 c_3}{2}  \right]  g_{\eta \pi^+\pi^-\gamma} &{\rm with} ~~
  
 \displaystyle g_{\eta \pi^+\pi^-\gamma} =
\epsilon + \left \{1 -\frac{A_\pm}{2} - \frac{3\lambda_0^2}{4}
\right \}  \sin{\delta_P}~.
\end{array}
\label{BoxEta-2}
\ee 

It is worthwhile noting that ${\bf i/}$ The dependency upon $c_1-c_2$ drops out
when summing up the $APPP$ and $VPPP$ contributions, ${\bf ii/}$ If one cancels
out the symmetry breaking contributions, $T^{NR}(\eta)$ remains non--zero and
corresponds to the Wess-Zumino-Witten (WZW) term \cite{WZ,Witten}.

On the other hand, the $T^{R1}(\eta)$  contributions to the $T(\eta)$ amplitude can be written
($m^2=a g^2 f_\pi^2$)~:
\be
\hspace{-1cm}
\left \{
\begin{array}{ll}
T^{R1}(\eta)=  \displaystyle  c_3~\frac{ie m^2}{8 \pi^2 f_\pi^3} 
  \left[ \frac{T^0_\rho(\eta)}{D_\rho(s)} +  
  \frac{T^0_\omg(\eta)}{D_\omg(s)} + \frac{T^0_\phi(\eta)}{D_\phi(s)}  \right] 
  \\[0.5cm]
 \displaystyle  T^0_\rho(\eta)=\epsilon + \frac{2\beta(s)}{z_A} 
   \cos{\delta_P}+  3 \left[1 -\frac{3\lambda_0^2}{4} - \frac{A_\pm}{6}
 + \frac{\alpha(s)}{3} +2 \xi_3\right] \sin{\delta_P}
  \\[0.5cm]
  \displaystyle  T^0_\phi(\eta)= -\left[ \frac{2 \beta(s)}{z_A} \right] \cos{\delta_P}  
  \\[0.5cm]
  \displaystyle  T^0_\omg(\eta) = -~\alpha(s)  \sin{\delta_P}
\end{array}
\right .
\label{BoxEta-3}
\ee 
where $D_\rho(s)$, $D_\omg(s)$ and $D_\phi(s)$ are the indicated inverse vector meson propagators;
they are parametrized as defined in Section 9 of \cite{ExtMod7}.   
Equations (\ref{BoxEta-3}) displays the dependency upon the  angles $\alpha(s)$ and $\beta(s)$ 
defining the dynamical vector meson mixing
(see Appendix \ref {HLS-dynMix}) and upon the parameter defined
by the kinetic breaking mechanism (see Appendix  \ref{HLS_PS}), once the Kroll conditions 
\cite{Kroll:2005}  are applied. It is worth remarking that $\rho^0$ is the only resonant contribution  
which survives when symmetry-breaking  terms are turned off. Moreover,
the $\omg$ and $\phi$ contributions are outside the phase space actually available in
the $\eta$ decay.

$T^{R2}(\eta)$, the second resonant contribution,  is produced by the
{\it non-anomalous} $\rho^\pm\eta \pi^\mp$ coupling purely generated by our breaking procedures
(see Equations (\ref{kroll-7d})) and by the $\omg \rho^\pm \pi^\mp$ term of the $VV\eta$ Lagrangian 
piece (see Appendix \ref{VVeta}). Setting~: 
$$s_{\pm 0}=(p_\pm +q)^2 ~~~,~~q= {\rm photon ~momentum}~,$$
it writes: 
\be
\left \{
\begin{array}{ll}
T^{R2}(\eta)=  \displaystyle c_3~  \frac{ie m^2}{8 \pi^2 f_\pi^3}
~ T^\pm_\rho (\eta) \left[ \frac{1}{D_\pm(s_{+0}) } + \frac{1}{D_\pm(s_{-0})} 
  \right] 
  \\[0.5cm] 
 \displaystyle  T^\pm_\rho(\eta)=\epsilon - \frac{A_\pm}{2}\sin{\delta_P} .
\end{array}
\right .
\label{BoxEta-4}
\ee    
The $D_\pm(s_{\pm 0})$'s denoting the inverse   $\rho^\pm$ propagators; the
$T^{R2}$ contribution, a pure product of symmetry breakings, 
 cancels out when all symmetries are restored.  Finally, the 3 amplitudes
 pieces just defined depend on the HLS parameter $c_3$.

At the chiral point~ $$s=s_{+0}=s_{-0}=0~,$$ the vector meson inverse propagators fulfill
\cite{ExtMod7}  $D_V(0)=-m_V^2$ with~:
\be
\begin{array}{llll}
m_{\rho_\pm}^2=m^2~~,& m_{\rho_0}^2=m^2(1+\xi_3)^2~~,  & 
m_\omg^2=m^2(1+\xi_0)^2~~,&m_\phi^2=m^2 z_V(1+\xi_0)^2~~.
\end{array}
\ee 
where  $m^2=ag^2 f_\pi^2$,  the conditions $\alpha(0)=\beta(0)=0$ being exactly fulfilled.

\section{The $\eta^\prime \to \pi^+ \pi^- \gamma$ Amplitude within BHLS$_2$ }
\label{BoxEtp}  
\indentB
The decay process $\eta^\prime \to \pi^+ \pi^- \gamma$ undergoes a quite similar treatment 
to those performed for the  $\eta \to \pi^+ \pi^- \gamma$ decay  
in the preceding Section and, so, one will avoid duplicating on the $\etp$ amplitude
the comments already stated  on the $\eta$ amplitude.
The three different kinds of contributions to the $\etp$ decay amplitude are~:
 \be
 \displaystyle 
T(\etp) = T^{NR}(\etp)+ T^{R1}(\etp) + T^{R2}(\etp) ~~.
\label{BoxEtp-1}
\ee
The first term, which gathers the $APPP$ and $VPPP$ contributions to the full 
amplitude $T^\etp$, is given by~:
\be
\hspace{-0.8cm}
\begin{array}{ll}
T^{NR}(\etp)=  \displaystyle  - \frac{ie}{4 \pi^2 f_\pi^3} 
  \left[ 1-\frac{3 c_3}{2} \right]  g_{\eta^\prime \pi^+\pi^-\gamma}
 ~~~{\rm with}~~~
 \displaystyle g_{\eta^\prime \pi^+\pi^-\gamma} =
\epsilon^\prime -   \left \{1 -\frac{A_\pm}{2} - \frac{3\lambda_0^2}{4}
\right \} \cos{\delta_P}
\end{array}
\label{BoxEtp-2}
\ee 
and does not depend on $c_1-c_2$. On the other hand, the contributions gathered in 
$T^{R1}(\etp)$  are given by~:
\be
\hspace{-1cm}
\left \{
\begin{array}{ll}
T^{R1}(\etp)=  \displaystyle  c_3  \frac{ie m^2}{8 \pi^2 f_\pi^3} 
  \left[ \frac{T^0_\rho(\eta^\prime)}{D_\rho(s)} +  
  \frac{T^0_\omg(\eta^\prime)}{D_\omg(s)} + \frac{T^0_\phi(\eta^\prime)}{D_\phi(s)} 
   \right] 
  \\[0.5cm]
\displaystyle  T^0_\rho(\etp)=\epsilon^\prime + \frac{2\beta(s)}{z_A} 
   \sin{\delta_P}-  3 \left[1 -\frac{3\lambda_0^2}{4} - \frac{A_\pm}{6}
  +\frac{\alpha(s)}{3} +2 \xi_3\right] \cos{\delta_P}
  \\[0.5cm]
  \displaystyle  T^0_\phi(\etp)= -\left[ \frac{2 \beta(s)}{z_A} \right] \sin{\delta_P}  
  \\[0.5cm]
  \displaystyle  T^0_\omg(\etp) = +~\alpha(s)  \cos{\delta_P}
\end{array}
\right .
\label{BoxEtp-3}
\ee
where, as for the $\eta$ decay,  only the $\rho^0$ term is ${\cal O} (\delta^0=1)$
in breakings. Finally~:
\be
\left \{
\begin{array}{ll}
T^{R2}(\etp)=  \displaystyle  c_3~ \frac{ie m^2}{8 \pi^2 f_\pi^3}
 ~T^\pm_\rho (\etp) \left[ \frac{1}{D_\pm(s_{+0}) } + \frac{1}{D_\pm(s_{-0})} 
  \right] 
  \\[0.5cm] 
 \displaystyle  T^\pm_\rho(\etp)=\epsilon^\prime + \frac{A_\pm}{2}\cos{\delta_P} .
\end{array}
\right .
\label{BoxEtp-4}
\ee 
which is purely ${\cal O} (\delta)$.

 The $\omg$ contribution in the $\etp$ decay must be visible 
in high statistics data samples (like \cite{BESIII_BOX_OK})
and worth to compare with its lineshape in the 
$e^+e^- \ra \pi^+ \pi^-$ annihilation. 
Regarding the $\phi$ contribution, it is  somewhat outside of
the allowed phase space -- by $\simeq 60$ MeV. Finally, the influence 
of higher vector mesons, especially the first radial excitation
$\rho^\prime$, are outside the
HLS scope; global fit properties may reveal their actual influence,
w.r.t. the broken HLS context.

\section{BHLS$_2$ and the WZW Box Anomalies}
\label{WZW}
\indent \indent 
Traditionally, the amplitudes associated with the
box anomalies are derived from the Wess-Zumino-Witten
(WZW) Lagrangian \cite{WZ,Witten}~:
\be
\displaystyle 
{\cal L}_{WZW}= -i\frac{N_c e}{3 \pi^2 f_\pi^3}
~\epsilon^{\mu \nu \alpha \beta} A_\mu
\mathrm{Tr} \left [Q \pa_\nu P \pa_\alpha   P \pa_\beta   P\right ]
\,.
\label{WZW-1}
\ee
\noindent where $P$ is the bare pseudoscalar meson $U(3)$ matrix.
This Lagrangian differs from the anomalous
$APPP$ Lagrangian piece of the HLS model
(see Equation (\ref{CC1}))  by the factor
$$\left[1-\frac{3}{4}(c_1-c_2+c_4)\right ]$$.
 
The BHLS$_2$ $\eta/\etp$ decay amplitudes just defined are 
expected to coincide with their WZW analogs  at the chiral  point, 
where the HLS  $c_i$'s dependencies of the decay amplitudes should 
cancel out. Their expressions at the 
chiral point  ($s=s_{+0}=s_{-0}=0$) are given by\footnote{The coupling 
$\pi^0 \pi^+\pi^-\gam$ is involved
in the $e^+e^- \to \pi^0 \pi^+\pi^-$ annihilation \cite{ExtMod7,ExtMod8}.}~:
\be
\left \{
\begin{array}{ll}
\displaystyle 
T(\eta)=  \displaystyle - \frac{ie}{4\pi^2 f_\pi^3} 
\left [
\epsilon + \left \{1 -\frac{A_\pm}{2} - \frac{3\lambda_0^2}{4}
\right \}  \sin{\delta_P}
\right ]
\,, \\[0.5cm]
T(\etp)=  \displaystyle  - \frac{ie}{4 \pi^2 f_\pi^3} 
\left [
\epsilon^\prime -   \left \{1 -\frac{A_\pm}{2} - \frac{3\lambda_0^2}{4}
\right \} \cos{\delta_P} \right]
\,, \\[0.5cm]
T(\pi^0)=  \displaystyle + \frac{ie}{4\pi^2 f_\pi^3} 
 \left [
 \left \{ 1 -\frac{A_\pm}{2} - \frac{\lambda_0^2}{3}\right \}
 - \epsilon  \sin{\delta_P} + \epsilon^\prime  \cos{\delta_P}\right ]
 \,.
\end{array}
\right .
\label{WZW-2}
\ee
and coincide with those which can be directly derived 
from the WZW Lagrangian Equation (\ref{WZW-1}) after applying the breaking
procedures reminded in the Appendices. 

\section{$\eta/\eta^\prime$ Radiative Decays~: The BHLS$_2$ Dipion Mass  Spectra}
\label{MassSpectra-1}
\indentB
The amplitudes $T(\eta)$ and $T(\etp)$ allowing to describe -- within the
full EBHLS$_2$ framework \cite{ExtMod7,ExtMod8} -- the dipion mass 
spectra observed in  the $\eta/\eta^\prime$ radiative decays 
have been derived
in resp. Sections \ref{BoxEta} and \ref{BoxEtp}; both
should be multiplied by the function\footnote{The notations $\epsilon(\gam,q)$
for the photon polarization vector, $p^\pm$ 
 and  $q$  for the pion and photon momenta are generally understood. } 
$F(s,s_{0+})$ (see Equation (\ref{BoxTensor})). 
The differential decay widths can be written~:
\be
\displaystyle   \frac{d^2\Gamma_X}{ds ds_{0+}}= \frac{1}{(2 \pi)^3} 
\frac{1}{32 M_X^3} |T_X~F(s,s_{0+})|^2~~~,~X=\eta,\eta^\prime
\label{width-diff}
 \ee 
 in terms of resp. $s$, the ($\pi^+ \pi^-$) and $s_{0+}$, the ($\pi^+ \gam$) pair
  invariant masses  squared of the $\eta/\eta^\prime$ decay products. The accessible
  invariant mass spectra being functions of only $s$, this expression
  should be integrated over $s_{0+}$~: 
\be
\displaystyle   \frac{d\Gamma_X}{ds}= \frac{1}{(2 \pi)^3} 
\frac{1}{32 M_X^3} \int_{s_{min}}^{s_{max}}|T_X~F(s,s_{0+})|^2 ds_{0+} ~~~,~X=\eta,\eta^\prime
\label{width}
 \ee
where~:
\be
  \begin{array}{lll}
  \displaystyle  s_{min/max}= \frac{M_X^2+2 m_\pi^2-s}{2} \mp p_\pi  \frac{M_X^2-s}{\sqrt{s}}
& {\rm and~~}   \displaystyle  p_\pi=\frac{\sqrt{s-4 m_\pi^2}}{2}~~.
\end{array}
\label{int_limits}
 \ee
 
 Both  amplitudes $T(\eta)$ and $T(\etp)$,  generically referred to as 
 $T_X$, can be written~:
 \be
 \begin{array}{lll}
T_X(s,s_{0+})=  \displaystyle  R_X(s)+C_X G(s,s_{0+})
& {\rm with~~}
G (s,s_{0+})=   \displaystyle  \frac{1}{D_\rho(s_{0-})} +  \frac{1}{D_\rho(s_{0+})} ~~,
 \end{array}
 \label{Amp-redefinition}
 \ee
having defined $s_{0\pm}=(q + p^\pm)^2$  related by~:
$$s_{0-}-m_\pi^2=(M_X^2-s)-(s_{0+}-m_\pi^2)~~.$$
\noindent $R_X(s)$ collects  the contributions previously 
named $T^{NR}(X)$ and $T^{R1}(X)$ and is (by far) the dominant term, 
whereas\footnote{$C_X$ can be read off the relevant expressions
for $T^{R2}(X)$ given in Sections \ref{BoxEta} and \ref{BoxEtp}.}
 $T^{R2}(X) = C_X G(s,s_{0+})$ is only ${\cal O}(\delta)$ in breakings.
 
On the other hand, the $[F(s,s_{0+})]^2$ factor in Equation (\ref{width}) is~:
\be
  \begin{array}{lll}
    \displaystyle  [F(s,s_{0+})]^2=
 \frac{s}{4} (s_{0+}-m_\pi^2)(s_{0-}-m_\pi^2)
   - \frac{m_\pi^2}{4} (M_X^2-s)^2 
\end{array}
\label{tensor}
 \ee
 and can be solely expressed in terms of $s$ and $s_{0+}$ to perform
 the integration shown in Equation (\ref{width}). This leads to
 {\it predefine}  within the fitting code the following integrals~:
 \be
\hspace{-1.cm}
  \begin{array}{ll}
\displaystyle  I_1(s)= \int_{s_{min}}^{s_{max}}|F(s,s_{0+})|^2 ds_{0+}~~,&
\displaystyle  I_2(s)= 
\int_{s_{min}}^{s_{max}}|F(s,s_{0+})|^2|G (s,s_{0+})|^2ds_{0+} \\[0.5cm]
\displaystyle  I_3(s)= \int_{s_{min}}^{s_{max}}|F(s,s_{0+})|^2 
{\bf Re} \left [ G (s,s_{0+}) \right ]ds_{0+},&
\displaystyle  I_4(s)= \int_{s_{min}}^{s_{max}}|F(s,s_{0+})|^2 
 {\bf Im} \left [ G (s,s_{0+}) \right ] ds_{0+}
 \end{array}
 \label{predefine}
 \ee
Actually,  $I_1(s)$ can be integrated in closed form~:
 \be
 \displaystyle  I_1(s)=  \frac{(M_X^2-s)^3}{3}\frac{p_\pi^3}{\sqrt{s}}~
 \label{standard_term}
 \ee
with $p_\pi$ given in Equations (\ref{int_limits}). The 3 other functions
should be integrated numerically within the iterative procedure context
already running to
address the $e^+ e^- \to \pi^+ \pi^- \pi^0$ annihilation data
within the BHLS \cite{ExtMod3} or BHLS$_2$ \cite{ExtMod7,ExtMod8}  frameworks.
One then gets~:
 \be
\hspace{-0.5cm}
  \begin{array}{llll}
\displaystyle   \frac{d\Gamma_X}{ds}= \frac{1}{(2 \pi)^3} 
\frac{1}{32M_X^3} \left [|R_X (s)|^2~I_1(s)+C_X^2
I_2(s) + 2 C_X ~({\bf Re} \left [ R_X(s) \right ]~I_3(s)
+  {\bf Im} \left [R_X(s) \right ]~I_4(s))~\right ]
\end{array}
 \label{width2}
 \ee
In  the BHLS$_2$ approach, 
only leading order terms in the breaking parameters ${\cal O}(\delta)$
(as the $C_X$ term) are addressed
and then  terms of order ${\cal O}(\delta^2)$ -- like the $C_X^2 $ 
contribution -- can be neglected. 

The $ I_1(s)$ term in Equation 
(\ref{width2})  can be rewritten, for subsequent use in the text~:
 \be
  \begin{array}{llll}
\displaystyle   \frac{d\widetilde{\Gamma}_X}{ds}=
\Gamma_0(s) |R_X (s)|^2~, 
& \displaystyle {\rm with~~}\Gamma_0(s) =  \frac{s (M_X^2-s)^3 [\sigma_\pi(s)]^3}
{3 \cdot 2^{11} \pi^3 M_X ^3}~
&\displaystyle {\rm and~~}\sigma_\pi(s) =\sqrt{1-\frac{4 m_\pi^2}{s}}~~.
\end{array}
\label{width3}
\ee
\section{Pion form factor in the $\eta/\etp$ Radiative Decays}
\label{FSI}
\indentB
The study in \cite{Stollenwerk_BOX_th}, also referred to hereafter as SHKMW,  
has placed a valuable emphasis on the connection
between the pion vector form factor $F_\pi(s)$ -- as it comes
out of the $e^+e^- \to \pi^+ \pi^-$ annihilation process --
 and the dipion 
spectra from the $\eta/\etp \to \pi^+ \pi^- \gam$ radiative decays.
Further works have followed -- see,  for instance, 
\cite{Hanhart_BOX_th,Kubis:2015sga,Dai:2017tew,Holz:2022hwz,Holz:2022Err} for further references --
generally motivated by a better understanding of the $\eta$ and $\etp$
meson properties regarding their contributions to  the light-by-light (LbL) 
fraction of the muon anomalous magnetic moment $a_\mu$. 
\vspace{0.5cm}

{\bf i)} It is worthwhile to briefly outline how  this connection is 
  established \cite{Stollenwerk_BOX_th}. The pion
  vector form factor $F_\pi(s)$ and the  
 $P$--wave $\pi^+ \pi^-$ scattering amplitude $T_{\pi\pi}(s)$
are related by~:
\be
\displaystyle {\rm Im } \left [ F_\pi(s) \right ]= 
\sigma_\pi(s) \left [T_{\pi\pi}(s) \right ]^* F_\pi(s) \Theta(s-4 m_\pi^2)~,  
\label{width4}
\ee
valid along the energy region where the $\pi^+ \pi^-$ scattering is {\it elastic};
 $\sigma_\pi(s)$ has been defined just above.
 Therefore,  in this energy region,  the  pion vector form factor $F_\pi(s)$ 
and  the elastic scattering amplitude $T_{\pi\pi}(s)$ should carry equal phases. 
The Heaviside function indicates that $F_\pi(s)$
is real below the $2\pi$ threshold; the 
first {\it significant} inelastic
channel being $\omg \pi$, the validity range of Equation (\ref{width4}) practically
extends up to $\simeq 922$ MeV, much above the $\eta$ mass and slightly below 
the $\etp$ mass (by only 36 MeV). Stated otherwise, the phase-equality 
property holds over almost the whole HLS energy range  of 
validity ($\sqrt{s}\leq 1.05$ GeV). 

On the other hand, assuming the $\pi^+\pi^-$ scattering is elastic for all $s \ge 4 m_\pi^2$,
 the $P$--wave amplitude $T_{\pi\pi}(s)$ writes~:
 \be
\displaystyle T_{\pi\pi}(s) =\frac{\sin{\delta_{11}(s)} e^{i\delta_{11}(s)}}
				{\sigma_\pi(s)}
 \label{p-wave}
 \ee
in terms of the $P$--wave phase shift $\delta_{11}(s)$ and 
the solution to Equation (\ref{width4}) can be expressed  
in terms of the  Omn\`es function $\Omega(s)$ by~:
\be 
\displaystyle
F_\pi(s)= K(s) \Omega(s)~~,~{\rm where}~~~
  \Omega(s)=\exp{ \left(
 \frac{s}{\pi} \int_{4 m_\pi^2}^\infty \frac{dz}{z}
 \frac{\delta_{11}(z)}{z-s-i\epsilon}
\right )}  ~~,
\label{omnes}
\ee
$K(s)$ being some appropriate real--analytic function,  required
to be free of singularities over
 the physical region $s\ge 4 m_\pi^2$. This expression intends to factor
out the non--perturbative contribution to $F_\pi(s)$ which is  contained
in the $ \Omega(s)$ function, and so the remaining part $K(s)$ (perturbative in the sense of ChPT, at low energy) is expected to 
behave  smoothly and  be well approximable by a polynomial \cite{Stollenwerk_BOX_th}
along our region of interest (up to $\simeq m_\phi$). This smooth function $K(s)$ is process dependent, whereas $\Omega(s)$, being determined by the $\pi\pi$ final state re-scattering phase shifts represents a more universal part of the vector form factor.
It is shown in \cite{Hanhart_BOX_th} that   a first-degree polynomial
$K(s)=1+\alpha_\Omega s$ allows to reach a nice (linear) correlation 
up to $s\simeq 1$ GeV$^{2}$ between
the dipion spectrum from Belle \cite{Belle} and the $\Omega(s)$ functions
derived from the phase shift data from \cite{Garcia-Martin:2011iqs}, see also \cite{Leutwyler:2002hm}\cite{Colangelo_BOX_th}\cite{Colangelo:2006cd};
 a value  $\alpha_\Omega  \simeq 0.1$ GeV$^{-2}$  can be inferred from 
Figure 1 in \cite{Hanhart_BOX_th}. The deterioration of the linear 
behavior above $s \simeq m_\phi^2$ is, actually, not unexpected  
because of  rising inelasticities 
and of the high mass vector meson influence.

\vspace{0.5cm}

{\bf ii)} Assuming   the pion pair emerging from the $\eta/\etp$  radiative decays 
is  purely  Isospin 1 and $P$--wave \cite{WASA_BOX_OK,BESIII_BOX_OK},
its amplitude should carry the same 
analytic properties  than $F_\pi(s)$, {\it i.e.} they may only differ
by a real--analytic function, free of right-hand side singularities. Reference 
\cite{Stollenwerk_BOX_th}
thus proposes  to write the differential dipion spectra ~:
 \be
  \begin{array}{llll}
\displaystyle   \frac{d\overline{\Gamma}_X}{ds}=
\Gamma_0(s) |A_X P_X(s) F_\pi(s)|^2~, ~~(X=\eta/\etp)~,
\end{array}
\label{width5}
\ee
where $\Gamma_0(s)$ has been already defined in Equations (\ref{width3}) and
 the $A_X$'s being  appropriate normalization constants. The $P_X(s)$ functions ($P_X(0)=1$) are remaining correction factors specific of the $\eta$ and 
 $\etp$ radiative decays (in the present case, but are more generally dependent on higher inelastic effects existing at any $\pi\pi$ production vertex) which  could both be analyzed within the  Extended ChPT  context \cite{Kaiser_2000,leutw} (see also \cite{Bijnens_BOX_th}) and 
  are  free of right-hand side singularities.

As just argued regarding the pion vector form factor and its $K(s)$ factor,
 the $P_X(s)$ functions should satisfactorily be approximated by low
degree polynomials   \cite{Stollenwerk_BOX_th}. This is what is shown by
 the down most panel in Figure 1 of \cite{Hanhart_BOX_th} which, moreover,
 indicates that $P_\eta(s)=P_\etp(s)$ should likely hold.
 Of course, procedures to complement this approach by symmetry breaking 
 effects have also to be invoked, prominently the $\rho^0-\omg$ mixing 
 for the $\etp$ decay process -- but not only.

\vspace{0.5cm}       

{\bf iii)} The issue is now to relate $d\overline{\Gamma}_X$ 
(Equation (\ref{width5})) and $d\widetilde{\Gamma}_X$ (Equation (\ref{width3}))
within the HLS framework {\it when no breaking is at work}. Equivalently,
this turns out to check whether the $R_X(s)$'s and $F_\pi(s)$ (can)
carry the same  phase in this case.

Let us consider the pion vector form factor $F_\pi(s)$ as given
in \cite{ExtMod7}, discarding  terms of order $ {\cal O}(\delta)$
or higher in breaking parameters;  keeping only tree contributions
(loop corrections, like 
 the $\rho^0-\gam$ transition amplitude, are counted as $ {\cal O}(\delta)$)
 and dropping out the ${\cal L}_{p^4}$  contributions, one derives
($m^2=a g^2 f_\pi^2$, the unbroken $\rho^0$ HK mass)~:
\be 
\displaystyle
F_\pi(s)=  \left( 1 - \frac{a}{2} 
\left  [ 1 + \frac{m^2}{D_\rho(s)} \right ]
\right )  +  {\cal O}(\delta) ~~.
\label{omnes-2}
\ee
Similarly, the $R_X(s)$ functions in Equation (\ref{width3}) reduce to~: 
\be
\begin{array}{lll}
\displaystyle
R_\eta= - \frac{ie \sin{\delta_P}}{4 \pi^2 f_\pi^3}
\left (
1-\frac{3}{2} c_3 \left  [ 1 + \frac{m^2}{D_\rho(s)} \right ]
\right )~~,
&\displaystyle
R_\etp= + \frac{ie \cos{\delta_P}}{4 \pi^2 f_\pi^3}
\left (
1-\frac{3}{2} c_3 \left  [ 1 + \frac{m^2}{D_\rho(s)} \right ]
\right )
\end{array}
\label{omnes-3}
\ee
up to terms of ${\cal O}(\delta)$ in breaking parameters,

\vspace{0.5cm}

These Equations lead us to define a {\it no--breaking} reference by requiring~:
  
{\bf 1/} The holding of the  Vector Meson Dominance assumption which implies 
$a \equiv a_{VMD}=2$  within the generic HLS model  \cite{HLSOrigin,HLSRef}.
It is worthwhile reminding here (see Section 2 in \cite{ExtMod8} for details) that the
HLS parameter $a$ is not reachable by fit, once the BKY breaking 
(see Appendix \ref{HLS-BKY})  is at work; 
indeed, all Lagrangian terms of
interest for our physics depend on the product $a^\prime =a(1+\Sigma_V)$ and 
not on each
of these parameters separately; therefore one can freely fix $a=2$ and, then, the  
term $\delta a= a_{VMD} \Sigma_V$ is clearly\footnote{In the course of the fitting
procedure, it is as appropriate to either choose to fit $a$, fixing $\Sigma_V=0$
or fix $a$ and fit $\Sigma_V$; we choosed the first option. }
  ${\cal O}(\delta)$.

{\bf 2/} The universality of the $\rho$ phase implies that
$R_\eta(s)$, $R_\etp(s)$ and $F_\pi(s)$ share the same phase 
and, therefore, it requires the existence of an "unbroken" value 
 for $c_3$~: Indeed,
imposing $c_3^{ref}=2/3$ beside  $a_{VMD}=2$, one can derive
a satisfactory no--breaking reference as, one obtains~:
\be
\begin{array}{lll}
\displaystyle 
F_\pi(s) = - \frac{m^2}{D_\rho(s)}
&
\displaystyle
R_\eta= + \frac{ie \sin{\delta_P}}{4 \pi^2 f_\pi^3}
 \frac{m^2}{D_\rho(s)} ~~,
&\displaystyle
R_\etp= - \frac{ie \cos{\delta_P}}{4 \pi^2 f_\pi^3}
\frac{m^2}{D_\rho(s)} ~~.
\end{array}
\label{omnes-4}
\ee 
which should be complemented by ${\cal O}(\delta)$ contributions to account
for real data.

The issue becomes whether the values returned for $a$ and $c_3$ 
from fits to the (real) data differ little enough from
$a_{VMD}$ and $c_3^{ref}$  that  their differences
can be considered ${\cal O}(\delta)$ effects. 
For this purpose, one can refer to the latest  published
BHLS$_2$  standard fit results
collected in Table 10 of \cite{ExtMod8}, in particular, one finds~:
\begin{itemize}
\item 
$a=1.766 \pm 0.001$  which shows a  deviation 
$\delta a=0.244 $ from  $a_{VMD}=2$ corresponding to having $\Sigma_V=0.122$, 
\item 
$c_3= 0.742 \pm 0.003$ which deviates by $\delta c_3= 0.076$
from  $c_3^{ref}=0.667$,
\end{itemize}
focusing on  the favored solution $A_-$ \cite{ExtMod8} to the Kroll 
conditions (see Section \ref{Kroll-cnd}) -- the $A_+$
solution actually provides similar  values. Thus, $\delta a$ and $\delta c_3$
look small enough to be viewed as departures from  resp.
$a_{VMD}$ and $c_3^{ref}$ and treated as ${\cal O}(\delta)$ corrections, 
on the same footing as the manifest breaking parameters. 
To our knowledge, it is the first time that an identified physics condition can propose a constraint
on one of the FKTUY \cite{FKTUY} parameters, namely\footnote{Actually,
another condition comes out from the data in analyses performed within
the HLS context~:  $c_3=c_4$. } $c_3$.
\begin{figure}[!ht]
\begin{center}
{\includegraphics[height=12.5cm]{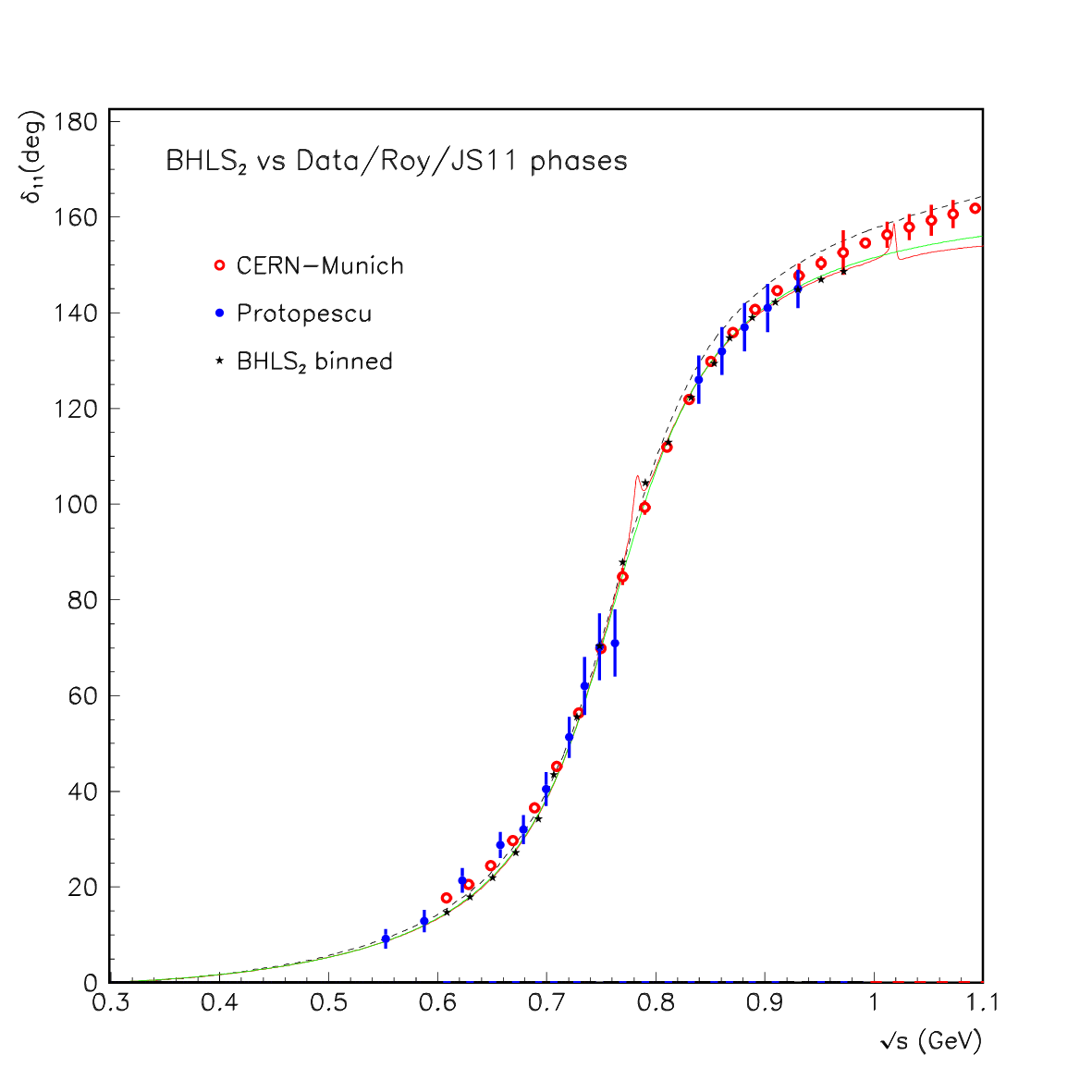}}
\end{center}
\begin{center}
\caption{\protect\label{Fig:phase_shift} The $\delta_{11}$ phase-shift
plotted as function of $\sqrt{s}$ . Besides the data points of 
\cite{Ochs1,Protopescu}, the dashed black curve is the solution
to the Roy Equations \cite{RoyEq}, the green full line shows 
the phase reconstructed in \cite{Fred11} and red full line
the BHLS$_2$ phase-shift exhibiting the $\omg$ and $\phi$ signals.
 The black stars show the smeared BHLS$_2$ spectrum (e.g. the red curve).}
\end{center}
\end{figure}

{\bf iv)} From what has been just argued, it is clear that, within the BHLS$_2$  
context, 
the $\eta/\etp \ra \pi^+\pi^- \gam$ decay amplitudes $T_X(s)$ reported in Sections
\ref{BoxEta} and \ref{BoxEtp} above can actually be written~:
\be
T_X(s)=B_X F_\pi(s) + {\cal O}(\delta) {\rm ~~~,~~ X=\eta/\etp~~,}
\label{omnes-5}
\ee
the $B_X$'s being definite constants depending on the breaking parameters.
$F_\pi(s)$ contains already manifest breaking terms like the
$\omg$ and $\phi$ signals with, however, different weights from
their analogs in the $T_X(s)$ amplitudes\footnote{For instance,
BHLS$_2$ predicts that the coupling ratio   
$\omg \pi \pi$  to  $\rho^0 \pi \pi$ is 3 times smaller in the $\etp$ radiative 
decay than in the pion vector form factor.}.

On the other hand, as shown in  \cite{ExtMod7}, yielding a fair
description of the data samples for $|F_\pi(s)|$ (see Figure 2 
and Table 3 in \cite{ExtMod7}),  BHLS$_2$  also leads  to a fair
account of the phase-shift $\delta_{11}(s)$ over its whole range of validity
 {\it without involving
any phase-shift data sample in its derivation}. This is illustrated 
 by\footnote{Reprinted from Figure 10 in \cite{ExtMod7}.} 
  Figure \ref{Fig:phase_shift} which reflects the fair accord reached   by
 the BHLS$_2$ prediction with the phase  derived from 
 the Roy Equations \cite{RoyEq} or the pion form factor phase of Reference \cite{Fred11} 
on the one hand,  and the experimental phase shift data  from \cite{Ochs1,Protopescu} 
on the other hand.
Moreover, the same BHLS$_2$ spectrum  was smeared over 10 MeV bins -- to mimic the Cern-Munich spectrum \cite{Ochs1} -- (black star symbols), to clearly show that the $\omg$ and $\phi$ signals cannot be manifestly observed in the existing data.

All this leads to the conclusion that the SHKMW modification \cite{Stollenwerk_BOX_th}
shown in Equation (\ref{width5})~:
$$F_\pi(s) \rightarrow A_X P_X(s)F_\pi(s)$$
also applies in the global BHLS$_2$ context. In
this case, this turns out to perform the change~:
$$ T_X(s) \Longrightarrow H_X P_X(s) T_X(s)$$
when using the amplitudes constructed in Sections \ref{BoxEta} and \ref{BoxEtp}. 
Our notations are connected with those in Reference \cite{Stollenwerk_BOX_th} 
by writing these\footnote{Actually, to be formally exact, 
Reference \cite{Stollenwerk_BOX_th}
writes $A=A_0 (1+\delta)$ for the $\eta$ meson decay, and 
$A^\prime=A_0^\prime (1+\delta^\prime)$ for the $\etp$ meson, as can be read 
around their Relations (9).}~: 
\be
A_X=A_X^0 H_X~~~,~~ H_X \equiv 1+\delta_X~~,~~X=\eta,\etp
\label{h_x} 
\ee
\noindent as the $A_X^0$ factors  are already accounted 
for in the $T_X$ amplitudes derived from
the BHLS$_2$ Lagrangian as shown below.

Then, the global character of the BHLS$_2$ fitting context\footnote{In this case,
its Reference set of data samples ${\cal H}_R$  \cite{ExtMod7,ExtMod8},
which already includes most of the existing pion form factor data samples will be supplemented with  the $\eta/\etp$ dipion spectra.}, ensures that 
the non-perturbative effects are suitably accounted for as reflected by 
Figure \ref{Fig:phase_shift}. 

From now on, 
 the $P_X(s)$ functions are chosen polynomials of the lowest possible degree
consistent with a satisfactory fitting.
Being   beyond the BHLS$_2$ scope, these functions are 
supplemented within the fit procedure  by performing the change~:
\be
\displaystyle  T_X(s) \Longrightarrow H_X P_X(s) T_X(s) 
~~~,~{\rm with~~}  P_X(0)=1
~~~,~X=\eta,\etp
\label{Polynom}
\ee
in Equation (\ref{width2}) above.   Practically, each term in the right-hand 
side of Equation (\ref{width2}) gets a factor of $|H_X P_X(s)|^2$, the coefficients
of which have to be derived by the global fit where the $[C_X]^2$ term can be
discarded as it is manifestly ${\cal O}(\delta^2)$.

\section{Fits of the $\eta/\etp$ Radiative Decay Spectra  within BHLS$_2$}
\label{HLS_eta_etp_Decays}
\indentB
 The reference set of data samples ${\cal H}_R$ included within the BHLS$_2$ 
framework has been presented several times and  recently in  
\cite{ExtMod7,ExtMod8}; it covers the
six $e^+ e^-$ annihilation channels to $\pi^+ \pi^-$, $K^+ K^-$, $K_L K_S$, 
$\pi^+ \pi^- \pi^0$, $\pi^0 \gam$,  $\eta \gam$, some more decay
widths (in particular $\pi^0/\eta/\etp \ra \gam \gam$) and, finally,
the dipion mass  spectrum in the $\tau \ra \pi \pi \nu$ decay.
These represent already the largest set of data 
(altogether 1366 data points) successfully submitted to a global fit,
as reflected by Table 9 in \cite{ExtMod8}; they will not be
discussed here anymore.  It is nevertheless relevant to remember that 
 ${\cal H}_R$  encompasses almost all existing samples
except  for the recent CMD-3 dipion data as already argued in footnote ${\ref{CMD3-pi}}$,
the KLOE08 \cite{KLOE08}, Babar \cite{BaBar,BaBar2} 
and the recent SND \cite{SND2020} dipion  spectra because of 
the  strong tension they exhibit with respect to the rest of the   
 (more than 60) ${\cal H}_R$  samples. This  issue has been
thoroughly reexamined in \cite{ExtMod8}.

The present study aims to include also the dipion spectra measured in the $\eta/\etp$ radiative decays  within the global BHLS$_2$ framework. 
  However, it is certainly cautious to avoid using {\it simultaneously} 
  the $\eta/\etp$  dipion spectra and the $\pi^+ \pi^- \pi^0$ annihilation data within global fits
 as long as a specific study has not assessed some clear statement about 
 process-specific corrections (accounted for by a polynomial)  in the latter channel\footnote{The studies \cite{Hoferichter:2019mqg,Hoferichter:2023bjm} and the fit results reported in 
 \cite{ExtMod7,ExtMod8}
 may as well indicate that polynomial corrections are small 
 or effectively absorbed in the parameter values returned by the fits.
 Anyway, this certainly deserves a devoted work \cite{Stamen:2022eda}.} and data.\\

 For the general model dependence of BHLS/BHLS$_2$ models and also the magnitude of possible biases in the fits, we rely on our previous studies (notably \cite{ExtMod7} and also \cite{ExtMod5}) that demonstrated those are limited with respect to other sources of systematics.\\
 More recently, our 2022 publication \cite{ExtMod8} (see for example Section 11.3 in there), which extends BHLS$_2$ and introduced a comprehensive treatment of the $[\pi^0,\eta,\eta']$ system includes also a detailed discussion on model dependencies in the new framework (partly by studying model variants)\footnote{On the $a_\mu$ determination question, since for this purpose we always end using well constrained, high global and partial probability fits which use \textit{all} the experimental data, this guarantees the fitted pion form factor is very close to the data, and hence that the $a_\mu$ is also very close to the integral model-independent determinations from other groups. This remains true when swapping KLOE and BaBar $\pi\pi$ data in the fit.}. This 2022 study, which in fact prefigured and allowed the present work, reached again the conclusion that model dependence exists but is not a strong source of systematics. It confirmed previous studies which explored systematic differences between BHLS and BHLS$_2$ type of models (and various subvariants), see for example Section 17 in \cite{ExtMod7}.\\
 We explore below eventual new sources of (model and others) systematics, again by studying the fit results dependencies when using model and hypotheses variations (in particular for the $P_X$ polynomials, but not only), in the same spirit than in our previous works.
 
 It is worthwhile to stress that all the  published
 dipion spectra  of the $\eta /\etp \ra \pi^+ \pi^- \gamma$ 
 decays carry an arbitrary normalization; so, {\it they only provide the spectrum lineshapes measured by the various experiments}.  It follows from this peculiarity that they allow to fit {\it only} the $P_X(s)$ polynomials
 and they are totally insensitive to the $H_X$ parameter values; 
 this issue will be addressed   by performing global fits where 
 the corresponding partial widths --
 taken from the Review of Particle Properties (RPP) \cite{RPP2022} --  
 are  also considered inside the fitting procedure.

\subsection{Available Dipion Spectra from the $\eta/\etp \ra \pi^+ \pi^- \gam$ Decays}
\label{exp-spectra}
\indentB
Measurements of the dipion spectrum in the $\etp \ra \pi^+ \pi^- \gam$ decay
started long ago -- as early as 1975 \cite{Grigorian_BOX} -- and  
several  experiments have collected samples of various (but low)
 statistics motivated by the $\simeq 20$ MeV shift reported for the $\rho^0$
 peak location compared to its value in $e^+e^- \ra \pi^+ \pi^-$ annihilations~:  
JADE \cite{JADE_BOX}, CELLO \cite{CELLO_BOX_OK}, TASSO \cite{TASSO_BOX_OK},
PLUTO \cite{PLUTO_BOX_OK}, TPC-2$\gamma$ \cite{TPC_2gamma_BOX},
ARGUS \cite{ARGUS_BOX_OK}, Lepton F \cite{LeptonF_BOX};  
the Crystal Barrel Collaboration published in 1997 the most precise 
spectrum \cite{Crystal_Barrel_BOX_OK}  carrying  7400 events.
The breakthrough has come from the BESIII Collaboration
\cite{BESIII_BOX_OK} which published a  970\,,000 event spectrum 
by 2017. 

The formerly collected samples have been examined and their
behavior is briefly reported below. Dealing with the uncertainty 
information provided with these $\etp$ samples is generally straightforward,
except for the BESIII dipion spectrum \cite{BESIII_BOX_OK}  for which
a spectrum for the energy resolution is provided. It is accounted for
by replacing within the minimization procedure the  genuine model function
value  by that of   its convolution with the resolution function, 
assuming the provided resolutions are the standard deviations
of Gaussians; the net effect of the BESIII energy resolution information
deserves to be shown (see below).  

The BESIII 
data \cite{KLOE_BOX_OK} are provided as two 112 data point spectra, 
the former giving the numbers of $\etp$ event candidates in 10 MeV bins ($N_{evt}^i$), 
 the latter the estimated numbers of background events ($N_{bkg}^i$) 
 within the same bins.  One has  provided our global fitting code with the
$N_{signal}^i= N_{evt}^i-N_{bkg}^i$ spectrum; we have  assumed the original
 distributions Poissonian and fully correlated by attributing to
$N_{signal}^i$  an uncertainty  $\sigma_i=\sqrt{N_{evt}^i}+ \sqrt{N_{bkg}^i}$;
it is shown below that these specific assumptions allow a fair dealing with 
the BESIII spectrum \cite{BESIII_BOX_OK}.

On the other hand, the reported dipion spectrum observed in the parent
$\eta \ra \pi^+ \pi^- \gam$ decay
has undergone much less measurements. Beside  former 
spectra\footnote{The numerical content of these spectra can only
be derived from the paper Figures.} from Layter {\it et al.}
\cite{Layter_BOX} and Gormley  {\it et al.} \cite{Gormley_BOX_OK},
 WASA-at-COSY reported for a 14\,,000 event spectrum \cite{WASA_BOX_OK} 
whereas the KLOE/KLOE2 Collaboration has collected a 205\,,000 
event spectrum \cite{KLOE_BOX_OK}; it should be noted that the WASA dipion 
spectrum is given  with only statistical errors.

\vspace{0.5cm}

It is worth stressing again that the normalization of all these spectra  
being arbitrary, the theoretical (absolute) distribution scales provided by 
the BHLS$_2$ Lagrangian are lost when normalizing to the specific
scale  of each  data set when fitting; stated otherwise these data samples 
only allow to address the fit of the $P_X(s)$ functions ($X=\eta,\etp$)
and {\it not} of the $H_X$ constants which are cancelled out
when normalizing the model functions to the experimental spectra.   

\subsection{$\eta/\etp$  Experimental Spectra~: Fits in Isolation}
\label{fits-1}
\indentB
The first exercise   is thus to explore the degree issue
for the $P_X(s)$ polynomials and so, does not need to deal with 
complications due to keeping the constant $H_X$ within the fit procedure. 
Therefore, fits have been performed,
 supplementing the Reference data set of samples  ${\cal H}_R$ 
by either of the  experimental $\etp$ or $\eta$ spectra. In this Section, 
one only reports on using the $A_-$ BHLS$_2$ variant\footnote{Nevertheless, 
the most relevant results obtained 
using the $A_+$  BHLS$_2$  variant are summarized in the following Subsections.} 
\cite{ExtMod8} which will be our working BHLS$_2$  version.

Regarding the $P_\etp(s)$ 
polynomial, the results given in the Table just below\footnote{The fits which
provide these results have been performed with our Reference set amputated from the
$ e^+e^- \ra \pi^+ \pi^- \pi^0$ annihilation data. The number of BESIII data points and
the total number of fitted  data points are given by the $N$ values within parentheses.} 
 focus on only the BESIII $\etp$ sample (112 data points) \cite{BESIII_BOX_OK}; 
indeed, because of their statistics,  all the other $\etp$ dipion spectra, 
including the Crystal Barrel one \cite{Crystal_Barrel_BOX_OK}, do not exhibit
 any clear sensitivity to the $P_\etp$  degree and may easily accommodate 
 $P_\etp\equiv 1$.
\begin{center}
\begin{tabular}{ ||c||c|c|c| } 
 \hline
\hhhv 
$P_\etp(s)$  degree & 1 &   2 &  3\\ 
 \hline
\hhhvw $\chi^2_{BESIII}~(N=112) $ 	& 160 	   &  99    	& 98 \\ 
\hhhvw $\chi^2_{TOT}~(N=1187)$ 		& 1167     & 1097   	& 1096 \\ 
\hhhvw Probability			& 44.7\%   & 90.6\% 	& 90.8 \% \\ 
 \hline
\end{tabular}
\end{center}
This clearly points out that, thanks to the statistics reached by the
BESIII Collaboration, the first 
degree for $P_\etp(s)$ can be  excluded ($<\chi^2>= 1.43$) and the third degree  
is obviously useless. 

Regarding the $\eta$ data,  complementing ${\cal H}_R$ with
 the KLOE/KLOE2 sample (59 data points) \cite{KLOE_BOX_OK} alone or together with 
 the WASA one (37 data points) \cite{WASA_BOX_OK},
 the picture returned by the fits is  much less conclusive as a first-degree $P_\eta(s)$  
 provides\footnote{\label{wasa}
 Note that $\chi^2$(WASA) is always overestimated because
 of an incomplete reported experimental error  information.}
$\chi^2(KLOE/KLOE2)=55$ and $\chi^2$(WASA)$=45$, and a second degree $P_\eta(s)$ 
yields $\chi^2(KLOE)=51$ and $\chi^2$(WASA)$=51$ with similar fit probabilities,
both at the 90\% level, as just above. The choice of a minimal degree
has been preferred for $P_\eta(s)$.

Therefore, in the following, when different, the polynomials  
$P_\eta(s)$  and $P_\etp(s)$, are definitely chosen, the former first degree, 
the latter
second degree.  The  polynomial coefficients returned by the 
global fits performed with the $A_-$
BHLS$_2$ variant are discussed below and given in Table \ref{Table:T3}. 

It is worthwhile noting that the degradation of the fit quality observed when
assuming a first degree $P_\etp(s)$ is essentially carried  by the
the  BESIII $\etp(s)$  data sample itself, with a quite marginal influence on the standard 
channels of the BHLS$_2$ framework and on the $\eta$ dipion spectra.
This  emphasizes the robustness of the BHLS$_2$ Lagrangian.
\vspace{0.5cm}

In order to lighten the forthcoming discussion, let us comment right now 
on the formerly collected 
 $(\eta/\etp)$ dipion spectra listed in the Subsection above which
have also been analyzed within the BHLS$_2$ context; 
they quite generally yield stable   $\chi^2/N$ values. Some
of them return  large  $\chi^2/N$ values  
from the global fit procedure, namely, those from
TPC-2$\gam$ (69/13), LEPTON-F (45/20) and 
Layter {\it et al.} (60/15). 
Most of these former samples, however,  are getting
reasonable $\chi^2/N$ values, typically
8/12 (TASSO), 15/21 (CELLO), 23/18 (PLUTO), 20/15 (ARGUS), 11/17
(CRYSTAL BARREL\footnote{Its data point at 812.5 MeV, soon identified
as   outlier, being dropped out; see footnote 21 in \cite{ExtMod1}.}), 
13/14 (Gormley {\it et al.}) but have a quite negligible impact on the
issues examined in the present study. Therefore one focuses
 on  the high statistics data samples from BESIII and KLOE/ KLOE2;
the case for the WASA data set may be nevertheless commented$^{\ref{wasa}}$.

\subsection{ The $\eta/\etp$ Experimental Spectra~: Analysis within the BHLS$_2$ Context}
\label{fits-2}
\indentB Table \ref{Table:T2} collects the relevant fit quality information 
derived  when running
global fits within the $A_-$ BHLS$_2$ variant. The first data column
gives the fit information in a global fit performed\footnote{\label{3pi}  
The  $ e^+e^- \ra \pi^+ \pi^- \pi^0$  annihilation data are  switched off.} 
by discarding the $\eta/\etp$  to provide the BHLS$_2$ reference fit pattern; 
using the full ${\cal H}_R$, one would have found the
numbers given in the last data column of  Table 9 in \cite{ExtMod8}. 
The second and third data columns report on the fits performed  by
including $\eta/\etp$ dipion spectra within the fit data set ${\cal H}_R$
under the conditions indicated in the top line of  
Table  \ref{Table:T2}. 

\begin{table}[!phtb!]
\begin{center}
\begin{minipage}{0.9\textwidth}
\begin{tabular}{|| c  || c  | c | c ||}
\hline
\hline
\hhhv $\chi^2/N_{\rm pts}$ Fit Configuration ($A_-$) &  \hhhv no $\eta/\etp$ Spectra &  \hhhv $P_\eta(s) \ne P_\etp(s)$ &  \hhhv $P_\eta(s) \equiv P_\etp(s)$\\
\hline
\hline
 \hhhv NSK  $\pi^+\pi^-$ (127)     	&  $137/127$    	& $139/127$      & $140/127$   \\
\hline
 \hhhv KLOE $\pi^+\pi^-$ (135)		& $141/135$    		& $140/135$      & $140/135$   \\
\hline
 \hhhv Spacelike $\pi^+\pi^-$ (59)   	& $64/59$    		&  $64/59$       & $64/59$   	\\
\hline
 \hhhv  $\etp$ BESIII (112)   		& $\times$      	&  $100/112$     & $102/112$ 	\\
\hline
 \hhhv  $\eta$ KLOE/KLOE2 (59)		& $\times$   		&  $57/59$      & $55/59$  	\\
\hline
 \hhhv Total $\chi^2/N_{\rm pts}$	& 995/1075		&  1156/1246	& $1154/1246$	\\
\hline
 \hhhv Fit Probability			&  88.6 \%		&    89.7\%  	&  90.6\%     	\\
\hline
\hline
\end{tabular}
\end{minipage}
\end{center}
\caption {
\protect\label{Table:T2} Fit properties of selected dipion data sample sets 
using the  $A_-$   BHLS$_2$  variant.
The fit reported in the  first data column is free of $\eta/\etp$ dipion influence.
The second data column corresponds to fitting with  independent $P_\eta(s) $ and $ P_\etp(s)$, whereas
the third data column reports on the fit  where $P_\eta(s) \equiv P_\etp(s)$ has been  imposed. The  
 $\chi^2/N_{\rm pts}$ value for the WASA sample, fitted or not, is in the range$^{\ref{wasa}}$ 
 $(44-47)$ for 37 data points. }
\end{table}

The fit information concerning the $ e^+e^- \ra \pi^+ \pi^- $ annihilation data collected in scan mode 
(with different detectors at the various Novosibirsk facilities) is displayed in the first
 data line  (the exact sample content behind
the wording NSK is explained in \cite{ExtMod8}, for instance). The line KLOE stands for the merging of the KLOE10 \cite{KLOE10} and  KLOE12 \cite{KLOE12} data samples. The spacelike vector pion form factor data 
merges the NA7 and Fermilab samples  \cite{NA7,fermilab2}.

Taking  the first data column of Table \ref{Table:T2} as a reference, one can
clearly conclude that the fit quality obtained when using the  $\eta/\etp$ dipion 
spectra is unchanged and fairly good.  Indeed, the
$\chi^2$ increase of the NSK set of scan data samples is obviously negligible
and those of the ISR
data collected under the name KLOE and  spacelike data are unchanged. The description
of the data samples in the other channels from BHLS$_2$ (not shown) is also  
unchanged\footnote{Their variations are always  $\chi^2$  unit fractions.}.

Regarding the Triangle Anomaly sector, the $\chi^2$ information for the
$\pi^0/\eta/\etp \ra\gam \gam$ decays  are~:
\be
\left \{
\begin{array}{lll}
\displaystyle {\rm BHLS}_2~~A_- ~~{\rm variant~with~} P_\eta(s) \ne P_\etp(s) :& 
(\chi^2_{\pi^0},\chi^2_{\eta},\chi^2_{\etp}) =( 1.08,0.01,3,33) \\[0.5cm]
\displaystyle {\rm BHLS}_2~~A_- ~~{\rm variant~with~}  P_\eta(s) \equiv P_\etp(s) :& 
(\chi^2_{\pi^0},\chi^2_{\eta},\chi^2_{\etp}) =( 0.73,0.03,4.77) 
\end{array}
\right .
\label{2gamma} 
\ee
\noindent Thus, the  RPP width \cite{RPP2022} for $\pi^0 \ra \gam \gam$ is reproduced 
at the $(0.9\div 1)~\sigma$ level
and the one for $\eta \ra \gam \gam$ is reconstructed at nearly  its RPP value;
the width for $\etp \ra \gam \gam$ is found in the range $(1.8\div 2.2)~\sigma$,
somewhat larger but still acceptable. 

On the other hand and more importantly~: Comparing the second and third data columns of
Table \ref{Table:T2} obviously substantiates the  SHKMW conjecture  
 \cite{Stollenwerk_BOX_th} about the uniqueness of
the $P_X(s)$ function, e.g. $P_\eta(s) \equiv P_\etp(s)$. One may also note the slight
improvement generated by having stated $P_\eta(s) \equiv P_\etp(s)$; 
this should be due to having
provided its curvature to $P_\eta(s)$ which in turn lessens the (already marginal)
tension between the KLOE/KLOE2 and BESIII data samples.

\begin{figure}[!ht]
\begin{center}
{\includegraphics[height=14.cm]{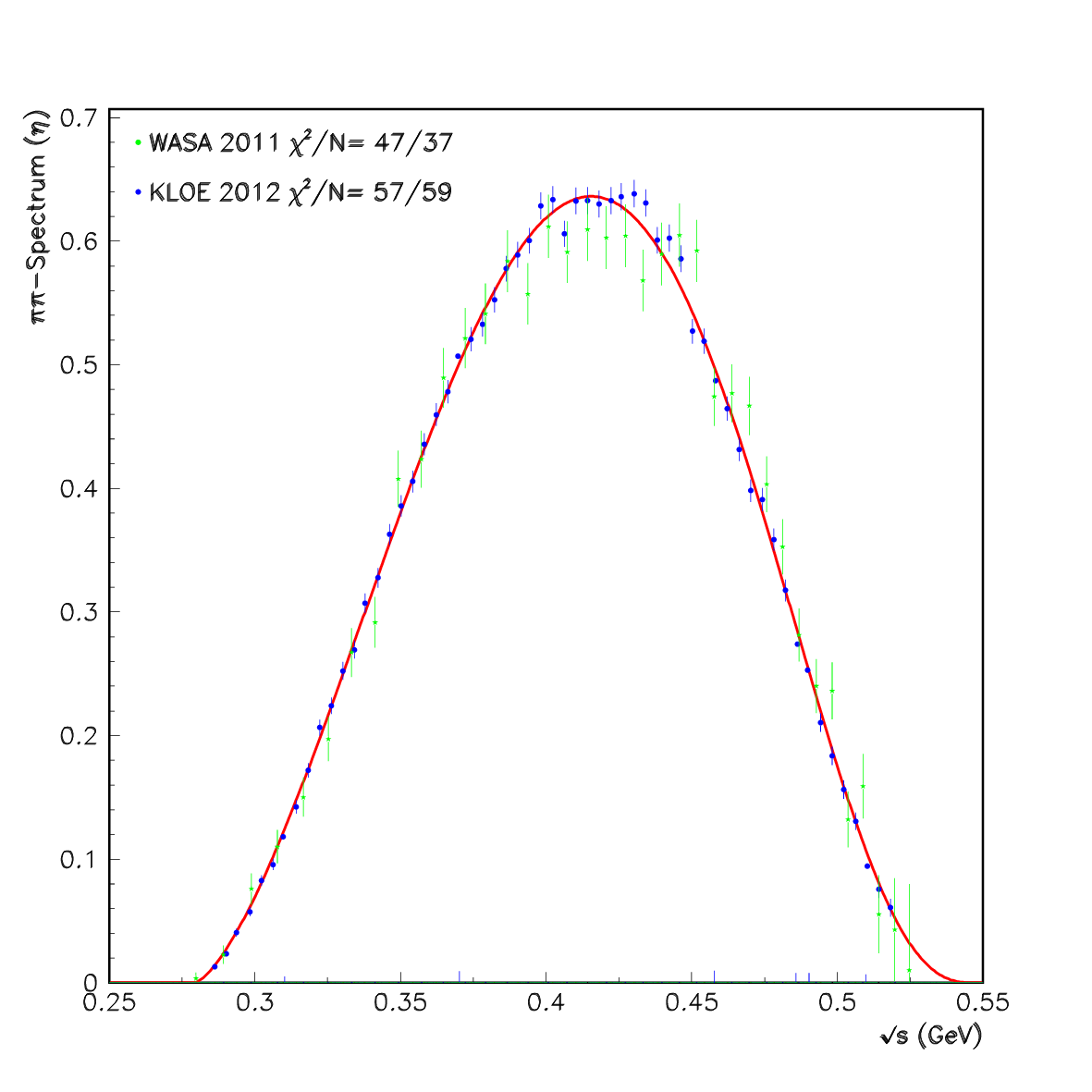}}
\end{center}
\begin{center}
\caption{\protect\label{Fig:eta_spectrum} The  dipion invariant mass spectrum in the
$\eta \ra \pi^+ \pi^-\gam$ decay. The blue data points are the KLOE/KLOE2
spectrum, the green ones display the WASA  spectrum. The red
curve is the BHLS$_2$ fit leaving free the $P_\eta(s)$ polynomial. 
 Vertical units are arbitrary. }
\end{center}
\end{figure}

Before going on with solely using the $A_-$ variant of 
the BHLS$_2$ Lagrangian, it is worthwhile reporting on its $A_+$ 
variant behavior.  
Let us limit oneself to reporting on the $A_+$  variant best fit performed 
assuming $P_\eta(s) \equiv P_\etp(s)$ second degree;  one obtains  
$\chi^2/N({\sc BESIII})=110/112$, and the $\eta$ dipion spectrum  from the 
KLOE/KLOE2 Collaboration yields this ratio at 54/59; for its part, 
the unfitted WASA sample 
yields 49/37. The  global fit  probability is 51.5\%  only,
to be compared to 90.6 \% for the global fit performed under the $A_-$ variant 
reported in Table \ref{Table:T2}.

This drop in probability is noticeable and its reason deserves to be identified;
 indeed,  the $\chi^2({\sc BESIII})$
increases by "only" 8 units, whereas  the $\chi^2$ for the $\eta$ 
dipion spectra are almost unchanged compared to Table \ref{Table:T2}.
Moreover,  the usual BHLS$_2$ channels also benefit from $\chi^2$'s 
comparable in magnitude to their $A_-$  analogs. 
Surprisingly, the single place where the disagreement blows up is in 
the $\gam \gam$ decays as~: 
$$(\chi^2_{\pi^0},\chi^2_{\eta},\chi^2_{\etp}) =(29.92,0.34, 0.08)~~~,$$
e.g. the $\pi^0 \ra \gam \gam $ partial width is at more than $5 \sigma$ 
from its accepted value \cite{RPP2022}, which is by far 
too large to be acceptable. Indeed, this implies that  the $A_+$
fit central value for the $\pi^0 \ra \gam \gam$ partial width is
 reconstructed at 70\% of its present RPP value \cite{RPP2022};
 this should  be brought in balance with  the $A_-$ variant  which yields
 this partial width  reconstructed  5\% larger than the expected value
 ($7.8$ eV).

Therefore,
the $A_+$ variant unexpectedly  exhibits a strong tension between the Triangle
and Box Anomaly sectors of the BHLS$_2$ Lagrangian,  whereas the $A_-$ variant behaves 
smoothly in both sectors. Therefore, from now on, 
one will focus on the $A_-$ variant of BHLS$_2$ which becomes our  
Reference model;  results derived using the $A_+$ variant 
are no longer reported  {\it except when explicitly stated}.

\vspace{0.5cm}
 
Regarding the $\eta$ spectra, Figure \ref{Fig:eta_spectrum}
shows an almost perfect account of the KLOE/KLOE2 spectrum~: the BHLS$_2$ spectrum
matches the dipion spectrum from KLOE/KLOE2 \cite{KLOE_BOX_OK} on the whole energy
range, except  for a marginal issue in the 0.45 GeV energy region. Even if 
its $\chi^2$ value is 
acceptable, the WASA spectrum \cite{WASA_BOX_OK} may look somewhat distorted
with respect to its KLOE/KLOE2 partner, clearly favored by BHLS$_2$ 
expectations$^{\ref{wasa}}$.

\begin{figure}[!ht]
\begin{center}
{\includegraphics[height=14.cm]{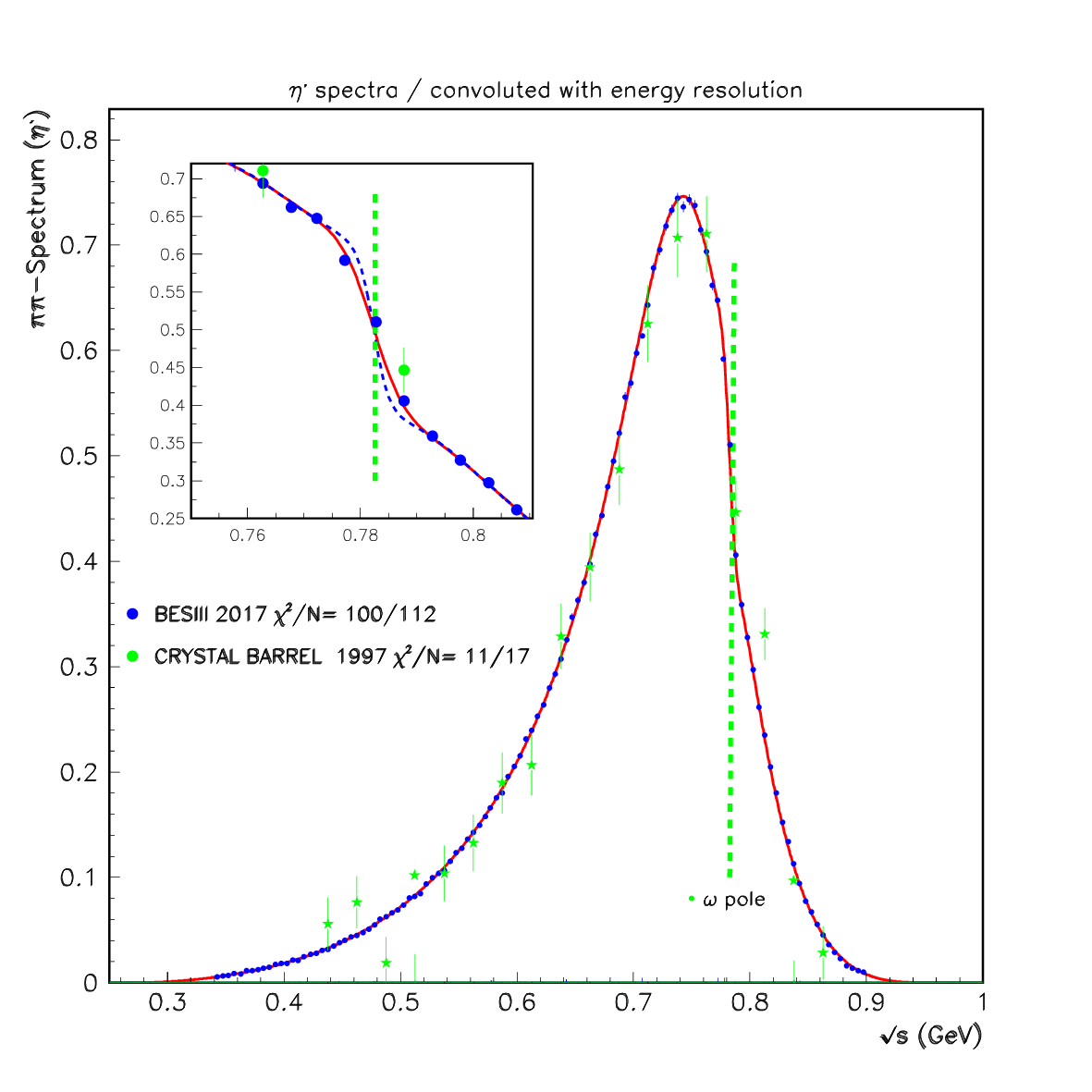}}
\end{center}
\begin{center}
\caption{\protect\label{Fig:etp_spectrum} The  dipion invariant mass spectrum in the
$\etp \ra \pi^+ \pi^-\gam$ decay. The blue data points are the BESIII spectrum, the
green ones are those from Crystal Barrel. The red curve is the fit function, {\it i.e.}
the convolution of the BHLS$_2$ model function with the energy resolution function assumed Gaussian;
the blue curve is the underlying BHLS$_2$ model function itself. Both curves 
superimpose 
over the whole energy range except for the $\rho-\omg$ drop-off region. 
Vertical units are arbitrary. }
\end{center}
\end{figure}

Regarding the $\etp$ spectrum, Figure \ref{Fig:etp_spectrum} shows a noticeably
fair accord between the BHLS$_2$ modeling and the BESIII spectrum \cite{BESIII_BOX_OK}
all along the energy range. 
The vertical green dotted lines locate the $\omg$ mass and so the 
$\rho-\omg$ drop--off region, otherwise magnified in the inset. Here, one can
observe the effect of convoluting the BHLS$_2$ model function
with energy resolution Gaussians as provided by the BESIII Collaboration~: It does
perfectly what it is supposed to do, {\it i.e.} soften  the drop--off to its right 
lineshape with, moreover, a noticeable accuracy. On the rest of the spectrum,
the convoluted curve and the underlying model one superimpose 
on each other within the thickness of the curves.  One should also state that
no tension in the $\rho-\omg$ drop-off region is observed in the fits with
any of the other dipion spectra submitted to the fit.

\begin{figure}[!ht]
\begin{center}
{\includegraphics[height=15.cm]{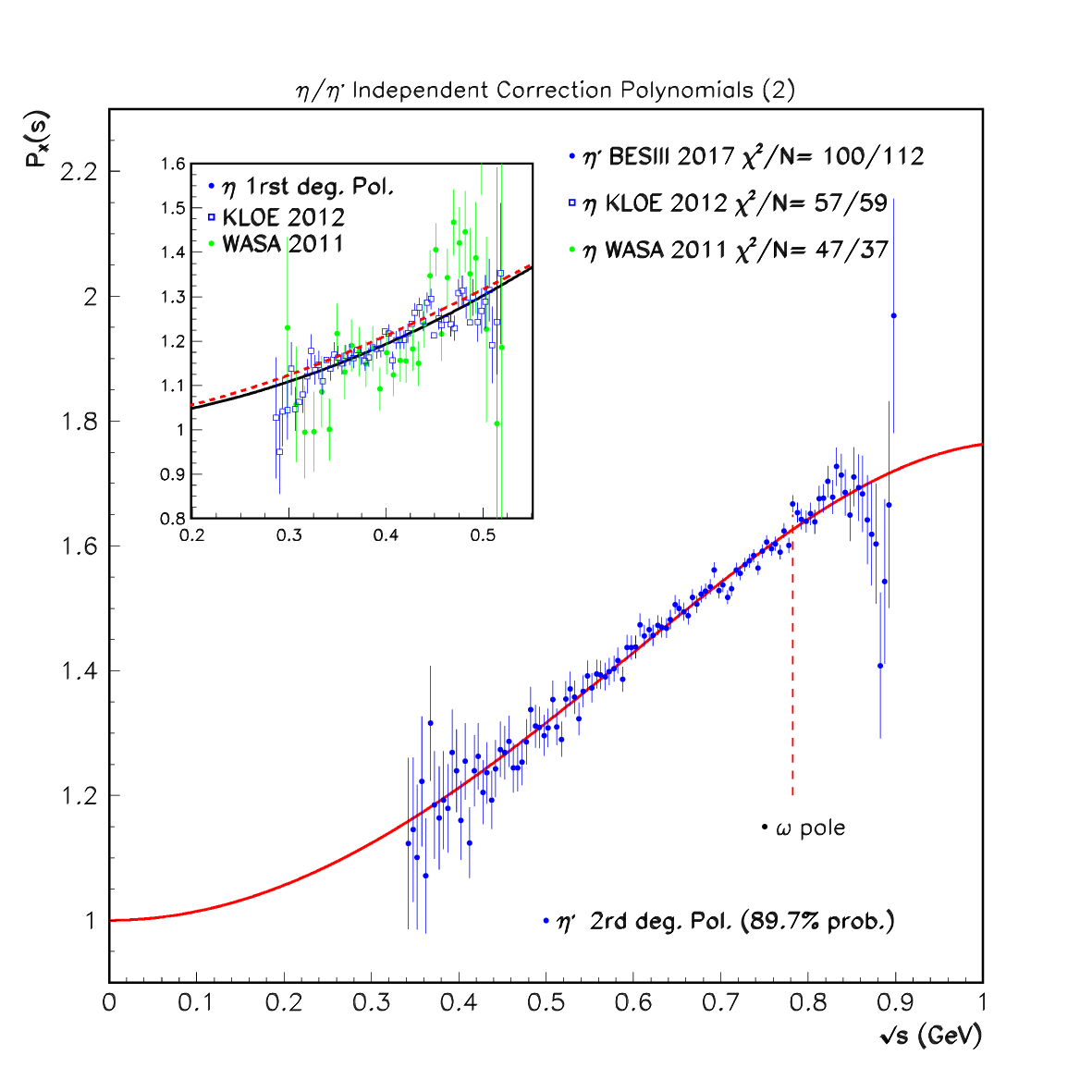}}
\end{center}
\begin{center}
\caption{\protect\label{Fig:polynomials_diff} The  $\overline{P}_\etp(s)$ and, 
in the inset, the $\overline{P}_\eta(s)$ spectra (Equation (\ref{polynom-1}). The full red curve 
and full black curve superimposed to  resp. $\overline{P}_\etp(s)$ and  
$\overline{P}_\eta(s)$ are resp. the $P_\etp(s)$ and $P_\eta(s)$ 
polynomials returned by the fits. The dashed red curve in the inset is also 
$P_\etp(s)$, but superimposed to the $\overline{P}_\eta(s)$ spectrum.
Some pieces of fit information are also displayed.}
\end{center}
\end{figure}

It is useful to consider the spectra\footnote{It is, of course, understood that, when dealing 
with the BESIII $\etp$ dipion sample, $d\Gamma_{theor}(s)/d \sqrt{s}$ is, actually, 
the convolution product of the model function with the BESIII energy resolution 
function.}~:

\be
 \displaystyle 
 \overline{P}_X(s)= \left [ \frac{d\Gamma_{exp}(s)}{d \sqrt{s}}/
 \frac{d\Gamma_{theor}(s)}{d \sqrt{s}} \right ]_X ~ P_X(s)~~, ~~ X=\eta,\etp
\label{polynom-1}
\ee
 to illustrate the behavior of the  $P_X(s)$ polynomials under
the two assumptions discussed above.
As the bracketed term in Equation (\ref{polynom-1}) 
fluctuates around 1 and reflects the experimental uncertainty spectrum, the
$\overline{P}_X(s)$ spectrum looks  an appropriate experimentally based evaluation 
of its  corresponding model function $P_X(s)$.
\begin{figure}[!ht]
\begin{center}
{\includegraphics[height=15.cm]{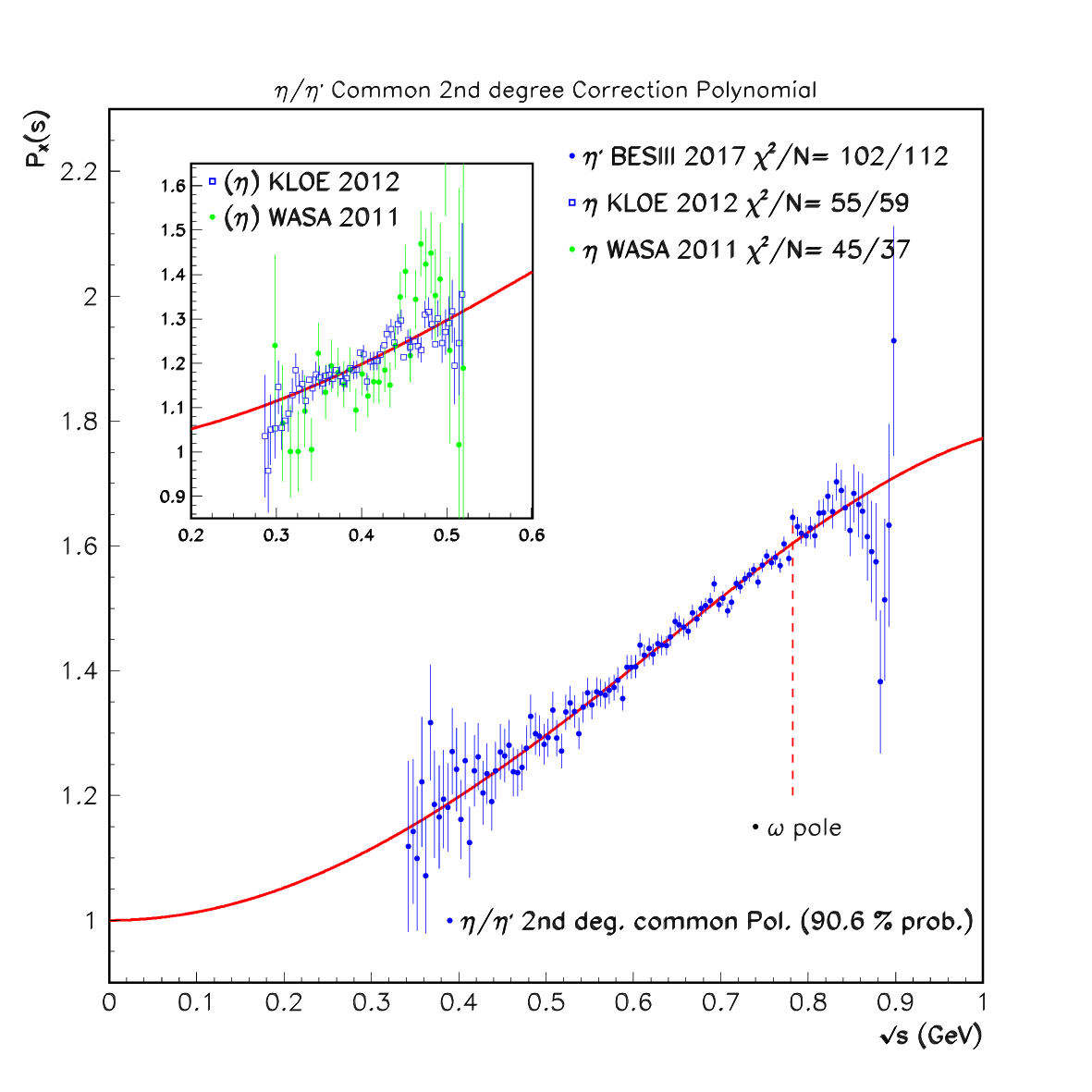}}
\end{center}
\begin{center}
\caption{\protect\label{Fig:polynomials_common} The  $\overline{P}_\etp(s)$ and, 
in the inset, the $\overline{P}_\eta(s)$ spectra  (Equation (\ref{polynom-1}).
 The full red curve superimposed on the $\overline{P}_\etp(s)$ and, 
in the inset, the $\overline{P}_\eta(s)$ spectra is their common  fit function $P_X(s)$.
The $\omg$ pole location is indicated. Some pieces of fit information are also displayed.
}
\end{center}
\end{figure}

Figure \ref{Fig:polynomials_diff} displays the  $\overline{P}_\etp(s)$ 
and $\overline{P}_\eta(s)$ spectra defined  just above in the case of BESIII, 
KLOE/KLOE2 and
WASA spectra together with their model partners  $P_\etp(s)$ (second degree) 
 and $P_\eta(s)$ (first degree). As could be
inferred from the fit properties shown in Table \ref{Table:T2},  $P_\etp(s)$
(the red dashed curve in the inset) is also a good evaluation for $\overline{P}_\eta(s)$.

Figure \ref{Fig:polynomials_common} also displays the $\overline{P}_\etp(s)$ 
and $\overline{P}_\eta(s)$ spectra for the BESIII, KLOE/KLOE2 and
WASA data samples, but together with their common model fit function 
denoted  $P_X(s)$,
a second-degree polynomial. As reflected by the fit information reminded in the body
of the Figure,
one has reached a fair simultaneous parametrization of the $\eta$ and $\etp$ dipion 
spectra
by  only supplying  the BHLS$_2$ model amplitudes with a single second-degree
polynomial $P_X(s)$  fulfilling $P_X(0)=1$.
\subsection{$P_X(s)$ : BHLS$_2$ Fit Results versus Others}
\label{fits-3}
\indentB 
The top bunch in Table \ref{Table:T3} displays the values returned for the polynomial coefficients of~:
\be
\displaystyle P_\eta(s) = 1 + \alpha_1 s~~~{\rm and}~~ P_\etp(s) = 1 + \alpha_1^\prime s + \alpha_2^\prime s^2~~.
\label{polynom-2}
\ee
\indent
When using the same polynomial for the $\eta$ and $\etp$ spectra, it is second degree and denoted $P_X(s)$.
It should be noted that the coefficients for $P_\etp(s)$ (second data column)  and $P_X(s)$
(third data column) carry numerical values close to each other, {\it i.e.} at $\simeq 1~\sigma$ from
each other for both
the first and second-degree coefficients\footnote{
It might be useful to provide, for completeness,  
the covariances when  $P_\eta(s) \ne P_\etp(s)$~:
Using obvious notations, they are
$<\delta  \alpha_1~\delta  \alpha_1^\prime>=-0.005$, 
$<\delta  \alpha_1~\delta  \alpha_2^\prime>=-0.026$ and 
$<\delta  \alpha_1^\prime~\delta  \alpha_2^\prime>= -0.812$. }.
In the case of having  a (single) common function  $P_X(s)$, the covariance is  
 $<\delta\alpha_1^\prime~\delta  \alpha_2^\prime>=-0.746$.

Regarding the systematics~: In the BHLS$_2$ approach, the statistical 
and systematic uncertainties provided
by the experiments together with their spectra are carefully embodied  
within the fitting code without
any modification; so our reported uncertainties automatically merge 
both kinds of experimental errors. 
\begin{table}[!phtb!]
\begin{center}
\hspace{-1.8cm}
\begin{minipage}{0.9\textwidth}
\begin{tabular}{|| c  || c  | c || c | c ||}
\hline
\hline
\hhhv Fit Parameter Value  &  \hhhv no $\eta/\etp$  &  \hhhv $P_\eta(s) \ne P_\etp(s)$ &  \hhhv
$[A_-]~:$  $P_X(s)$   &  \hhhv $[A_+]~:$ $P_X(s)$ \\
\hline
\hline
 \hhhv $\alpha_1^\prime$  (GeV$^{-2}$)   	& $\times$    		& $~~1.388 \pm 0.072$      & $~~1.326 \pm 0.053$  & $~~0.953 \pm 0.065$ \\
\hline
 \hhhv $\alpha_2^\prime$  (GeV$^{-4}$) 		& $\times$    		& $-0.607 \pm 0.055$      & $-0.553\pm 0.048$   & $-0.511\pm 0.052$  \\
\hline
 \hhhv  $\alpha_1$  (GeV$^{-2}$)   		& $\times$    		& $1.169 \pm 0.063$       & $\times$   	& $\times$\\
\hline
\hline
 \hhhv  $a_{HLS}$   			 	& $1.789  \pm 0.001$    &  $1.842\pm0.001$    	 & $1.821 \pm 0.001 $   & $1.830 \pm 0.001 $ \\
\hline
 \hhhv  $(c_3+c_4)/2$   			& $0.756  \pm 0.005$    &  $0.773\pm 0.005$    	 & $0.772\pm0.004$ 	& $0.819\pm0.007$ \\
\hline
\hline
 \hhhv Fit Probability		&  88.6 \%		&    89.7\%  	&  90.6\%     &  51.4\% 	\\
\hline
\hline
\end{tabular}
\end{minipage}
\end{center}
\caption {
\protect\label{Table:T3} The parameter values from the $A_-$  BHLS$_2$ variant fit. 
 The first data column reports on the fit using the usual set of data samples 
${\cal H}_R$,  excluding the 
$e^+ e^-$ annihilation to $3 \pi$ data. The second and third data columns report on the fits performed on
the same amputated  ${\cal H}_R$ sample set,  completed with the  $\eta/\etp$ dipion spectra under the conditions
indicated in the top line of the Table ($P_X(s)=P_\eta(s) \equiv P_\etp(s)$). 
The fair probability values can be emphasized. The last data column
displays the fit results when using the $A_+$ variant.
 }
\end{table}
On the other hand, the last two data lines in Table \ref{Table:T3} clearly illustrate that 
$\delta a= a-2$ and $\delta c_3= c_3-2/3$ remain consistent with expectations, {\it i.e.}
they can be regarded as ${\bf  O}(\delta)$ breaking parameters. The other fit parameter values are given in Table \ref{Table:T3bis} displayed in the next section \ref{BriefPar}; they are  scrutinized in order to detect eventual hints of effects spoiling the BHLS$_2$ model fit in the $3 \pi$ channel -- 
where correction polynomials are \textit{not} implemented by now.

\begin{itemize}
\item
{\bf j/} Regarding the $P_\eta(s)$ polynomial, it is worth comparing
our numerical value for $\alpha_1$ with those available in the literature.
 The first published evaluation (GeV$^{-2}$) of  $\alpha_1$ is the one from the WASA-at-COSY
Collaboration $\alpha_1 = 1.89 \pm 0.25_{stat} \pm 0.59_{syst} \pm 0.02_{th}$ \cite{WASA_BOX_OK}, soon followed
by $\alpha_1 = 1.96 \pm 0.27_{fit} \pm 0.02_{F_\pi}$  \cite{Stollenwerk_BOX_th}; more precise evaluations
have been proposed\footnote{Introducing a possible  $a_2$ exchange, 
Reference \cite{Kubis:2015sga}  also reports for a smaller value 
($\alpha_1 =1.42\pm 0.06_{stat}$).} 
since (GeV$^{-2}$)~:
\be
\begin{array}{lll}
\displaystyle
\alpha_1 =1.32\pm 0.08_{stat} \pm 0.10_{syst} \pm 0.02_{th}\cite{KLOE_BOX_OK},&
\displaystyle
\alpha_1 =1.52\pm 0.06_{stat} & \cite{Kubis:2015sga}~~.
\end{array}
\label{alpha_1}
\ee 
Our own evaluation -- reported in Table \ref{Table:T3} -- is in good agreement ($\simeq 1 \sigma$) with the
KLOE/KLOE2 Collaboration result \cite{KLOE_BOX_OK}.

\item
{\bf jj/} As far as we know, there are only two evaluations  for the $P_\etp(s)$ coefficients
available in the literature, the former from the BESIII Collaboration \cite{BESIII_BOX_OK}~:
\be
\displaystyle
{\rm BESIII~:} 
\left \{
\begin{array}{lll}
\displaystyle
\alpha_1^\prime ({\rm ~~GeV}^{-2})= ~~0.992  \pm 0.039_{stat} \pm 0.067_{syst} \pm 0.163_{th}\\[0.5cm]
\displaystyle
\alpha_2^\prime ({\rm ~~GeV}^{-4})= -0.523  \pm 0.039_{stat} \pm 0.066_{syst} \pm 0.181_{th} 
\end{array}
\right \}
,
\label{alpha_bes3}
\ee 
\noindent the latter from the HHHK group \cite{Holz:2022hwz,Holz:2022Err}. In their Addendum Table 1, the HHHK group proposes quite comparable values (we quote here the values obtained by the likelihood method) ~:
\be
\displaystyle
{\rm HHHK~:} 
\left \{
\begin{array}{lll}
\displaystyle
\alpha_1^\prime = ~~0.714  \pm 0.055~~{\rm GeV}^{-2}~~,& 
\displaystyle
\alpha_2^\prime= -0.412  \pm 0.055 ~~{\rm GeV}^{-4}
\end{array}
\right \}
.
\label{alpha_hhhk}
\ee 
Here one is faced with a surprising pattern~: While the BESIII parametrization for $P_X(s)$ is far from the favored $A_-$ variant one reported in Table \ref{Table:T3}, it is in quite remarkable accord with the $A_+$ solution displayed in the last data column of Table \ref{Table:T3}; as BESIII does not  deal with the intrinsic relationship between the Box and the Triangle Anomalies, their modeling is not influenced by the $\pi^0 \ra \gam \gam$ partial width  issue identified
 in Subsection \ref{fits-2} just above.

\end{itemize}

As a matter of conclusion, within the BHLS$_2$ framework, 
it has been shown that  the conjecture $P_\etp(s)=P_\eta(s)$ 
is a valid statement at the (high) degree of precision permitted  by the spectra from
the BESIII  and KLOE/KLOE2 Collaborations.  Moreover, Table \ref{Table:T2}
exhibits fair  fit probabilities and does not reveal any noticeable tension among the dipion spectra from KLOE/KLOE2 and BESIII on the one hand and,
 on the other hand,  the other channels embodied within the BHLS$_2$ fit procedure and their data, especially the dipion spectra collected in $e^+ e^-$ annihilations\footnote{Let us remind that KLOE08 \cite{KLOE08}, Babar \cite{BaBar,BaBar2} and SND \cite{SND2020} dipion spectra have been discarded because of tensions with the ${\cal H}_R$ set.}.

\subsection{Brief Analysis of the BHLS$_2$ Parameters Values}
\label{BriefPar} 
\indentB Table \ref{Table:T3bis} collects the model parameter values of the BHLS$_2$ Lagrangian.
 In order to figure out the effect  of the $e^+ e^- \ra \pi^+ \pi^- \pi^0$ 
 annihilation data on the numerical results, its first data column\footnote{same as last column in Table 10 in \cite{ExtMod8}} displays the fit parameter values derived when they are considered, whereas the second data column provides the same information when they are excluded from the fit procedure. The third and fourth data columns report the fit results when the $\eta/\etp$ dipion spectra are included within the set of data samples ${\cal H}_R$ amputated from the 3-pion data.

\begin{table}[!phtb!]
\begin{center}
\hspace{-1.5cm}
\begin{minipage}{0.9\textwidth}
\begin{tabular}{|| c  || c  || c  | c | c ||}
\hline
\hline
\hhhv Fit Parameter   &  \hhhv $3 \pi$ spectra only  &  \hhhv no $\eta/\etp/3 \pi$ Spectra &  \hhhv $P_\eta(s) \ne P_\etp(s)$ &  \hhhv $P_\eta(s) \equiv P_\etp(s)$\\
\hline
 \hhhv  $a_{HLS}$   	&	$1.766 \pm 0.001$	& $1.789  \pm 0.001$    &  $1.842\pm 0.001$    		 & $1.821 \pm 0.001 $ \\
\hline
 \hhhv  $g $   	   	&	$ 6.954  \pm 0.002$ 	 & $ 6.334  \pm 0.001$    &  $6.236\pm0.001$    	 & $6.379 \pm 0.001 $ \\
\hline
 \hhhv  $(c_3+c_4)/2$   & $0.742  \pm 0.003$ 	& $0.756  \pm 0.005$    &  $0.773\pm 0.005$    	 		 & $0.772\pm0.004$ 	\\
\hline
 \hhhv  $\theta_P$ (degrees) & $-15.59 \pm 0.28$	& $-16.471 \pm 0.295$   &  $-17.614 \pm 0.282$   		 & $-17.433 \pm 0.282$  	\\
\hline
 \hhhv $\lambda_0$	&  $0.285\pm 0.009$     &  $0.325\pm 0.008$	&  $0.339\pm 0.008$			& $0.334\pm 0.008$	\\
\hline
 \hhhv $z_A $	        &  $1.406\pm 0.004$	&  $1.416\pm 0.015$	&  $1.418\pm 0.005$			& $1.415\pm 0.005$	\\
\hline
 \hhhv $z_V $		& $1.420\pm 0.001$      &  $1.375\pm 0.007$	&  $1.304\pm 0.001$			& $1.320\pm 0.001$	\\
\hline
 \hhhv $\Delta_A~\times 10^{2}$	 		&  $12.94\pm 4.91$	&  $12.191\pm 4.05$ &  $10.173\pm 5.39$	& $10.249\pm 5.428$	\\
\hline
\hline
 \hhhv  $\epsilon~\times 10^{2}$		& $3.62\pm 0.30$   	& $5.383\pm 0.440$&  $6.456 \pm 0.439$    & $6.385 \pm 0.411 $  	\\
\hline
 \hhhv  $\epsilon^\prime~\times 10^{2}$	        & $0.17\pm 0.27$   	& $-3.623\pm 0.711$&  $-6.809 \pm 0.581$  & $-7.021\pm 0.475$  	\\
\hline
 \hhhv $\xi_0~\times 10^{2}$			& $-6.838\pm 0.018$	& $1.178\pm 0.018$&  $1.119\pm 0.013$	  & $-0.538\pm 0.014$	\\
\hline
 \hhhv $\xi_3~\times 10^{2}$			& $1.496\pm 0.150$	& $6.082\pm 0.153$ &  $6.070\pm 0.136$	& $5.609 \pm  0.137$	\\
\hline
\hline
 \hhhv Fit Probability				&  83.5 \%		&  88.6 \%&    89.7\%  	&  90.6\%     	\\
\hline
\hline
\end{tabular}
\end{minipage}
\end{center}
\caption {\protect\label{Table:T3bis}
Fit parameter values based on the $A_-$  BHLS$_2$ variant   ~: The first data column reminds the parameter values when including the $3 \pi$ spectra only, and the second one provides the same information when the $3 \pi$ spectra are discarded from the fit procedure. The third and fourth data columns display the fit results when the $\eta/\etp$  spectra are included and the $3 \pi$ spectra excluded.
}
\end{table}

 Besides providing the parameter values themselves, the issue here is to reach an educated guess about unaccounted-for effects in the fit (like $P_X$ polynomial equivalent corrections) in the $e^+ e^- \ra \pi^+ \pi^- \pi^0$ annihilation  process~: some effects in this channel could be numerically invisible or be absorbed effectively by the other model parameters.

 First of all,  the last line in Table \ref{Table:T3b} clearly shows that 
 one always  reaches    fair accounts of the spectra submitted to the BHLS$_2$ global fit.
 Regarding the parameters collected in the top rows of the Table, one observes value differences beyond the reported fit uncertainty, however with magnitudes consistent with reasonable systematic effects.
 
 The parameters in the lower section of the table look less well behaved.  Indeed, regarding $\epsilon$, $\epsilon^\prime$ and $\xi_3$, the pieces of information derived by the three fits excluding the 3-pion data  are 
 consistent with each other, but not with the first column result. The values for $\xi_0$ look confusing and may only indicate large systematics.

This is in fact reproducing an enduring situation since our previous 2022 publication \cite{ExtMod8}: there we also noticed such variations in the Isospin breaking parameters, and also shifts from those same parameters estimations based on meson mass differences (see Section 17 and 21 in \cite{ExtMod8})\footnote{The sign difference between $\epsilon$ and $\epsilon'$ in the last columns of Table \ref{Table:T3bis} is related to the fact that we are using a fit that leaves $\epsilon$ and $\epsilon'$ free (corresponding to the A$_-$ solution in the last column of Table 10 and 13 in \cite{ExtMod8}), which allows for unlike signs (and has the best global fit probability in \cite{ExtMod8}, compared to the so-called 'Condition C' fit that constrains more $\epsilon$ and $\epsilon'$, and indirectly forbids a different sign).}. This was not the case for the mixing parameters (Section 20 in \cite{ExtMod8}) which behaved more robustly, and were close to other groups results.\\
While this situation is not pleasing, we have not investigated yet what could be the origin of these variations, because we feel that a) their understanding probably needs long investigations; b) they seem to have little influence on the central subject matter of the present work, which is focused primarily on our first shot at global fitting the $\eta/\eta'$ data in the HLS framework, and secondarily on the implications for the $a_\mu$ estimation, especially concerning the DR-LQCD discrepancy; c) the limited number of independent evaluations of these Isospin breaking parameters from other groups.\\
Good candidate explanations for this situation could be: the effect of higher order corrections, which may in part be accounted for by the fit and absorbed in these parameters; parametric ambiguities in the fit representation of the model\footnote{see discussion after Eq.\ref{omnes-3} for a simple example, it can certainly be more complex.}, particularly in the case of parameters which are more indirectly connected to fitted physics observables, or only in combination with other parameters\footnote{Indeed, parameters like $\epsilon$ and $\epsilon'$ enter in the mixing description with many others (see Eq.101 and subsequent in \cite{ExtMod8}).}. \\
In this context, the question of the 3-pion data (see 1st column), which is not used in the present work, is still open, and is complicated by the fact that no correction polynomial (\textit{à la} $P_X$) was used in \cite{ExtMod8}.\\
Nevertheless, there is no obvious hint of significant fit probability spoiling effects in the $e^+ e^- \ra \pi^+ \pi^- \pi^0$ annihilation  process, but it is clear that this process and the way it could be integrated in the fit certainly deserve more scrutiny in future work \cite{Stamen:2022eda}.


\vspace{0cm}

\subsection{The $T^{R2}(\eta/\etp)$ Terms in BHLS$_2$ ~: The Role of $\rho^\pm$ Exchanges}
\label{TR2}
\indentB  Thanks to the  breaking mechanisms \cite{ExtMod7,ExtMod8} which lead to 
the BHLS$_2$ Lagrangian,
the derived $\eta/\etp$ decay amplitudes  involve $\rho^\pm$ exchanges as
depicted in Figure \ref{Fig:vmd_tree}  by the diagram classes  (c1) and (c2). Relying
on previous works in the HLS context 
which have shown that $c_3=c_4$ is fairly well accepted by the
data,  this constraint is assumed; as a straightforward consequence 
\cite{HLSOrigin,HLSRef}
 all diagrams involving direct $AVP$ couplings -- all proportional to 
($c_3-c_4$) -- identically vanish and, therefore, the diagram class (c1) contributions
also do.  Nevertheless, the (c2) diagram class, also   
${\cal O}(\delta)$ in breakings, survives and participates to 
the decay amplitudes $T_\etp$ and $T_\eta$ at ${\bf O}(\delta$). Such contributions are not involved in the BHLS$_2$ vector pion form factor $F_\pi(s)$ expression \cite{ExtMod7}; they come naturally in the derivation
of the amplitude $T(\eta/\etp)$ and are not governed by an 
additional {\it ad hoc} parameter.

\vspace{0.5cm}

Even if ${\bf O}(\delta)$  corrections, the $T^{R2}(\eta/\etp)$ amplitudes
play a noticeable role within the  BHLS$_2$ context~: 
\begin{itemize}
\item
{\bf i/} They are necessary in order for the full amplitudes
$T(\eta/\etp)=T^{NR}(\eta/\etp)+T^{R1}(\eta/\etp)+T^{R2}(\eta/\etp)$ to
coincide with their analogs directly derived from the WZW Lagrangian \cite{WZ,Witten}
at the chiral point\footnote{One has previously defined $s=(p_+ + p_-)^2$, 
$s_{0+}=(p_++p_0)^2$ and  $s_{0-}=(p_- +p_0)^2$.} $s=s_{0+}=s_{0-}=0$.

Indeed, at the chiral point, the intensities  $T^{\pm}(\eta/\etp)$ of 
the $T^{R2}(\eta/\etp)$ amplitudes 
defined in Sections \ref{BoxEta} and \ref{BoxEtp} write~:
\be
\begin{array}{lll}
\displaystyle 
T^{R2}(\eta) = -\frac{ie c_3}{4 \pi^2 f_\pi^3} 
\left [ \epsilon -\frac{A_\pm}{2}\sin{\delta_P} \right]~~~& {\rm and~~~}
\displaystyle 
T^{R2}(\etp) = -\frac{ie c_3}{4 \pi^2 f_\pi^3} 
\left [ \epsilon^\prime +\frac{A_\pm}{2}\cos{\delta_P} \right]
\end{array} 
\label{WZW-3}
\ee
and manifestly depend on the FKTUY parameter \cite{FKTUY} $c_3$. The condition
for the amplitudes $T(\etp)$ and $T(\eta)$ to coincide with those derived from
the WZW Lagrangian (see Equations (\ref{WZW-2})) is that all dependencies 
upon the FKTUY parameters vanish at $s=s_{0+}=s_{0-}=0$; this condition cannot
be fulfilled if dropping out (artificially) the $T^{R2}(\eta/\etp)$ terms from
the full amplitude expressions  $T(\eta/\etp)$. 

\item
{\bf ii/} To identify the effects of  the  $T^{R2}(\eta/\etp)$ 
terms, fits have been performed by discarding them in the full
amplitudes and rather fit using 
$T(\eta/\etp)=T^{NR}(\eta/\etp)+T^{R1}(\eta/\etp)$. The fits have been 
performed by imposing the constraint $P_\eta(s)=P_\etp(s)$ and return
the results collected in the next Table.

\begin{center}
\begin{tabular}{ ||c||c|c|c| } 
 \hline
\hhhv 
  $T^{R2}(\eta/\etp)$   (off/on)      &  off  &  on \\ 
 \hline
\hhhvw $\chi^2_{BESIII}~(N=112)$    & 122  & 102  \\ 
\hhhvw $\chi^2_{KLOE/KLOE2}~(N=59)$ &  57  &  55 \\ 
\hline
\hhhvw $\chi^2_{TOTAL}~(N=1246)$    & 1187  & 1154 \\
\hhhvw Probability &  73.0\% & 90.6 \% \\ 
 \hline
\end{tabular}
\end{center}
The $\chi^2$ values indicate that $T^{R2}(\eta)$ can be safely neglected,
but also that discarding  $T^{R2}(\etp) $ is not safe. The $P_X(s)$ 
parametrization returned by the fit is~:
\be
\left \{
\displaystyle
{\rm A}_-/{\rm no~TR2~}: 
\begin{array}{lll}
\displaystyle
\alpha_1^\prime = ~~0.437  \pm 0.039~~{\rm GeV}^{-2}~,& 
\displaystyle
\alpha_2^\prime= -0.573  \pm 0.007 ~~{\rm GeV}^{-4}
\end{array}
\right \}
,
\label{alpha_noTR2}
\ee 
closer to the HHHK results \cite{Holz:2022hwz,Holz:2022Err} reminded
in  Expressions (\ref{alpha_hhhk}) than to those  
in Table \ref{Table:T3}. 
Therefore, it is clear from the  results collected in Table  \ref{Table:T3} 
and the other  presented ones that~:

\hspace{0.2cm}  {\bf 1/}  The $\eta$ dipion spectrum
is essentially insensitive to using or discarding the $T^{R2}$ term in its
parametrization, 

whereas
 
\hspace{0.2cm} {\bf 2/} 
The $\etp$ dipion spectrum parametrization is significantly degraded if 
its $T^{R2}$ component
is dropped out. This absence may explain the reported failure of the so-called 
"model--dependent" fit in \cite{BESIII_BOX_OK}.
\end{itemize}

As summary, one may conclude that,  once the polynomial correction
and the ${\cal O}(\delta)$ $T^{R2}$ contribution
 predicted by the kinetic breaking of BHLS$_2$ \cite{ExtMod8}
 are considered, the average $\chi^2$ per data point for
the $\eta/\etp$ dipion spectra can be considered optimum ($<\chi^2> \simeq 1$). 
Thus, at the level of precision permitted by the presently available
$\eta$ \cite{KLOE_BOX_OK}
and $\etp$ \cite{BESIII_BOX_OK} dipion spectra,
additional contributions  beyond those of the basic vector meson nonet -- like
the higher mass vector mesons
\cite{BESIII_BOX_OK} or  the $a_2(1320)$ exchanges
\cite{Kubis:2015sga} -- need not be invoked. 
\subsection{Dealing with the Absolute Scale of the $\eta/\etp$ dipion spectra}
\label{absolute_norm}
\indentB
Having determined the $\eta/\etp$ dipion spectrum lineshapes by fitting their
common factor $P_X(s)$ ($X=\eta/\etp$), it remains to derive  the value of the 
 $H_X$'s ($X=\eta/\etp$) to also have their absolute magnitudes. As already noted the value 
 of the $H_X$ 
constants can be derived by introducing the accepted values \cite{RPP2022}
for the $\Gamma(\eta/\etp \ra \pi^+ \pi^- \gam)$ partial widths into 
the fitting procedure. This can be  (and has been) done and global fits have
been performed in order to get the optimum values for the 
$\{H_\eta, ~H_\etp,~P_X(s)\}$ triplets.

\begin{table}[!phtb!]
\begin{center}
\hspace{-2.5cm}
\begin{minipage}{0.9\textwidth}
\begin{tabular}{|| c  || c  | c || c | c ||}
\hline
\hline
\hhhv Fit Parameter   &  \hhhv NSK+KLOE  &  \hhhv NSK+KLOE &   \hhhv NSK+BaBar  &  \hhhv NSK+BaBar \\
\hhhv ~~  &  \hhhv fit $P_X(s)$ only  &  \hhhv fit $P_X(s)$ \& $H_X$ &   \hhhv fit $P_X(s)$ only  &  \hhhv fit $P_X(s)$ \& $H_X$ \\
\hline
\hline
 \hhhv $H_\eta$     	   		& $\times$    		& $0.789 \pm 0.017$     & $\times$  		& $0.797 \pm 0.017$ \\
\hline
 \hhhv $H_\etp $  	   		& $\times$    		& $0.671 \pm 0.017$     & $\times$     		& $0.682 \pm 0.015$  \\
\hline
 \hhhv  $\alpha^\prime_1$  (GeV$^{-2}$) & $1.326 \pm 0.053$    	& $1.309 \pm 0.055$     & $1.248 \pm 0.058$   	& $1.241 \pm 0.041$\\
\hline
 \hhhv  $\alpha^\prime_2$  (GeV$^{-4}$) & $-0.553 \pm 0.048$    & -$0.562\pm 0.047$     & -$0.535 \pm 0.048$   	& $-0.560 \pm 0.037$\\
\hline
\hline
 \hhhv  $10^{10}\times a_\mu(\pi \pi)$  	& $490.09 \pm 0.89$    	& $490.15 \pm 0.89$    & $494.98 \pm 0.91$   	& $494.85 \pm 0.88$\\
\hline
\hline
 \hhhv  $(\chi^2/N)_{BESIII}$       	& 102/112     		&  99/112    	 	& 101/112 		& 99/112	\\
\hline
 \hhhv  $(\chi^2/N)_{KLOE/KLOE2}$   	& 55/59    		&  53/59    	 	& 55/59    		& 53/59		\\
\hline
 \hhhv  $(\chi^2/N)_{TOTAL}$	   	& 1154/1246    		&  1149/1248    	& 1346/1381   		& 1341/1383	 \\
\hline
  \hhhv Fit Probability			&  90.6 \%		&   92.3\%  		 &  55.9\%     		&  59.4\% 	\\
\hline
\hline
\end{tabular}
\end{minipage}
\end{center}
\caption {
\protect\label{Table:T3b} Main global fit results involving the KLOE+NSK and BaBar+NSK samples collected in $e^+ e^- \ra \pi^+ \pi^-$ annihilations.
On top are displayed the parameters involved in the correction polynomials (see text for details) followed by the contribution to $a_\mu(\pi \pi)$ of
the $[2 m_\pi, 1.0$ GeV] energy range. The lowest bunch provides statistical 
information relative to the corresponding global fits.
 }
\end{table}

However, regarding  the $\eta/\etp \ra \pi^+ \pi^- \gamma$ decays, each of the
published dipion spectra is solely given  by its lineshape; concerning
their normalization, they  are tightly related to their partial widths.
It happens that the single available 
"measurement" for each of these decays is the corresponding RPP piece of 
information \cite{RPP2022}. In this case, as just argued,
the values for $H_X$ ($X=\eta/\etp$)
can be derived through the fitting code appropriately modified to take
the partial widths into account, but also algebraically once the fit
to determine the $P_X(s)$ ($X=\eta/\etp$) function has been performed. 
In this case one has, using obvious notations ~:
\be
 \hspace{-0.5cm}
\left [\Gamma(\eta/\etp \ra \pi^+ \pi^- \gam) \right ]_{RPP}\equiv
\int \left [ \frac{d\Gamma_X(s)}{d\sqrt{s}} \right ]_{exp.} d\sqrt{s} =
H_X^2 \int \left [ \frac{d\Gamma_X(s)}{d\sqrt{s}} \right ]_{BHLS2} 
[P_X(s)]^2 d\sqrt{s}~~, 
\label{int}
\ee
\noindent the integrals being performed over the whole energy range of the 
$X=\eta/\etp$ decays and the fit values for the 
$\Gamma(\eta/\etp \ra \pi^+ \pi^- \gam)$ partial widths coincide with the
RPP pieces of information.

Two cases have been considered regarding the specific $e^+ e^- \ra \pi^+ \pi^-$ 
annihilation sample combinations involved;
the first one  is $\{{\cal H}_R +  \eta/\etp\}$
which corresponds to global fitting with the \{KLOE, NSK, BESIII, CLEO-c\}
combination. Correspondingly, the second case involves the
\{BaBar, NSK, BESIII, CLEO-c\} sample combination. The  
relevant fit results regarding the correction polynomials are summarized
in  Table \ref{Table:T3b}.

The average $\chi^2$ per point of the $\eta$ and $\etp$ dipion spectra  are clearly insensitive to using
either of the KLOE or BaBar $e^+ e^- \ra \pi^+ \pi^-$ annihilation data within the global fit procedure. 
The global fit probabilities  are instead
quite different and correspond to our previous BHLS$_2$ results \cite{ExtMod7,ExtMod8}.
This insensitivity to the KLOE versus BaBaR issue
is well reflected  by the fit results  collected in the 
top part of Table \ref{Table:T3b}: None  of the $P_X$ and $H_X$
 parameter central values is observed to differ by more than $1 \sigma$  
 in the various fit configurations.

Similarly, as the different $P_X(s)$ parameter  values
derived from fitting with the various sample combinations
look like statistical  fluctuations,
differences  observed between fitting only $P_X(s)$ or  
the $(P_X(s) \& H_{\eta/\etp})$ triplet  look like statistical fluctuations. 
 Moreover, 
 defining  $\delta_X=H_X-1$ and focussing, for instance, on the KLOE+NSK combination, one gets~:
 \be
 \delta_\eta= -0.211 \pm 0.017~~~, \delta_\etp= -0.329 \pm 0.017 
 \label{universal}
 \ee
which corresponds to resp. $\delta$ and $\delta^\prime$ as defined by
 Stollenwerk {\it et al.} \cite{Stollenwerk_BOX_th} for which these 
authors derived the values $\delta=-0.22 \pm 0.04$ and $\delta^\prime=-0.40 \pm 0.09$; 
these are clearly identical to our
$\delta_\eta$ and $\delta_\etp$ respectively. As a last remark, it should be noted that, once 
$P_X(s)$ is determined -- which implies that both $ \frac{d\Gamma_X(s)}{d\sqrt{s}}$  
and both BHLS$_2$ functions are   known, Equation (\ref{int}) implies 
that both $H_X$ are not free but are algebraically related.

\section{$\eta/\etp$ Decays~: The Muon Anomalous Magnetic Moment}
\label{MAMM} 
\indentB
The renewed interest\footnote{See, for instance, \cite{Hanhart_BOX_th,Kubis_Review} 
and  the references collected therein.} in the $\eta/\etp$ physics is intimately
related to   dealing with the Light-by-Light contribution to the  
anomalous magnetic moment (AMM) of the muon. As shown above and previously  
in \cite{ExtMod8}, the BHLS$_2$ approach can address accurately several  
topics related to the  $\eta/\etp$ 
physics  and its results are supported by fair probabilities; 
these probabilities faithfully reflect the actual behavior of
each of   the data samples within the global framework as the error
information provided with it  is embodied   
without any  {\it ad hoc} enlargement inside the fitting code.

\subsection{Accuracy of the $P_X(s)$ Parametrization}
\label{PX-param} 
\indentB It has been shown above that a single polynomial $P_X(s)$ 
allows to address simultaneously  both  the $\eta/\etp \ra \pi^+ \pi^- \gam$ decays 
within the BHLS$_2$ framework and that 
the second degree is quite satisfactory. 
The  $P_X(s)$ parametrizations derived using the $A_\pm$ variants of
BHLS$_2$   displayed in Table \ref{Table:T3} are based on the
choice of the largest set of  data samples collected in almost all physics channels 
covering the HLS energy region
(e.g.  up to the $\simeq \phi$ mass region) and {\it consistent 
with each other}.
It was also shown 
that the $A_-$ parametrization is  the best favored but, nevertheless, 
one found it relevant to also provide  the $A_+$ parametrization despite its 
(sole  real) identified  failure with the   $\pi^0$ lifetime 
(or partial width) that $A_+$ reconstructs at 
more than $5\sigma$ from its commonly accepted value \cite{RPP2022}.

\begin{table}[!phtb!]
\begin{center}
\hspace{-0.6cm}
\begin{minipage}{0.9\textwidth}
\begin{tabular}{|| c  || c  || c | c || c ||}
\hline
\hline
\hhhv Data Set  &  \hhhv $<\chi^2_{\pi\pi}>$  &  \hhhv $\alpha^\prime_1$ &  \hhhv $\alpha^\prime_2$   &  \hhhv Prob. (\%) \\
\hline
\hline
 \hhhv 	 ${\cal X}_\tau$ +  KLOE08 + $\eta/\etp$	  & $1.57$    	& $1.294 \pm 0.053$       & $-0.379 \pm 0.049$   	& $61.4$\%\\
\hline
 \hhhv  ${\cal X}_\tau$ +  BaBar + $\eta/\etp$ 	  & $1.20$      & $1.249 \pm 0.076$    	 & $-0.522 \pm 0.0.69$  & $39.6\% $ \\
\hline
 \hhhv ${\cal X}_\tau$ + NSK + $\eta/\etp$  	  & $0.98$    	& $1.314 \pm 0.054$      & $-0.606 \pm 0.052$   & $96.6\%$ \\
\hline
 \hhhv ${\cal X}_\tau$ +  KLOE + $\eta/\etp$         & $0.99$    	& $1.341 \pm 0.054$      & $-0.525\pm 0.050$    & $92.4\%$  \\
\hline
\hline
\hhhv ${\cal H}_R$ + $\eta/\etp$ & $1.07$       & $1.326 \pm 0.053$      &  $-0.553\pm0.048$   &  $90.6\%$ \\	
\hline
\hline
\hhhv  ${\cal X}_\tau$ + $\eta/\etp$  & $\times$    & $1.453\pm 0.060$    	 & $-0.792\pm 0.065$ 	& $96.3\%$ \\
\hline
\hline
\end{tabular}
\end{minipage}
\end{center}
\caption {
\protect\label{Table:T4} The $P_X(s)$ parameter values from the $A_-$  BHLS$_2$ variant fit. The first 
column indicates which is the data set combination
 submitted to the global fit. $<\chi^2_{\pi\pi}>$ indicates 
the average $\chi^2$ of the timelike $F_\pi(s)$ data points of the sample named in the first column.  
$\alpha^\prime_1$ and  $\alpha^\prime_2$ are the coefficients of resp. the first and second-degree terms of
$P_X(s)$. The last data column displays the probability of the corresponding global fit.
 }
\end{table}

 In this Section, one aims to emphasize  the reliability 
 of the  $A_-$  parametrization by examining carefully how the $P_X(s)$ 
 parameter values evolve while using the various dipion spectra collected
 in $e^+e^-$ annihilations which are known to  exhibit --   sometimes severe --
  inconsistencies among  themselves.

A possible fit bias in the parametrizations reported in Table \ref{Table:T3} is the choice of
the  dipion data samples, because of their mutual consistency;
this issue is examined first. For this purpose, it is useful to define and name
some sets of data samples in order to ease the reading.

Basically, the data samples$^{\ref{3pi}}$ common to the sets of data samples 
 presently embodied within the BHLS$_2$ based fit procedure are the 
 \{$(\pi^0/\eta)\gam$, $K_L K_S$, $K^+K^-$\} $e^+e^-$ annihilation channels, 
the dipion spectra from the $\tau$ decay provided by the
 ALEPH, CLEO, and BELLE Collaborations
and the pion and kaon spacelike spectra	from NA7\cite{NA7} and Fermilab
\cite{fermilab2}; let us, for clarity, name this basic 
set ${\cal X}_\tau$. 

Regarding the available $e^+e^- \ra \pi^+ \pi^-$ annihilation spectra, one has
distinguished four groups\footnote{As the more recent dipion spectra from BESIII 
\cite{BES-III,BESIII-cor} and Cleo-c \cite{CESR} accommodate easily any of the  
groups we are listing, they would not be conclusive and have been put aside 
for clarity; regarding the SND20 spectrum  \cite{SND20} deeply analyzed 
in our \cite{ExtMod8}, we have proceeded likewise.} 
(two of which being, actually, a one sample "group")~: 
{\bf 1/} The scan data collected under the name NSK (see \cite{ExtMod4} for its 
content),  {\bf 2/} the KLOE ($\equiv$ KLOE10+KLOE12)  \cite{KLOE10,KLOE12} 
ISR data sample group, {\bf 3/} the KLOE08 ISR sample \cite{KLOE08} 
and {\bf 4/} the Babar  one \cite{BaBar,BaBar2}. For definiteness,
the largest set of data samples found consistent with each other and referred
to here and before \cite{ExtMod7,ExtMod8} as ${\cal H}_R$ gathers the sets 
${\cal X}_\tau$, NSK and KLOE just listed. Finally, the set of dipion spectra 
from the $\eta/\etp \ra \pi^+ \pi^- \gam$ decays
 \cite{BESIII_BOX_OK,KLOE_BOX_OK}  is referred to as  $\eta/\etp$.

\begin{figure}[!pht]  
\hspace{-2.cm}
\begin{minipage}{20.cm}
\begin{center}
{\includegraphics[height=9.cm]{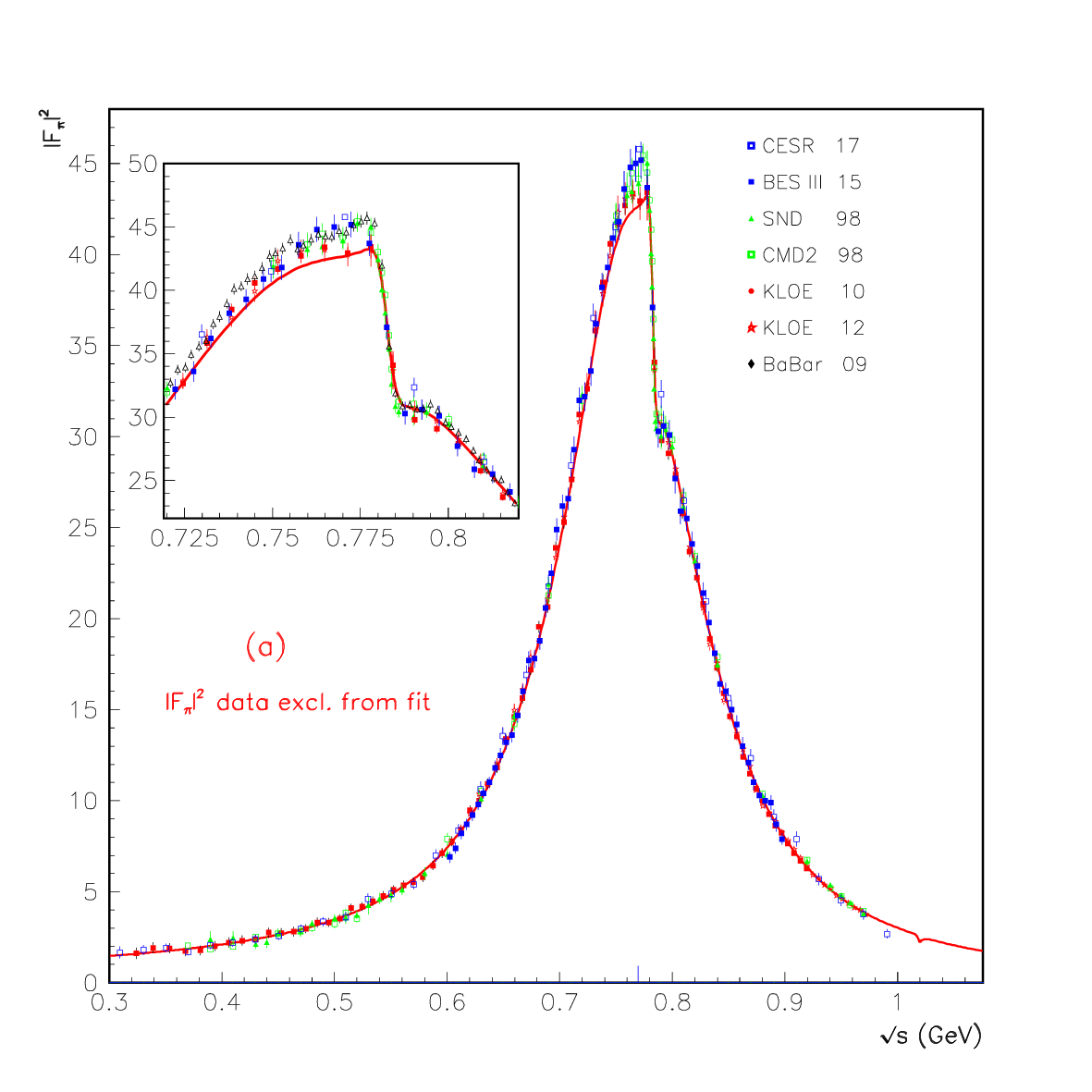}}
{\includegraphics[height=9.cm]{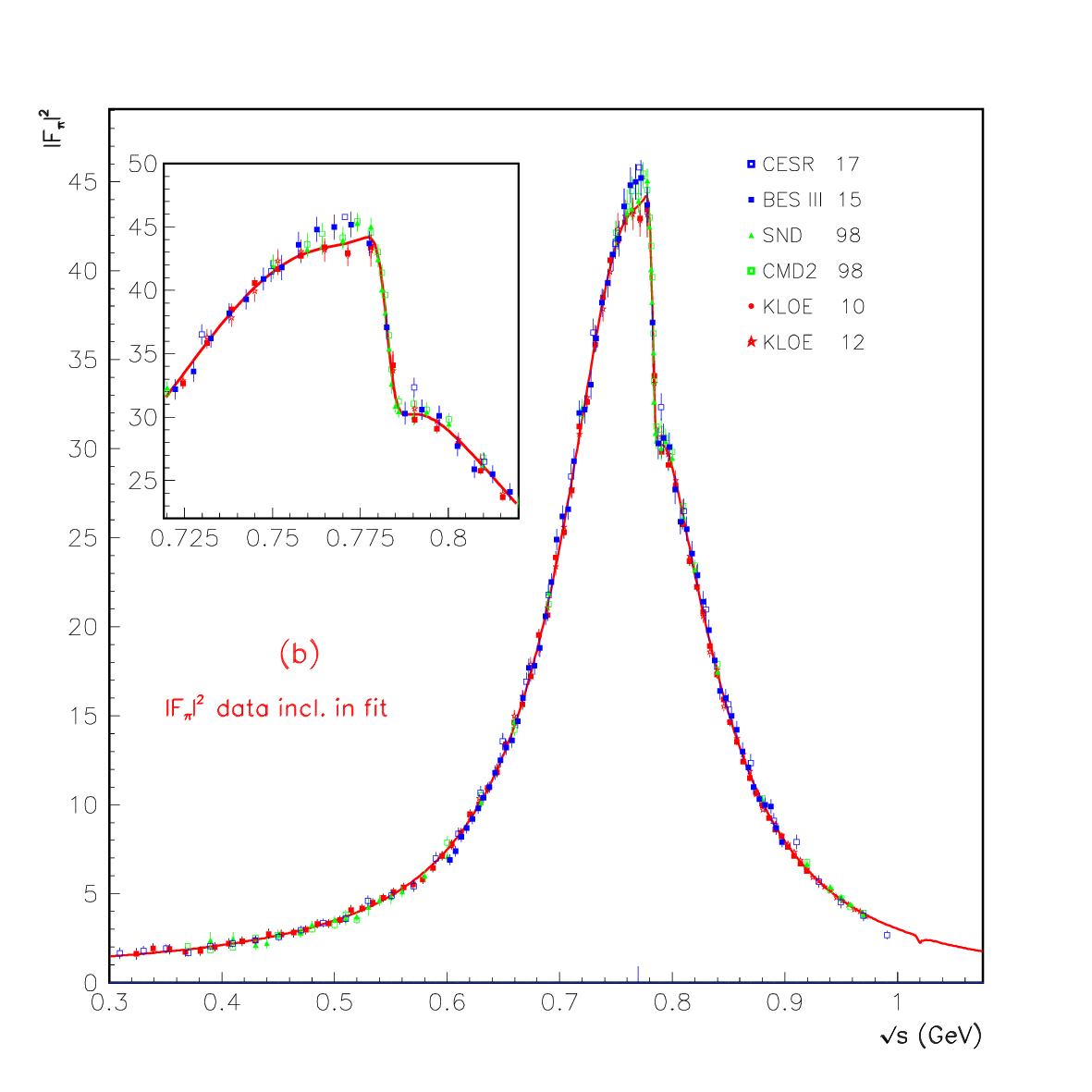}}
\end{center}
\end{minipage}
\begin{center}
\caption{\protect\label{Fig:etp_fpi}
The curve displayed in the left-side panel (a) is the pion form factor {\it predicted} by
fitting the data sample set $\{{\cal X}_\tau+\eta/\etp\}$ and, superimposed, the {\it unfitted} pion form factor spectra (including those from BaBar).
The right-hand side panel (b) shows the pion form factor derived from 
fitting  the full $\{{\cal H}_R+\eta/\etp\}$ data sample set  which includes the KLOE and NSK pion form factors (but not the BaBar spectrum). See the text for comments.
 }
\end{center}
\end{figure}

The four top lines in Table \ref{Table:T4} display the coefficient values of the first ($\alpha^\prime_1$) and 
second degree ($\alpha^\prime_2$) terms of the polynomial $P_X(s)$; as indicated in its first column,
the corresponding fits differ from each other only by the exact content of
$e^+e^- \ra \pi^+ \pi^-$ annihilation spectra sample set submitted to the minimization procedure. 
Whatever   the fit quality, reflected by its corresponding $<\chi^2_{\pi\pi}>$ value and its  probability, 
the different values derived for $\alpha^\prime_1$ as for $\alpha^\prime_2$  
are not distant by more 
than $(1\div 2) \sigma$ from each other. It should also be remarked that the parameter values
derived in the fit for \{${\cal H}_R$ + $\eta/\etp$ \}  -- which includes  the KLOE and NSK data sets together --
are intermediate between those involving the   KLOE and  NSK  sample sets separately.  
Therefore, the large spread of probabilities  between the fits involving NSK and/or KLOE 
and those involving BaBar or KLOE08, does not produce a significant change in the
determination of the common $\eta/\etp$ function $P_X(s)$.

The last line in Table \ref{Table:T4} displays the $P_X(s)$ coefficients returned by a fit
excluding the  $e^+e^- \ra \pi^+ \pi^-$ annihilation spectra.  The linear term coefficient $\alpha^\prime_1$ 
is never found  distant by more than $\simeq 2 \sigma$ from  the other corresponding values displayed in the same 
Table. In contrast, the curvature coefficient $\alpha^\prime_2$  exhibits a  $\simeq (4 \div 5) \sigma$ departure
  regarding  the other reported fit values.  
  Relying on Figures \ref{Fig:polynomials_diff} and \ref{Fig:polynomials_common},
  one expects  the second degree term ($\alpha^\prime_2$)
  to mostly  affect  the $\rho^0-\omg$ energy region.  This piece 
  of information renders it interesting to compare the pion form factor 
  {\it predicted}  by the fit of the  $\{ {\cal X}_\tau+\eta/\etp\}$ 
  set\footnote{Supplemented by the phase information between the $\rho$
  and $\omg$ propagators or by the product of branching fractions 
  ${\cal B}(\omg \ra e^- e^+) \times {\cal B}(\omg \ra \pi^+ \pi^-)$ available
  in the RPP \cite{RPP2022}.}  with the
  $e^+e^- \ra \pi^+ \pi^-$ annihilation data and the fit results derived when 
  fitting  the $\{ {\cal H}_R+\eta/\etp\}$ set. This is the purpose of
  Figure~\ref{Fig:etp_fpi}.

 Comparing the curve in both panels of   Figure \ref{Fig:etp_fpi}, 
 the overall agreement between both fits is fairly good, except for the magnitude  at the very  $\rho^0$ peak location which may look somewhat underestimated\footnote{
 Nevertheless, the lineshape  is in good correspondence with those of the KLOE12 spectrum included in the $\{{\cal H}_R+\eta/\etp\}$ sample set, but slightly smaller than the others.} by the $\etp$ dipion spectrum.  Instead, the drop-off location and its intensity are  fairly well predicted by  the $\{ {\cal X}_\tau+\eta/\etp\}$  sample set.
 This behavior deserves to be confirmed by new precise $\etp$ dipion spectra, complementing \cite{BESIII_BOX_OK}. So, while within the BHLS$_2$ framework the $\etp$ decay is accounted for largely independently of the $e^+e^- \ra \pi^+ \pi^-$ annihilation process, this does not prevent its prediction for $F_\pi(s)$ to exhibit a relatively fair accord  with the (fully independent) $e^+e^- \ra \pi^+ \pi^-$ annihilation spectra. 

Still, this accord is not perfect, and one has to contrast the good compatibility of the $\eta/\eta'$ samples with the rest of the reference data set (as reflected in the high corresponding fit probability), the also high probability of the $\{ {\cal X}_\tau+\eta/\etp\}$ fit (without the annihilation data contribution), and the visible pion form factor deviation causing the $a_\mu$ shift. It is difficult to be categoric at the moment about the nature of this effect on the form factor, which could be due to a variety of causes\footnote{We rechecked our calculations, the various fit informations, and other aspects but found no anomaly, or hints about what could explain the shift. Of course relatively straightforward causes of the deviation are almost surely excluded by the fact that the full data sets fit works quite well.}, also because as discussed in Section \ref{taudata} below it is sometimes difficult to interpret restricted data set fits, which have a more exploratory and confirmatory nature than 'full' fits, and in principle one should trust more our full data set fits, which indeed have no difficulty using the usual data sets in association with the $\eta/\eta'$ data.

In any case, it seems that the $a_\mu$ value from valid fits where the $e^+e^- \ra \pi^+ \pi^-$ annihilation data are, to some extent, replaced by the $\eta/\eta'$ data are much closer to the dispersive estimations than to the LQCD ones. This could indicate that part of the DR-LQCD discrepancy may not depend on experimental or analysis artifacts in the various experiments, since we witness a similar effect for the newly HLS fitted $\eta/\eta'$. If the present study is taken at face value, the DR-LQCD puzzle seems in fact bolstered, but on the other hand, tentative solutions based on potential purely experimental-level problems could appear less favored (to be clear : we also trust LQCD results).

Based on the current knowledge, other specific interpretations for the DR-LQCD conundrum are speculative (possible effects beyond the Standard Model ? $\pi\pi$ loop differences in $e^+e^-$ annihilation channel ? See also below \ref{taudataamu}), and in this light the same can be stated about the FNAL measurement discrepancy~\cite{FNAL:2023} (and even more speculative are eventual causal links between the three $a_\mu$ determinations).\\

\begin{figure}[!pht]  
\begin{center}
{\includegraphics[height=14.5cm]{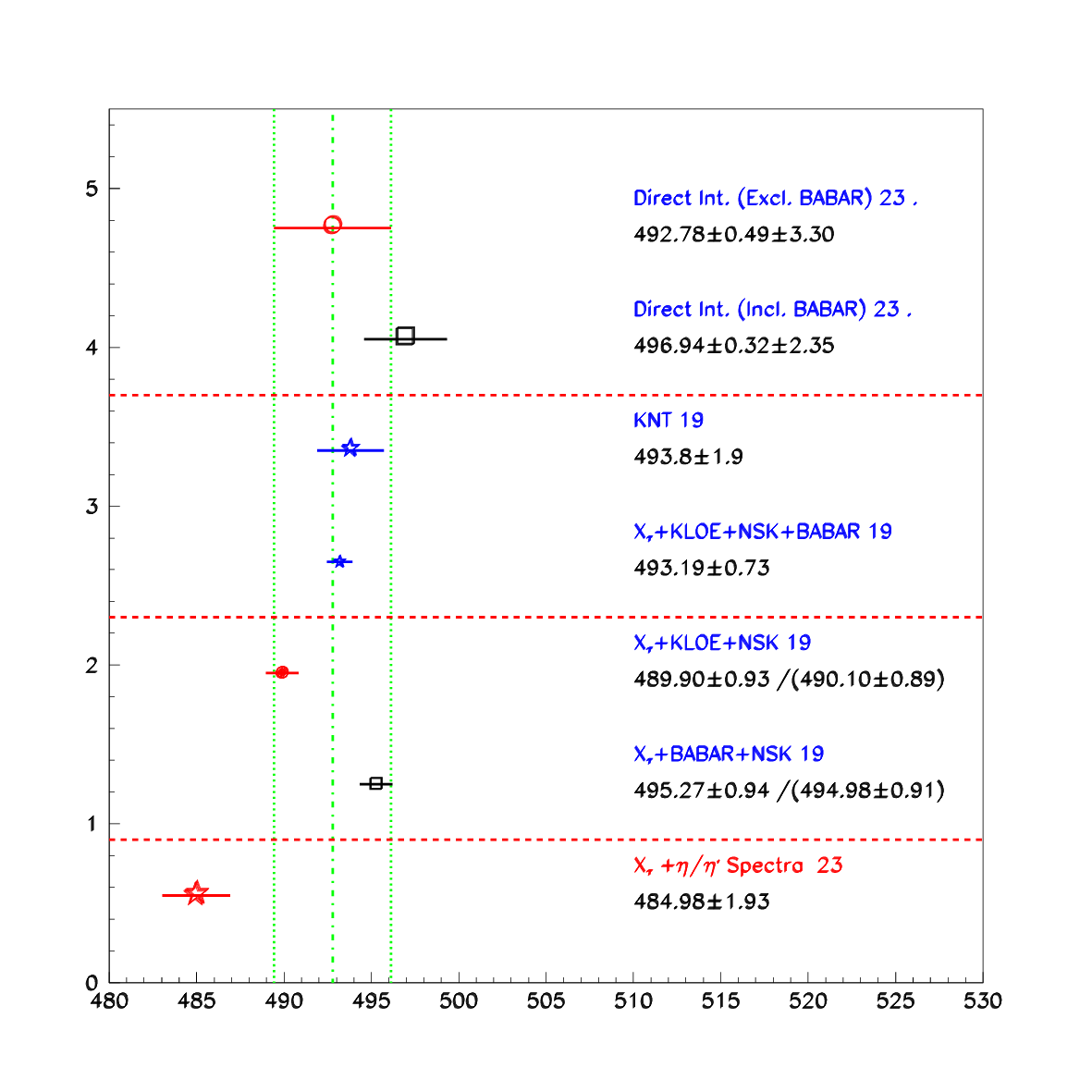}}
\end{center}
\begin{center}
\caption{\protect\label{Fig:amu_pipi} $a_\mu(\pi\pi, \sqrt{s} < 1.0 $ GeV) in units of $10^{-10}$
for various data sample combinations. The top two data points display the values derived
by a direct integration  of all the dipion spectra (when including BaBar, KLOE is excluded)~\cite{alphaQED}. 
The point tagged
by KNT19 \cite{Teubner3} is a usual (external) reference; the following point is derived using 
BHLS$_2$ with the indicated (largest) content of $e^+e^- \ra \pi^+ \pi^-$  spectra. 
The  two following points show our fit results for two indicated combinations of data samples; 
within parentheses, one also displays the results
obtained by also including the $\eta/\etp$ samples within the global fit procedure. 
The small magnitude of the BHLS$_2$ derived uncertainties should be noted (see text).
The downmost entry in this Figure exhibits the prediction derived for $a_\mu(\pi\pi, \sqrt{s} < 1.0 $ GeV)
when all annihilation to dipion data are discarded from the fit. The growth of its uncertainty reflects
the drastic reduction of the statistics involved in the corresponding fit. }
\end{center}
\end{figure}

Regarding the function $P_X(s)$, awaiting for other theoretical estimates of 
it, one can conclude that our favored $P_X(s)$ parametrisation\footnote{
\label{FSI_alternative} The $P_X(s)$ polynomial may well be interpreted as the lowest order 
terms of the Taylor expansion of a more complicated function which does not behave 
as fast as a power law; for instance, one has checked that  the function 
$U(s) = 1+0.5 \log{(1+4 s)}$ ({\it i.e.} with no free parameter) 
gives results identical to those derived using the second-degree polynomials $P_X(s)$.
Indeed the  probability returned by the  fit of the \{${\cal H}_R+ \eta/\etp$\} data sample set is
then 91.7\%, and the average $\chi^2$'s per data point are quite favorable~: For instance, 
1.08 for NSK, 1.04 for KLOE, 0.92 for the BESIII $\etp$ spectrum, and 0.90 for the $\eta$ spectrum
from KLOE/KLOE2 .} derived from fitting (${\cal H}_R$ + $\eta/\etp$) provides already a 
reliable one and  benefits from resp. a $\simeq 3 \%$ and $\simeq 10 \%$
precision for resp. the linear  and the curvature terms.

\subsection{The $\eta/\etp$ Spectra and HVP Estimates}
\label{ETP-HVP1} 
\indentB 
The purpose of Figure \ref{Fig:amu_pipi} is to figure out the overall picture of the estimates for $a_\mu(\pi\pi,\sqrt{s}<1.0$ GeV) which emerges from the present work. 
The top bunch data points displays the values for $a_\mu(\pi \pi, \sqrt{s} \leq 1 ~{\rm GeV})$ in units of $10^{-10}$ derived by direct integration of the dipion data   taking all dipion spectra, but either excluding the Babar spectrum or excluding the KLOE spectra; the reason to proceed this way is related to inconsistencies occurring when fitting the pion form factors \cite{ExtMod8} as reported since a long time \cite{BM_roma2013}.

The point showing the KNT19  result \cite{Teubner3}, the usual 
reference \cite{WhitePaper_2020},  is followed by  the evaluation derived from
the  BHLS$_2$ global fit involving ${\cal X}_\tau +KLOE+NSK+BABAR$ sample set
which contains  the same $e^+ e^- \ra  \pi^+ \pi^-$ dipion spectra\footnote{It 
should be reminded that the corresponding fit probability is low \cite{ExtMod8}
 (11.4\%), reflecting the KLOE--BaBar tension.}  as KNT19. 
The central values derived for  $a_\mu(\pi \pi, \sqrt{s} \leq 1 ~{\rm GeV})$ are 
substantially identical, reflecting the fact that the normalization uncertainty
treatment used to derive the KNT19  evaluation is similar to our own \cite{ExtMod5}.
The BHLS$_2$ uncertainty is however much improved (by a factor of $\simeq$2),  as can be expected from having 
performed a (more constraining) global fit; indeed, within a global context,
in contrast with KNT19 and others who treat the dipion spectra in a standalone mode, 
one benefits from also involving  the $\tau$ dipion spectra and all non-$\pi \pi$
final state spectra which  play as an increased statistics for all the
channels involved by the underlying HLS context, in particular the $\pi\pi$ one. Therefore,
 comparing KNT19 and our evaluation illustrates  that the  BHLS$_2$ Lagrangian approach does 
 not generate biases and that the difference in the central values is  essentially
 due to the data samples chosen to derive motivated physical conclusions.

The top two data points of the lowest bunch substantiate numerically the amplitude of  the tension 
between using ${\cal X}_\tau +KLOE+NSK$ and  ${\cal X}_\tau +BABAR+NSK$; both agree with the direct
integration results and exhibit  a $\simeq 5.4 \times 10^{-10}$ distance between 
 their evaluations  of $a_\mu(\pi \pi, \sqrt{s} \leq 1 ~{\rm GeV})$. In both cases, 
 the first number displayed is the evaluation  derived by a standard BHLS$_2$ fit and is 100\% consistent with the results published in \cite{ExtMod8}.\\
 The number within parentheses instead displays the result obtained when adding the $\eta/\etp$ data set defined above to resp. ${\cal X}_\tau +KLOE+NSK$ and ${\cal X}_\tau +BABAR+NSK$.  One should note that the fit probabilities are unchanged when adding the $\eta/\etp$ data set and reflect fairly good fits~:
 $88.7\% \ra 90.6\%$ for ${\cal X}_\tau +KLOE+NSK (+\eta/\etp)$,
 $47.2\% \ra 55.9\%$  for ${\cal X}_\tau +BABAR+NSK (+\eta/\etp)$.
 This illustrates that there is no tension between the $e^+ e^- \ra  \pi^+ \pi^-$ dipion spectra and those derived from the $(\eta/\etp)$ decays as the probability difference between the fits involving the two data sample sets is not degraded by  including the $(\eta/\etp)$ samples. The corresponding central values are very similar too, confirming that there are no bad surprises with the $\eta/\etp$ data and the new lagrangian terms describing them, at least in the ${\cal X}_\tau +KLOE+NSK$ and  ${\cal X}_\tau +BABAR+NSK$ case.

\subsection{The case for the $\tau$ data}
\label{taudata}

\subsubsection{Fitting the $\tau$ data}
\label{taudatafit}
Among the data sets combinations that look interesting to explore and fit, and in addition to the ones mentioned above, the quasi 'standalone' fitting of the $\tau$ decay specific data, in the spirit of what was achieved in our previous \cite{ExtMod3}, \cite{ExtMod4} and \cite{ExtMod5}\footnote{see Section 7.1 and Fig.4 in the last reference for example.} for the BHLS model, seems in order.\\

Since the inception of the BHLS$_2$ type of models \cite{ExtMod7}, we are unable to reach a satisfactory convergence (and good global probability) for such a fit, using the same type of complementary simple ($\pi\pi$ independent) RPP data that was used for BHLS (with some tentative variations in these data choices, and also in the MINUIT fit handling). We made additional tries\footnote{recently, for some of them.} by adding in the fitted data set spacelike $\pi\pi$ data (the idea being to constrain the fit behavior around $s=0$, while mostly preserving independence from the annihilation $\pi\pi$ information, due also to the limited statistical weight of the spacelike data sets), but to no avail.
So, whereas in the BHLS framework the working quasi 'standalone' $\tau$ fit allowed us to nicely reconstruct the full pion vector form factor\footnote{by also using some simple complementary RPP informations, see details in our previous \cite{ExtMod4}, Section 4.1.} (including in the near spacelike region), and vice versa for the dipion spectra in $\tau$ decays, using only the $\tau$ data for the former and only the annihilation $\pi\pi$ data for the latter, BHLS$_2$ does not seem to behave in the same way.\\

While this could seem disappointing, the importance of not being able to fit the $\tau$ data in 'standalone' should be put into perspective. Indeed, several considerations are relevant:

\begin{itemize}
\item {\bf i/} Compared to BHLS, in BHLS$_2$ more general methods are employed to break symmetries (covariant derivative, kinetic breaking for example), while those allow for a more consistent and more powerful way to parametrize the breakings, 
it seems that BHLS$_2$ is differently constrained in the $\tau$ sector. This could have the
unfortunate effect of spoiling the BHLS$_2$ standalone $\tau$ fit convergence (due to secondary false minima, for example).

\item {\bf ii/} While the standalone $\tau$ fit does not work in BHLS$_2$, we stress that the global fit, involving the full range of usual data sets (plus the $\tau$  and the new $\eta/\eta'$ data), is well behaved and returns good probabilities, as reported above. This means BHLS$_2$, when correctly constrained, describes adequately the data, and that the involved data sets, including the $\tau$, are statistically compatible\footnote{The BHLS $\tau$ standalone fit was demonstrated at a time (around 2012) where various groups were doing and discussing this type of $a_\mu$ estimations (in particular due to the '$\tau$ puzzle'), so some effort was invested into the study of this fit.}

\item {\bf iii/} While it is advisable to perform partial data fits (like the $\tau$ one, and also the ${\cal X}_\tau + \eta/\eta'$ fit), there is no guarantee that fits will continue to converge while discarding data sets (without changing the model). It is difficult with relatively complicated and parametrically intricated models like BHLS$_2$ to predict the minimal set of data allowing sound convergence, and in practice one resorts to the empirical method of trying to fit, hoping for good convergence and probability\footnote{This is indeed an example of some type of model dependence, and probably of the price to pay for having a global model encompassing all sorts of data.}. So, while many workers use partial fits to explore sensitivities (of $a_\mu$ for example) relative to the data set composition, in some cases the meaning of those fits should be pondered. The minimal acceptability requirement is of course that partial fits behave correctly and yield probabilities comparable to the full data set fits.

\end{itemize}

Apart from the well known variations in the fitted data sets (like replacing KLOE by BaBar data, Fig. \ref{Fig:amu_pipi}), in this work we explored two interesting partial fits: the ${\cal X}_\tau + \eta/\eta'$ one and the standalone $\tau$ one. The former fit has good convergence and good probability, which is not the case for the latter (here we confirmed our previous observations). This is why we kept the ${\cal X}_\tau + \eta/\eta'$ fit and discarded the incomplete $\tau$ one, and give no $a_\mu$ estimation from this last fit.\\
Again, we emphasize that the role of these partial fits should not be overestimated: the real benchmarks for data compatibility and model adequation are the full fits.\\
A worry could be that, in full fits, data sets with lower statistical weight may be of little influence when fitted together with large and precise data sets (like BaBar, KLOE, etc). In these regards and in the $a_\mu$ determination case at least, we can think of two reasons to be reassured: a) we consider only fits with high probability, meaning high \textit{global} $\chi^2$ probability which practically guarantees that the fitted pion form factor will be close to the data, and hence to the usual dispersive estimations; b) we require also good \textit{local} (partial) $\chi^2$ probability in the various sectors/data sets in the function to be minimized, which \textit{de facto} eliminates full fits which describe poorly a particular data set, and also allows sometimes to detect discrepancies between data sets.\\
Still, in view also of the recent renewed interest for the $\tau$ decay data, we have not completely abandoned the idea of demonstrating a $\tau$ quasi-standalone BHLS$_2$ fit, but the present lack of success in this endeavor certainly indicates an unknown level of difficulty.

\subsubsection{$\tau$ data and $a_\mu$}
\label{taudataamu}
To broaden the discussion on the $\tau$ decay data, we remind the comment conveyed in Subsection 21.3 of~\cite{ExtMod7}, where we put forward the idea that the QCD-QED interference at work in the neutral current (NC) $\pi^+\pi^-$ channel may cause a shift of the data with respect to the LQCD result. While in the neutral channel (NC) process $e^+e^- \to \pi^+\pi^-$, experiments measure the photon propagator, at leading order in lattice QCD one is calculating the pure QCD hadronic-current two-point correlator, free of external QED corrections. In the charged current (CC) $\tau^\pm \to \pi^\pm \pi^0 \nu_\tau$-decay, in contrast, one measures the $W$ propagator in the quasi-static limit and the $\rho^+$ meson does not mix with a photon-exchange (also no HVP subtraction correction needs to be applied), and the hadronic blob exhibited in the $\tau$ data is naturally much more directly related to the LQCD HVP than the one in the NC data. Actually, some recent $\tau$ data-driven analyses~\cite{Miranda:2020wdg,Masjuan:2023qsp,Masjuan:2023yam} find results $a_\mu^{HVP-LO}[\tau] = 704.1^{+4.1}_{4.0}\,\times 10^{-10}$ which also is close to the dispersive results~\cite{DavierHoecker} $a_\mu^{HVP-LO}[\tau] = 705.3\pm4.5 [689.8\pm5.2]\,\times
10^{-10}$ (result excluding $\tau$ data in brackets).\\
When combining $e^+e^-$ and $\tau$ data in HVP calculations, usually, the $\tau$ data are corrected towards matching the $e^+e^-$ ones, where in the latter the QCD-QED mixing is inherent. A part of the QCD-QED mixup is the $\rho^0-\gamma$~\cite{Fred11} mixing. Besides having corrected the $\tau$-spectra by the commonly accepted Isospin breaking effects, other corrections like the $\rho^0-\gamma$ mixing or the mass and width shift between $\rho^0$ and $\rho^\pm$ can be applied to the I=1 component of $\pi^+\pi^-$ data in order to reduce the QED contamination of the latter\footnote{In HLS however, and as a consequence of using the extended BKY isospin breaking mechanism, we showed that the data is better described by the induced difference in the universal vector coupling constant $g$ appearing in the anomalous and non-anomalous parts of the Lagrangian (see \cite{ExtMod3}, Section 12 for example).}. Belle~\cite{Belle}, in 2008 already, after applying standard (commonly accepted) Isospin breaking corrections, has obtained the $\tau$ data-driven result $a_\mu^{\pi\pi}[2 m_\pi, 1.8~{\rm GeV}]=(523.5\pm1.5({\rm exp.})\pm2.6({\rm Br.})\pm2.5({\rm isospin}))\times 10^{-10}\: (\tau$ data from ${\rm Belle})$ which was compared to $e^+e^-$ data-driven NSK result
$a_\mu^{\pi\pi}[2 m_\pi, 1.8~{\rm GeV}]=(504.6\pm3.1({\rm exp.})\pm0.9.6({\rm rad}))\times10^{-10}\: (e^+e^-:\,{\rm CMD2,\,SND})$. The difference is $18.9 \times 10^{-10}$, and if applied to dispersive result $a_\mu^{HVP-LO}[e^+e^-] = (694.79 \pm 4.18)\,\times 10^{-10}$~\cite{alphaQED}, one gets $713.7 \times10^{-10}$ and we note that with $a_\mu^{HVP-LO} = 718.2 \times10^{-10}$ we would have the coincidence of theory and experiment in the muon $g-2$: $a_\mu^{exp.}=a_\mu^{th.}$. Recently, also the CMD-3 Collaboration~\cite{CMD3prelim19}, for the range 
0.327 to 1.2 GeV, obtains $a_\mu^{had;LO}(2\pi,\,{\rm CMD3}=(526.0\pm4.2)\,\times 10^{-10}$, to be compared to $(506.0\pm 3.4)\,\times 10^{-10}$ reported in the White Paper document. Again the difference shifts the WP value $a_\mu^{HVP-LO}[e^+e^-] = (693.1\pm4.0)\times 10^{-10}$ to $(713.1\pm4.7)\times 10^{-10}$.  
We note that the CMD-3 result, clearly at variance with all previous $e^+e^- \to \pi^+\pi^-$ results has been obtained by applying a new RLA-based generalized vector dominance model (GVDM)  correction~\cite{Ignatov:2022iou}, whereas other experiments apply scalar QED to perform QED corrections. Using older $\pi\pi$ data, our HLS global fits (standalone or not) showed no incompatibility between the $\tau$ data and the annihilation $\pi\pi$ data, which seems to shift the problem away from a pure $\tau$ data incompatibility with other main data sets but rather towards a questionning of the relation between $a_\mu$ and the pion form factor as we know it.

\subsection{$\eta/\etp$ Based Evaluations of the HVP}
\label{etp_based_hvp}
\indentB 
If, as conjectured long ago  \cite{Stollenwerk_BOX_th},  
an accurate enough determination of the function $P_X(s)$  can be provided (by Extended
ChPT \cite{Kaiser_2000,leutwb,leutw}, possibly),  dipion spectra from the $\etp$ decay
may provide a new way to estimate the dipion contribution to the muon HVP up to $\simeq 1$ GeV. 
The present work has shown that phenomenology is able to provide already 
a function $P_X(s)$ carrying a noticeable precision and, moreover,
it has also been   shown that a unique function accommodates easily   the available
$\eta$ and $\etp$ high precision  dipion spectra simultaneously.

Indeed, within the BHLS$_2$ context \cite{ExtMod7,ExtMod8},  the amplitudes
for  the $\eta/\etp \ra \pi^+ \pi^- \gam$ decays and for the $e^+e^-  \ra \pi^+ \pi^-$ annihilation
proceed from  the same Lagrangian and do not call for  a special treatment of their common dominant 
neutral $\rho$ meson signal. Moreover, once the $P_X(s)$ effects   are factored out,
the derivation of both amplitudes from the same Lagrangian is  unchanged.

On the other hand,  discrepancies revealed by comparing with each other the dipion spectra
collected in scan mode (NSK)   and
the various samples collected in ISR  mode by KLOE  \cite{KLOE08,KLOE10,KLOE12} and Babar
 \cite{BaBar,BaBar2} has not found a really satisfactory solution;   
the recent  SND20 \cite{SND20}   -- and even more, presumably, the new CMD3 \cite{CMD-3:2023alj} 
data -- seems rather to darken the picture.

Therefore, getting high statistics dipion spectra independent of the $e^+ e^-$ annihilation 
mechanism, carrying different kinds of systematics, may helpfully contribute to a more 
satisfactory understanding of the crucial $\pi^+ \pi^-$ contribution to the muon HVP.

For the time being, the limited number of high statistics 
$\eta$ \cite{KLOE_BOX_OK} and $\etp$ \cite{BESIII_BOX_OK}
dipion spectra allow us to already derive the prediction for
$a_\mu(\pi\pi, \sqrt{s} < 1.0 $ GeV)  displayed in the bottom of Figure
\ref{Fig:amu_pipi}, namely~:
\be 
a_\mu(\pi \pi, \sqrt{s} \leq 1 ~{\rm GeV}) = (484.98 \pm 1.93) \times 10^{-10}
\label{eta_etp_pred}
\ee
with a 96.3 \% fit probability, and is distant from its estimate
based on fitting ${\cal H}_R$ data sample 
set\footnote{It is interesting to note that the distance between this prediction and 
the solution derived using NSK+KLOE is almost equal to the distance
between the NSK+KLOE and NSK+BaBar  solutions.} by $2.6 \sigma$ (conservatively).  Therefore, additional
high statistics $\eta/\etp$ data samples can put more light on the issue,
clearly located in the $\rho^0-\omg$ invariant mass region. 

\section{Concluding Remarks}
\label{conclusion}
\indentB
The present work has shown that, besides the already reported $e^-e^+$ annihilation spectra, some
decay modes (especially  the $P \ra \gam \gam$ ones) or $\tau$ dipion spectra 
\cite{ExtMod7,ExtMod8},
 BHLS$_2$ can encompass the dipion spectra from the $\eta$ and $\etp$ decays; however, 
 to reach this result, one has to invoke a (common) correction polynomial -- not a part of the HLS model --
as inferred by the SHKMW group in \cite{Stollenwerk_BOX_th}. 
                                              
 In this context, BHLS$_2$ offers   
 a fairly good  simultaneous fit of the $\eta$ and $\etp$ dipion spectra together
 with the $e^+ e^-$ annihilations into 
 $ \pi^+ \pi^-/K \overline{K}/\pi^0 \gamma /\eta \gamma$
final states and the $\tau^\pm \ra \pi^\pm \pi^0 \nu_\tau$ decay also
addressed by BHLS$_2$ framework in our previous \cite{ExtMod7,ExtMod8}.
 
This proves that, once the $P_X(s)$ correction is accounted for, the BESIII
$\etp$ spectrum \cite{BESIII_BOX_OK} does not need more information
that  those already present in BHLS$_2$ to get a satisfactory picture;
the picture is found as fair  for the $\eta$ spectrum reported in  
\cite{KLOE_BOX_OK} -- and, actually,  even for those in \cite{WASA_BOX_OK}.
The role of the charged $\rho$ meson -- a natural feature of BHLS$_2$, 
\cite{ExtMod8},  never considered elsewhere -- has been shown to provide a fair   
treatment of the $ \etp \ra \pi^+ \pi^- \gam$ dipion spectrum.

This turns out to state that most of the parameters needed to write out
the relevant decay amplitudes are not free but numerically shared with  the other
channels embodied within the  same BHLS$_2$ framework. This is an additional step
in the proof that a unified Effective Lagrangian can fairly describe the low-energy physics up to and including the $\phi$ mass region.                                                                               

\vspace{0.4cm} 
 
 
 One has first shown that the  $\eta$ and $\etp$ dipion spectra are 
 well-fitted with specific low-degree polynomials supplementing the 
 amplitudes derived from the  BHLS$_2$ Lagrangian. In a second step,
 it has been proved that, actually,  the same  second-degree polynomial
 $P_X(s)$  is involved in the considered $\eta$ and $\etp$ decays as inferred in  
 \cite{Stollenwerk_BOX_th}. 
 As already noted,  the $\rho^\pm$ exchange implied by the
  kinetic breaking  defined in \cite{ExtMod8} is shown to enhance  the global fit quality.
  The polynomial coefficients have been derived from our fits with fair precision
  and found that they remain stable when varying the fit conditions (see Table \ref{Table:T4}).  

\vspace{0.4cm}

It should be noted that the picture revealed by comparing both panels of 
Figure \ref{Fig:etp_fpi} suggests that the traditionally
used dipion spectra carry a lineshape compatible with the $\etp$ dipion spectrum.
Thus, higher statistics on this can be a helpful tool in the present controversy concerning the dispersive approaches and LQCD, due to the different systematics affecting the $\etp$ dipion spectrum, certainly independent of those involved in the $e^-e^+ \ra \pi^+ \pi^-$ annihilation.
At its level of accuracy, the present $\etp$ dipion spectrum  \cite{BESIII_BOX_OK} rather favors the DR prediction, as shown in Figure~\ref{Fig:amu_pipi}; this could indicate that the DR-LQCD $a_\mu$ discrepancy is not entirely related to experimental biases, since it appears also in the independent $\eta/\eta'$ data, but could rather be a misunderstood effect in dispersive estimations for processes involving pion pairs\footnote{It should be noted also that, in our global fit, the present $\eta/\eta'$ data do not impact the appraisal of the BaBar/KLOE discrepancy.}. However, better statistics and a finer binning
in the $\rho^0-\omg$ energy region look mandatory for a competing estimate of the muon $a_\mu(\pi^+ \pi^-, \sqrt{s} < 1.0~{\rm GeV})$. This may motivate our colleagues to enlarge the available $\etp$ dipion sample by analyzing the already existing data or by collecting new samples at other detectors.

\section*{Acknowlegements}
We gratefully acknowledge Andrzej Kupsc, Uppsala University, for having
provided the KLOE/ KLOE2 and the BESIII dipion spectra; 
additional information on these has also been quite helpful.  The CNRS/IN2P3 Computing 
Center (Lyon - France) is also gratefully acknowledged  for having 
provided the computing and data-processing resources needed for this work.

\section*{\Large{Appendices}}
\appendix
\section{Brief Outline of the HLS/BHLS$_2$ Approach}    
\label{HLS_unbrk}
\indent \indent For the reader's convenience, it looks worth to avoid
too much cross--references and briefly collect here the various ingredients which 
participate in the definition and working of our symmetry-broken Hidden Local 
Symmetry (HLS) model which is spread out into several references. The HLS model admits a
non-anomalous  sector \cite{HLSOrigin} and, beside, an anomalous one \cite{FKTUY}
--- see also \cite{HLSRef}. To make this approach a successful tool
in its physical realm, the HLS model should undergo, symmetry breaking
mechanisms.   The salient features of the broken version named BHLS$_2$
which underly the present study can be found, reminded, or defined, 
in\footnote{  For full details the interested reader is 
referred to these  articles, where former references can also be found.}
 \cite{ExtMod7,ExtMod8}. As it grounds the present study, the anomalous sector
 of the HLS model \cite{FKTUY,HLSRef} is mostly discussed in the body of the text. 
 
\subsection{The Unbroken Non-Anomalous HLS Lagrangian}
\label{HLS_org}
\indent \indent  The non--anomalous HLS Lagrangian is a generalization
of the ChPT Lagrangian \cite{GL1,GL2} which can be written \cite{HLSRef}~:
\be
\displaystyle
{\cal L}_{\rm chiral}=\frac{f_\pi^2}{4} {\rm Tr} \left [ \pa_\mu U~  \pa^\mu U^\dagger \right ]
=-\frac{f_\pi^2}{4} {\rm Tr} \left [ \pa_\mu \xi_L ~ \xi_L^\dagger - \pa_\mu \xi_R ~ \xi_R^\dagger
\right ]^2\,,
\label{eq1-1}
\ee
where $f_\pi$ (= 92.42 MeV) is the pion decay constant and~:
\be
\displaystyle \xi_{R/L}(x)=\exp{\left [ \pm iP(x)/f_\pi \right]}~~\Longrightarrow
U(x)=\xi_L^\dagger(x) \xi_R(x)\,, 
\label{eq1-2}
\ee
when working in the so-called unitary gauge  which removes a scalar field term  in the definition of
$\xi_{R/L}(x)$; $P(x)$ is the usual pseudoscalar (PS) field matrix.
 Ignoring in this reminder the weak sector \cite{HLSRef,ExtMod7}, the 
HLS approach turns out to replace  in  Equation (\ref{eq1-1}) the usual derivative  by the
covariant derivative~:
\be
\displaystyle
 D_\mu \xi_{R/L} =\pa_\mu \xi_{R/L} -ig V_\mu \xi_{R/L} +ie \xi_{R/L} A_\mu Q\,,
\label{eq1-3}
\ee
where $A_\mu$ is the photon field, $Q={\rm Diag}[2/3,-1/3,-1/3]$  the quark charge matrix and
$V_\mu$ is the  vector field matrix; the expressions for $P$ 
and\footnote{In the $V$ matrix the $\rho$, $\omg$ and $\phi$ fields correspond
to the so-called ideal fields. }  $V$ are the usual
ones -- fulfilling the $U(3)$ flavor symmetry --
and can be found in \cite{HLSRef,Heath,ExtMod3}, for example. In this way, the
first HLS Lagrangian piece named ${\cal L}_A$ is derived from Equation (\ref{eq1-2}). However,
a second piece -- ${\cal L}_V$ --  can be defined which vanishes in the inverse substitution 
$D_\mu \ra \partial_\mu$. The two pieces write~:
\be
\displaystyle
\begin{array}{lll}
\displaystyle
{\cal L}_A=-\frac{f_\pi^2}{4} {\rm Tr} \left [ D_\mu \xi_L ~ \xi_L^\dagger - D_\mu \xi_R ~ \xi_R^\dagger
\right ]^2~~,&
\displaystyle
\displaystyle
{\cal L}_V=-\frac{f_\pi^2}{4} {\rm Tr} \left [ D_\mu \xi_L ~ \xi_L^\dagger + D_\mu \xi_R ~ \xi_R^\dagger
\right ]^2\epo
\end{array}
\label{eq1-4}
\ee
and the full non--anomalous HLS Lagrangian writes~:
\be
\displaystyle {\cal L}_{\rm HLS}={\cal L}_A +a {\cal L}_V\,,
\label{eq1-5}
\ee
where $a$ is a free parameter specific to the HLS approach \cite{HLSRef}. 
This (unbroken) HLS Lagrangian can be found expanded in \cite{Heath}. 

\subsection{Breaking the HLS Lagrangian I~: The BKY Mechanism}
\label{HLS-BKY}
\indent \indent The first breaking mechanism for the HLS Lagrangian has been  
proposed in \cite{BKY}; one uses a modified version of it  given in \cite{Heath} in
order to avoid identified undesirable properties of the original proposal \cite{BGP}. 
Originally, the
BKY mechanism was intended to only break the $U(3)$ symmetry of the HLS Lagrangian; 
it has been extended following the lines of \cite{Hashimoto} to also cover isospin breaking effects.

Defining   $L=D_\mu \xi_L ~ \xi_L^\dagger$ and 
$R=D_\mu \xi_R ~ \xi_R^\dagger$, the (modified and extended) BKY breaking 
is implemented  in the BHLS$_2$ framework by modifying Equations (\ref{eq1-4}) 
as follows~:
 \be
\displaystyle
\begin{array}{lll}
 \displaystyle
{\cal L}_A=-\frac{f_\pi^2}{4} {\rm Tr} \left [ (L - R) X_A\right ]^2
~~,&
 \displaystyle
{\cal L}_V=-\frac{f_\pi^2}{4} {\rm Tr} \left [ (L + R) X_V\right ]^2
~~~,
 \end{array}
\label{eq1-6}
\ee
where the constant matrices $X_{A/V}$ provide departures from the unit matrix; they
have been parametrized as $X_{A/V} =$Diag$(q_{A/V},y_{A/V},z_{A/V})$.
In  practice, one prefers setting
$q_{A/V}=1+(\Sigma_{A/V}+ \Delta_{A/V})/2$ and 
$y_{A/V}=1+(\Sigma_{A/V}- \Delta_{A/V})/2$.
As $z_{A}$ and $z_{V}$ are affecting the $s\overline{s}$ entries, their departure from 1 
can be (and are found) large compared 
to $q_{A/V}$ and $y_{A/V}$ -- which refer to resp. the $u\overline{u}$ and
$d\overline{d}$ entries \cite{ExtMod3,ExtMod7,ExtMod8}. 

Within the BHLS$_2$ context opened in  \cite{ExtMod7}, it has been shown
that the diagonalization of the vector meson mass term implies
$\Delta_{V}=0$; on the other hand, it has also been proved  \cite{ExtMod7}
 that $\Sigma_{V}$ is actually out of reach and can be fixed to zero
 without any loss of generality. Therefore the BKY-breaking mechanism
 introduces 3 free parameters~: $z_A$ and $\Delta_A$ tightly related
 with the ratio $f_K/f_\pi$ and $z_V$ with the Higgs--Kibble $\phi$
 meson mass.
\subsection{Breaking the HLS Lagrangian II~: The Covariant Derivative (CD) Breaking}
\label{HLS-CD}
\indent \indent The main ingredient in the HLS approach is the covariant derivative
as displayed in Equation (\ref{eq1-3}), complemented when relevant by $W$
and $Z^0$ terms \cite{HLSRef}. Thus, a relevant breaking mechanism can be
chosen  affecting the covariant derivative itself; this can be done by replacing
Equation (\ref{eq1-3}) by~:
 \be
\displaystyle
D_\mu \xi_{R/L} =\pa_\mu \xi_{R/L} -ig \left[V_\mu^I +\delta V_\mu \right ] \xi_{R/L} +ie \xi_{R/L} A_\mu Q\,,
\label{eq1-9}
\ee
where $\delta V_\mu$ can be chosen to break the $U(3)_V$ symmetry in a 
controlled way. Breaking the universality of the vector coupling $g$ is
an interesting tool; {\it a priori} one may think that breaking nonet symmetry
({\it i.e.} along the Gell--Mann matrix $T^0$) can be performed independently
of breaking the $SU(3)_V$ symmetry ({\it i.e.} along the Gell--Mann matrix $T^8$); the diagonalization of the vector meson mass  term as well as the expected values of the pion and kaon form factors at the chiral point prevent such a freedom of choice \cite{ExtMod7}.

Identifying the field combinations associated with each of the canonical Gell--Mann $T_a$ 
$U(3)$ matrix basis, one is led to define the following components which can participate
to $\delta V_\mu$ separately or together~:
\be
\left \{
\displaystyle
 \begin{array}{l}
\displaystyle \delta V_\mu^0 =\frac{\xi_0}{\sqrt{2}}
\left[ \frac{\sqrt{2} \omg_\mu^{I} + \Phi_\mu^{I}}{3}\right]{\rm Diag} [1,1,1]
\,,\\[0.5cm]
 \displaystyle
\delta V_\mu^8 =\frac{\xi_8}{\sqrt{2}}
 \left[ \frac{\omg_\mu^{I} - \sqrt{2} \Phi_\mu^{I}}{3\sqrt{2}}\right]{\rm Diag} [1,1,-2] \,,\\[0.5cm]
\displaystyle
 \delta V_\mu^3 =\frac{\xi_3}{\sqrt{2}} \left[\frac{\rho_I^0}{\sqrt{2}}
 \right]{\rm Diag} [1,-1,0]\,,
\end{array}
\right .
\label{eq1-10}
\ee
in terms of the usual ideal field combinations; the CD-breaking term is
 $$\delta V_\mu=\delta V_\mu^0+\delta V_\mu^8+\delta V_\mu^3~~.$$

The (free) breaking parameters $\xi_0$, $\xi_8$ and  $\xi_3$ are  only requested to be real
in order that $\delta V_\mu$ is hermitian as $V_\mu^I$ itself. Clearly, $\delta V_\mu^0$ defines
a breaking of the nonet symmetry down to $SU(3)_V \times U(1)_V$,  $\delta V_\mu^8$ rather
expresses the breaking of the  $SU(3)_V$ symmetry, while $\delta V_\mu^3$ is related to
a direct  breaking of Isospin symmetry in the vector sector.

As mentioned above, it happens that the $\xi$ parameters introduced by
Equations (\ref{eq1-10}) should fulfill \cite{ExtMod7} $\xi_0=\xi_8$ and so, that the CD breaking 
only involves 2 new free parameters. This means  that within BHLS$_2$, one cannot
solely break nonet symmetry, which should be accompanied by a $SU(3)$ breaking of 
 same intensity.

\subsection{Breaking the HLS Lagrangian III~: Dynamical Vector Meson Mixing}
\label{HLS-dynMix}
\indent \indent  The unbroken HLS Lagrangian already exhibits couplings for 
$\rho_I/\omg_I/\phi_I \ra K^+K^-/K^0 \overline{K}^0$ transitions; this property is 
naturally transferred to all its broken versions. This implies that, at one loop
order, the $\rho^0/\omg/\phi$ squared mass matrix exhibits non--diagonal entries
and thus, the ideal vector fields are no longer mass eigenstates.

At one loop order, the squared mass matrix of the $\rho^0/\omg/\phi$ system 
can be written~:
\be
\displaystyle M^2(s)=M_0^2(s)+\delta M^2(s)\,,
\label{eq1-28}
\ee
where the dependence upon the momentum squared $s$ flowing through the vector lines
is made explicit. 
After the $BKY$ and $CD$ breakings just sketched, the vector meson masses 
write\footnote{One should note that within BHLS$_2$
the charged and neutral $\rho$ mesons carry different masses as
$ m_{\rho^\pm}^2= \displaystyle m^2 ~(1+\Sigma_V)$.}~:
\be
\left \{
\begin{array}{ll}
\displaystyle m_{\rho^0}^2= \displaystyle m^2 \left [
1+\Sigma_V + 2 ~\xi_3\right ]
\,,\\[0.5cm]
\displaystyle m_{\omg}^2= \displaystyle m^2 \left [
1+\Sigma_V + \frac{4}{3} ~\xi_0 +  \frac{2 }{3} ~\xi_8\right ]
=m^2 \left [
1+\Sigma_V +2 ~\xi_0\right ]
\,,\\[0.5cm]
\displaystyle m_{\Phi}^2  =
\displaystyle m^2 ~z_V  \left[ 1+\frac{2}{3}~\xi_0 + \frac{4}{3}~\xi_8 \right]
=  m^2 ~z_V  \left[ 1+2~\xi_0 \right]
\epo\\[0.5cm]
\end{array}
\right .
\label{eq1-26-b}
\ee
in terms of the  various breaking parameters;
 $\Sigma_V$ has been kept for convenience. The $M_0^2(s)$ matrix occurring in Equation
(\ref{eq1-28}) thus writes~:
\be
\displaystyle M_0^2(s)={\rm Diag} (m_{\rho^0}^2+ \Pi_{\pi\pi}(s),m_\omg^2,m_\phi^2)\epo
\label{eq1-29-b}  ~~~
\ee
and is diagonal; $\Pi_{\pi\pi}(s)$ is the pion loop and includes the $\rho \pi^+ \pi^-$ 
coupling squared. 

 The expression for $\delta M^2(s)$ is slightly more involved. 
Having defined the ($\rho^0,~\omg,~\phi$) renormalized fields, generally
indexed by $R$ ({\it i.e.} those
which diagonalizes the vector meson mass term), one can derive the 
${\cal V}_R^i \rightarrow {\cal V}_R^j$ transitions ($i,j=\rho^0,~\omg,~\phi$).
For this purpose, having defined  $\Pi_\pm(s)$ and $\Pi_0(s)$,  resp.
the {\it amputated} charged and neutral kaon loops, the transition amplitudes 
($i,j=\rho^0,~\omg,~\phi$) write~:
\be
\displaystyle
\delta M^2_{i, j}(s)=g^i_{K^+ K^-}g^j_{K^+ K^-}\Pi_\pm(s)+
 g^i_{K^0 \overline{K}^0}g^j_{K^0 \overline{K}^0}\Pi_0(s)
\label{eq1-30d}
\ee
where the $g_{K \overline{K}}$ coupling constants are displayed in
 Section 10 of \cite{ExtMod7}. 
 
 The {\it physical} $\rho^0,~\omg,~\phi$ are the eigenvectors of the
 full squared mass matrix $M^2(s)$; they are related to their {\it renormalized}
 partners by~:
  \be
 \left (
\begin{array}{ll}
\rho_{R}\\[0.5cm]
\omg_{R}\\[0.5cm]
\Phi_{R}\\[0.5cm]
\end{array}
\right )
 = \left (
\begin{array}{ccc}
\displaystyle 1     &\displaystyle -\alpha(s) &  \beta(s) \\[0.5cm]
\displaystyle \alpha(s) & \displaystyle 1 & \displaystyle \gamma(s) \\[0.5cm]
\displaystyle  -\beta(s) & \displaystyle  -\gamma(s)  & \displaystyle 1
\end{array}
\right )
 \left (
\begin{array}{ll}
\rho_{Phys}\\[0.5cm]
\omg_{Phys}\\[0.5cm]
\Phi_{Phys}\\[0.5cm]
\end{array}
\right )
\label{eq1-33}
\ee
The 3 complex angles occurring here are combinations of the $\delta M^2(s)$
matrix elements and of the eigenvalues of the full $M^2(s)$ matrix, as displayed
in Subsection 10.2 of  \cite{ExtMod7}.

It is worth remarking that the dynamical mixing just sketched has provided
 the first solution \cite{taupaper,ExtMod3} to the long-standing  puzzle
 "$e^+e^-$  versus $\tau$" \cite{DavierHoecker,DavierHoecker3,CapriMB} as it generates a $s$--dependent difference between the $\rho^\pm - W^\pm$ and $\rho^0 - \gamma$ transition amplitudes.
 
\subsection{The Kinetic Breaking and the [$\pi^0,~\eta,~\etp$] System}   
\label{HLS_PS} 
\indent \indent This Section mostly aims at reminding notations 
used in the body of the paper; these essentially deal with the
 pseudoscalar meson (PS)  sector of the HLS model.
 
The full pseudoscalar meson kinetic energy term of the
BHLS$_2$ Lagrangian \cite{ExtMod8} writes~:
\be
 \displaystyle {\cal L}_{kin}^\prime = {\rm Tr} \left [ \pa P_{bare}
   X_A \pa P_{bare}  X_A\right ]
 + 2 ~ \{ {\rm Tr} \left [X_H \pa  P_{bare} \right ] \}^2  ~~.
   \label{kin12}
\ee
where $P_{bare}$ is the  PS {\it bare} field matrix.
The first term is already  broken by the BKY mechanism applied to the
${\cal L}_{A}$ HLS Lagrangian piece (see Equation (\ref{eq1-6}) in Appendix
\ref{HLS_unbrk}) and the second one expresses  the so-called kinetic breaking generalizing
the 'tHooft mechanism \cite{tHooft}. It has been shown in \cite{ExtMod8} that
an appropriate choice for the $X_H$ matrix is~:
\be
\displaystyle  
 X_H = \lambda_0 T_0 +\lambda_3 T_3+\lambda_8 T_8
\label{kin10}
\ee
in terms of the canonical $U(3)$ Gell-Mann matrices ($T_0=I/\sqrt{6}$,
${\rm Tr} [T_aT_b] = \delta_{ab}/2$) with 
real $\lambda_i$ coefficients in close correspondence with the 
CD breaking term $\delta V$ affecting the vector sector (see Appendix \ref{HLS-CD}). 
This choice manifestly allows for  Isospin Symmetry breaking, nonet symmetry
breaking (the so--called 't~Hooft term \cite{tHooft}) and $SU(3)$ breaking.

It is useful to introduce the vector of PS fields~:
 \be
\begin{array}{lll}
\displaystyle 
{ \cal V}_{any}=(\pi^3_{any},\eta^0_{any},\eta^8_{any})~~& {\rm where~~} any=(bare,~R1,~R)
\end{array}
\label{kin5c}
\ee
to clarify the component indexing. 

The diagonalization of the kinetic energy Equation (\ref{kin12}) which leads 
from the {\it bare} PS fields to their renormalized  partners (hereafter 
indexed by $R$) is performed in 2 steps. The intermediate  step 
(from {\it bare} to  to $R1$ fields) turns out to diagonalizing 
${\rm Tr} \left [ \pa P_{bare}   X_A \pa P_{bare}  X_A\right ]$ and  
to define the $W$ transformation matrix~: 
  \be
\displaystyle W =
\left (
\begin{array}{ccc}
\displaystyle 1 & \displaystyle -\frac{\Delta_A}{\sqrt{6}} & \displaystyle  -\frac{\Delta_A}{2\sqrt{3}}   \\[0.5cm]
 \displaystyle - \frac{\Delta_A}{\sqrt{6}}   & \displaystyle B  & \displaystyle A \\[0.5cm]
\displaystyle  -\frac{\Delta_A}{2\sqrt{3}}     & \displaystyle A& \displaystyle C 
\end{array}   
\right )
\label{kin4}
\ee
which depends on the BKY breaking parameter $\Delta_A$ and via~:
 \be
\begin{array}{lll}
\displaystyle A=\sqrt{2} \frac{z_A-1}{3z_A} ~,& B= \displaystyle \frac{2
z_A+1}{3z_A} ~,& C=\displaystyle \frac{z_A+2}{3z_A}
\end{array}
\label{kin5}
\ee
on the other BKY breaking parameter $z_A$ (see Appendix \ref{HLS-BKY} above).

In order to achieve the diagonalization of the (full) kinetic energy term of
the BHLS$_2$ Lagrangian, one still has to
define the linear transform which relates the intermediate $R1$ and final $R$ renormalized PS
fields (see Equation (28) in \cite{ExtMod8}). Given the (co-)vector~: 
 \be
\displaystyle
a^t = 
\left (~\lambda_3, \lambda_0 B + \lambda_8 A, 
\lambda_0 A + \lambda_8 C \right )~~,
\label{kin5d}
\ee
one can then prove \cite{ExtMod8} that 
Equation (\ref {kin12}) becomes canonical (at first order in breakings)
when expressed in terms of the ${ \cal V}_R$ fields defined by~:
 \be
\displaystyle
 {\cal V}_{bare} = W \cdot \left [1- \frac{1}{2}a \cdot a^t \right ] \cdot {\cal
 V}_{R}~~.
   \label{kin20}
\ee

However, the ${\cal V}_{R}$ fields are not still the PS mass eigenstates denoted
 by the triplet ($\pi^0,~\eta,~\etp$). One  expects these {\it physical} states
 to be related with the ${ \cal V}_R$ fields via  a 
 3--dimensional rotation and thus 3 angles. Adopting 
 the Leutwyler parametrization \cite{leutw96}, one has~:
\be
\left (
\begin{array}{l} 
\displaystyle \pi^3_R \\[0.5cm]
\displaystyle \eta^8_R \\[0.5cm]
\displaystyle  \eta^0_R
\end{array}
\right )
=
\left (
\begin{array}{ccc} 
\displaystyle 1 & \displaystyle -\hvar & \displaystyle -\hvar^\prime \\[0.5cm]
\displaystyle \hvar \cos{\theta_P}+\hvar^\prime \sin{\theta_P} &
\displaystyle \cos{\theta_P} &\displaystyle \sin{\theta_P} \\[0.5cm]
\displaystyle -\hvar \sin{\theta_P}+\hvar^\prime \cos{\theta_P} &
\displaystyle -\sin{\theta_P} &\displaystyle \cos{\theta_P}  
\end{array}
\right )
\left (
\begin{array}{l} 
\displaystyle \pi^0 \\[0.5cm]
\displaystyle \eta \\[0.5cm]
\displaystyle  \etp
\end{array}
\right )
 \label{current12-1}  
\ee
to relate the $R$ fields which diagonalize the kinetic energy to the physical
({\it i.e. } mass eigenstates) neutral PS fields. The three angles 
($\hvar$, $\hvar^\prime$ and even $\theta_P$)
are assumed  ${\cal O}(\delta)$ perturbations; nevertheless, 
for clarity, the so-called third mixing angle  \cite{WZWChPT}
is not treated as manifestly small.

On the other hand, the "angles" $\hvar$ and $\hvar^\prime$ are related
with the light quark masses  and it is worth stating that they  are expected 
likesign (see the discussion in \cite{ExtMod8}).

\section{Erratum~: The VPP/APP interaction pieces in BHLS$_2$}
\label{Erratum}
\indent \indent  It is  worthwhile to list the $VPP$ and $APP$
interaction terms of the BHLS$_2$ Lagrangian, corrected when needed, related with the
 present study, {\it i.e.} the charged and neutral
pion fields, the $\eta$ and $\etp$ mesons. We have~:
\be
\begin{array}{ll}
 \displaystyle  
 {\cal L}_{\pi^-\pi^+}=&  \displaystyle ie 
 \left[ 1 -\frac{a}{2} (1+ \Sigma_V) \right] A \cdot \pi^- \parsym \pi^+
 + \frac{i a g}{2} (1+ \Sigma_V) \left[ 1+\xi_3 \right] \rho^0_I \cdot\pi^- \parsym \pi^+
 \\[0.5cm]
\displaystyle  {\cal L}_{\pi^0 \pi^\pm}= & \displaystyle 
\frac{iag}{2} (1+\Sigma_V)  (1-\frac{\lambda_3^2}{2})\left[
\rho^- \cdot \pi^+ \parsym \pi^0   - \rho^+ \cdot  \pi^- \parsym \pi^0 
\right ]\\[0.5cm]
\displaystyle  {\cal L}_{\eta \pi^\pm}= & \displaystyle 
- \frac{iag}{2} \left [
\left \{ \frac{1}{2\sqrt{3}} \Delta_A + \frac{\lambda_3 \tlambda_8}{2} \right \} \cos{\theta_P}
-\left \{ \frac{1}{\sqrt{6}} \Delta_A + \frac{\lambda_3 \tlambda_0}{2} \right \} 
\sin{\theta_P}
+\epsilon \right ]
\\[0.5cm]
\hspace{-2.5cm} ~& \displaystyle
\left[ 1+ \Sigma_V \right]  \left[
\rho^- \cdot \pi^+ \parsym \eta   - \rho^+ \cdot \pi^- \parsym \eta
\right ]\\[0.5cm]
\displaystyle  {\cal L}_{\eta^\prime \pi^\pm}= & \displaystyle 
-\frac{iag}{2} 
 \left [
\left \{ \frac{1}{\sqrt{6}} \Delta_A + \frac{\lambda_3 \tlambda_0}{2} \right \}
\cos{\theta_P}
+ 
\left \{ \frac{1}{2\sqrt{3}} \Delta_A + \frac{\lambda_3 \tlambda_8}{2} \right \} 
\sin{\theta_P}
+\epsilon^\prime \right ]
\\[0.5cm]
 \hspace{-2.5cm} ~& \displaystyle
\left[ 1+ \Sigma_V \right] \left[
\rho^- \cdot \pi^+ \parsym \eta^\prime   - \rho^+ \cdot \pi^- \parsym \eta^\prime
\right ]
 \end{array}
\ee

The last 2 Lagrangian pieces supersede the corresponding formulae
displayed in Equations (45) of \cite{ExtMod8}; they were given for
completeness but unused. In the present study, they should be considered.

In the expressions above, the kinetic breaking parameters occur;
beside $\lambda_3$, one also has~:
 \be
\begin{array}{ll}
 \displaystyle \tlambda_0= \lambda_0 B + \lambda_8 A~,
&  \displaystyle \tlambda_8= \lambda_0 A + \lambda_8 C
 \end{array}
 \label{kin14}
\ee
where $A$, $B$ and $C$ have also been reminded in the Appendix \ref{HLS_PS} just above.
On the other hand, we have chosen here to keep the $\Sigma_V$ parameter for  
clarity. However, in \cite{ExtMod8} it has been shown that it is out 
of reach and can be fixed to zero without any loss of generality.

\section{$A_\pm$ Solutions~: The $AAP$ and  $VVP$ Lagrangians}
\label{AAP-VVP}
\indentB It is worthwhile displaying the anomalous BHLS$_2$ Lagrangian
pieces associated with the so-called triangle anomalies, having imposed
the Kroll Conditions \cite{Kroll:2005}, examined in full  details in 
\cite{ExtMod8} and briefly sketched in Section  \ref{Kroll-cnd}.
Using obvious notations, these anomalous pieces are derived from 
\cite{FKTUY,HLSRef}~:
 \be
\left \{
\begin{array}{lll}
{\cal L}_{VVP}=& \displaystyle - \frac{N_c g^2}{4 \pi^2 f_\pi} ~c_3
 \epsilon^{\mu \nu \alpha \beta}{\rm Tr}[ \partial_\mu V_\nu \partial_\alpha V_\beta P] \\[0.5cm] 
 {\cal L}_{AAP}=& \displaystyle - \frac{N_c e^2 }{4 \pi^2 f_\pi} ~(1- c_4)
 \epsilon^{\mu \nu \alpha \beta}\partial_\mu A_\nu \partial_\alpha A_\beta{\rm Tr}[Q^2 P]\\[0.5cm]
 {\cal L}_{AVP}=& \displaystyle - \frac{N_c ge }{8 \pi^2 f_\pi} ~(c_4- c_3)
 \epsilon^{\mu \nu \alpha \beta}\partial_\mu A_\nu
 {\rm Tr}[ \{ \partial_\alpha V_\beta , Q \} P] 
 \end{array}
\right .
\label{QQ1}
\ee
The phenomenology examined so far with the broken variants of the HLS
model never led to consider a non-zero  $c_3-c_4$; therefore, one assumes
$c_3=c_4$ which turns out to discard   ${\cal L}_{AVP}$  Lagrangian piece.

Unless otherwise stated
the neutral  vector fields displayed here are the so-called
 ideal combinations generally named $\rho^I$, $\omg^I$ and $\phi^I$. 
 The transformation which connects
the {\it bare} vector fields to their {\it physical} partners is treated
in \cite{ExtMod7} and briefly reminded in Appendix \ref{HLS_unbrk} above.

We also remind here the definition for $\delta_P$~:
\be
\left \{
\begin{array}{ll}
\displaystyle \sin{\delta_P} =  \frac{1}{\sqrt{3}}
\left ( \sqrt{2} \sin{\theta_P}- \cos{\theta_P} 
\right ),&
\cos{\delta_P}= \displaystyle \frac{1}{\sqrt{3}}
\left ( \sqrt{2} \cos{\theta_P} +  \sin{\theta_P}\right )
\end{array}
\right . 
\label{AA25}
\ee
and ($d_\pm \equiv \pm 1$)~:
\be
A_\pm= \Delta_A+ d_\pm \lambda_0^2~~.
\label{Apm}
\ee
used below.
\subsection{The $AAP$ Lagrangian}
\label{AAP}
\indent \indent The $AAP$ Lagrangian defined in the header just above
where $Q$ is the quark charge matrix and $P$  the $U(3)$ symmetric matrix
of the bare pseudoscalar fields is given for definiteness. Defining~:

\be
\hspace{-2.5cm}
\left \{
\begin{array}{lll} 
\displaystyle g_{\pi^0\gam \gam}=&\displaystyle  ~~\frac{1}{6} \left \{
1 - \frac{5}{6}A_\pm - \frac{\lambda_0^2}{3} \right \} \\[0.5cm]
~&\displaystyle   -\displaystyle 
 \frac{\epsilon}{18 z_A}
\left \{5 z_A\sin{\delta_P}+\sqrt{2}\cos{\delta_P}
\right\} -\displaystyle  \frac{\epsilon^\prime}{18 z_A}
\left \{\sqrt{2}\sin{\delta_P}- 5 z_A\cos{\delta_P}
\right\}\,,\\[0.6cm] 
\displaystyle g_{\eta \gam \gam}=&\displaystyle -\frac{\epsilon}{6}
-\frac{\sqrt{2}}{18 z_A} \cos{\delta_P}
+\frac{1}{12} \left \{
 A_\pm + \frac{5}{6} (3\lambda_0^2 -4)
\right \} \sin{\delta_P}
 \\[0.5cm] 
 \displaystyle g_{\etp \gam \gam}=&\displaystyle -\frac{\epsilon^\prime}{6}
-\frac{\sqrt{2}}{18 z_A} \sin{\delta_P}
-\frac{1}{12} \left \{
 A_\pm + \frac{5}{6} (3\lambda_0^2 -4) \right \}  \cos{\delta_P}
\end{array}
\right . 
\label{AA26}
\ee
the coupling constants for the physical mesons $P_0 \gam \gam$ ($P_0=\pi^0 ,\eta,\etp$)
are given by:
 \be
\displaystyle 
G_{P_0 \gam \gam}= -\frac{3 \alpha_{em}}{\pi f_\pi} (1-c_4) g_{P_0\gam \gam},
 \label{AA26-2}
\ee
and the $AAP$ Lagrangian can also be written~:
 \be
\displaystyle 
{\cal L}_{AAP_0}= G_{P_0 \gam \gam}~~  P_0 ~  \epsilon^{\mu \nu \alpha \beta}  \pa_\mu A_\nu \pa_\alpha A_\beta ~~
{\rm for~each~of}~~~~ P_0 = \pi^0 ,\eta,\etp~~.
\label{AA26-3}
\ee

\subsection{The $VVP$ Lagrangian}
\indent \indent The $VVP$ Lagrangian is given by~:

\be
\displaystyle 
{\cal L}_{VVP}= -\frac{3 g^2}{4 \pi^2 f_\pi}~c_3 
~\epsilon^{\mu \nu \alpha \beta} \mathrm{Tr} \left [ \pa_\mu V_\nu \pa_\alpha V_\beta   P \right ]
~~~, ~~~~ \displaystyle C=-\frac{N_c g^2 c_3}{4 \pi^2 f_\pi}\,,
\label{BB1}
\ee

\subsubsection{The $VV\pi$ Lagrangians}
\label{VVpi_lags}
\indent \indent The $VV\pi$ Lagrangians relevant for our phenomenology 
 are given by~:

\be
\hspace{-1.cm}
\begin{array}{ll}
\displaystyle
{\cal L}_{VVP}(\pi^\pm)=&
\displaystyle
\frac{C}{2} \epsilon^{\mu \nu \alpha \beta} \Biggl\{
 \left [
 \left ( 1+\frac{2 \xi_0+\xi_8}{3} \right )
\partial_\mu \omg_\nu^I
+\frac{\sqrt{2}}{3} (\xi_0-\xi_8)
\partial_\mu \phi^I_\nu 
 \right] \times \left[ \partial_\alpha \rho^+_\beta \pi^- +\partial_\alpha \rho^-_\beta \pi^+
 \right]\Biggr\}
 \end{array}
\label{AC1}
\ee
and~:
\be
\hspace{-1.cm}
\begin{array}{ll}
\displaystyle
{\cal L}_{VVP}(\pi^0)=&
\displaystyle \frac{C}{2}
\epsilon^{\mu \nu \alpha \beta} \Biggl\{ 
G_0 \partial_\mu \rho^I_\nu \partial_\alpha \omg^I_\beta 
+G_1 \left [2 \partial_\mu \rho^-_\nu \partial_\alpha \rho^+_\beta +
 \partial_\mu \rho^I_\nu \partial_\alpha \rho^I_\beta +
 \partial_\mu \omg^I_\nu \partial_\alpha \omg^I_\beta
\right]  \Biggr. \\[0.5cm]
~~& \displaystyle \hspace{2. cm} \Biggl . + G_2  \partial_\mu \phi^I_\nu \partial_\alpha \phi^I_\beta
+ G_3   \partial_\mu \rho^I_\nu  \partial_\alpha \phi^I_\beta  \Biggr\}~\pi^0
 \end{array}
\label{AC2}
\ee
where~:
\be
\left\{
\begin{array}{ll}
\displaystyle G_0=\left [
1-\frac{\lambda_0^2}{3} + \frac{2 \xi_0+\xi_8}{3} +\xi_3 \right] \\[0.5cm]
\displaystyle G_1 = -\frac {A_\pm}{4} 
+\frac{1}{2} \left [ \epsilon^\prime \cos{\delta_P} -  \epsilon  \sin{\delta_P} \right ]
\\[0.5cm]
\displaystyle G_2 = 
-\frac{1}{z_A \sqrt{2}} \left [ \epsilon^\prime \sin{\delta_P} +  \epsilon  \cos{\delta_P} \right ]
\\[0.5cm]
\displaystyle G_3 = \frac{\sqrt{2}}{3} (\xi_0-\xi_8)
\end{array}
\right .
\label{AC3}
\ee
Actually, one imposes $\xi_0=\xi_8$, so that, always,  $G_3 =0$.
\subsubsection{The $VV\eta$ Lagrangian}
\label{VVeta}
\indent \indent The $VV\eta$ Lagrangian is given by~:
\be
\hspace{-1.cm}
\begin{array}{ll}
\displaystyle
{\cal L}_{VVP}(\eta)=&
\displaystyle \frac{C}{2}
\epsilon^{\mu \nu \alpha \beta} \Biggl\{
K_1 \partial_\mu \rho^-_\nu \partial_\alpha \rho^+_\beta +
 K_2 \partial_\mu \rho^I_\nu \partial_\alpha \rho^I_\beta +
 K_3  \partial_\mu \omg^I_\nu \partial_\alpha \omg^I_\beta +
 K_4  \partial_\mu \phi^I_\nu \partial_\alpha \phi^I_\beta  \Biggr. \\[0.5cm]
 ~~& \displaystyle \hspace{2. cm} \Biggl . + 
 K_5  \partial_\mu \omg^I_\nu  \partial_\alpha \phi^I_\beta +
 K_6 \partial_\mu \rho^I_\nu \partial_\alpha \omg^I_\beta  \Biggr\}~\eta
 \end{array}
\label{AD1}
\ee

Having defined\footnote{Referring to \cite{ExtMod8},  
 the Kroll conditions turn out to fix $H_1=0$.}~:
\be
\left\{
\begin{array}{ll}
\displaystyle H_2 = 
\frac {1}{8} \left [ 3\lambda_0^2 -4 \right]
~~,&
\displaystyle H_3 = -\frac{\sqrt{2}}{6 z_A} \left [(3+2 \xi_0+4 \xi_8) \right] 
\end{array}
\right \}
\label{AD2}
\ee
the $VV\eta$ couplings become~:
\be
\left\{
\begin{array}{ll}
\displaystyle K_1= 2  H_2 \sin{\delta_P}~~,&
\displaystyle K_2 = (H_2 - \xi_3) \sin{\delta_P}
  \\[0.5cm]
\displaystyle K_3 =  
\left[ H_2 - \frac{2 \xi_0+\xi_8}{3} \right] \sin{\delta_P}
~~,&
\displaystyle K_4 = H_3 \cos{\delta_P}
  \\[0.5cm]
\displaystyle K_5 = 
-\frac{(\xi_0 -\xi_8)}{3 z_A} \left [ 2 \cos{\delta_P} + z_A \sqrt{2} \sin{\delta_P}
 \right]
~~,&
\displaystyle K_6 = \frac{A_\pm}{2} \sin{\delta_P} - \epsilon
\end{array}
\right .
\label{AD4}
\ee
Actually, similarly to just above, the $K_5$ term drops out in the practical BHLS$_2$
context.
\subsubsection{The $VV\etp$ Lagrangian}
\label{VVetp}
\indent \indent The $VV\etp$ Lagrangian is given by~:
\be
\hspace{-1.cm}
\begin{array}{ll}
\displaystyle
{\cal L}_{VVP}(\etp)=&
\displaystyle \frac{C}{2}
\epsilon^{\mu \nu \alpha \beta} \Biggl\{
K^\prime_1 \partial_\mu \rho^-_\nu \partial_\alpha \rho^+_\beta +
 K^\prime_2 \partial_\mu \rho^I_\nu \partial_\alpha \rho^I_\beta +
 K^\prime_3  \partial_\mu \omg^I_\nu \partial_\alpha \omg^I_\beta +
 K^\prime_4  \partial_\mu \phi^I_\nu \partial_\alpha \phi^I_\beta  \Biggr. \\[0.5cm]
 ~~& \displaystyle \hspace{2. cm} \Biggl . + 
 K^\prime_5  \partial_\mu \omg^I_\nu  \partial_\alpha \phi^I_\beta +
 K^\prime_6 \partial_\mu \rho^I_\nu \partial_\alpha \omg^I_\beta  \Biggr\}~\etp
 \end{array}
\label{AE1}
\ee
the $VV\etp$ couplings being~:
\be
\left\{
\begin{array}{ll}
\displaystyle K^\prime_1= -2  H_2 \cos{\delta_P}
~~,&
\displaystyle K^\prime_2 = -(H_2 - \xi_3)\cos{\delta_P}
  \\[0.5cm]
\displaystyle K^\prime_3 = -\left[ H_2 - \frac{2 \xi_0+\xi_8}{3} \right]  \cos{\delta_P}
~~,&
\displaystyle K^\prime_4 = H_3 \sin{\delta_P} 
  \\[0.5cm]
\displaystyle K^\prime_5 = 
-\frac{(\xi_0 -\xi_8)}{3 z_A} \left [- z_A \sqrt{2} \cos{\delta_P} +2 \sin{\delta_P}
 \right]
~~,&\displaystyle K^\prime_6 = 
-\frac{A_\pm}{2}   \cos{\delta_P}
 -\epsilon^\prime
\end{array}
\right .
\label{AE2}
\ee
where also the $K^\prime_5$ term drops out in the practical BHLS$_2$ context, 
where $\xi_0 =\xi_8$. The $H_i$ functions occurring here have been defined in our
 previous paper and have been reminded in the Subsection
just above --  $H_1$ vanishes thanks to having requested the Kroll Conditions.

One should also note that the $VV\etp$ couplings are related
to the $VV\eta$  couplings and can be derived herefrom by making in the $VV\eta$  couplings~:
$$\{\sin{\delta_P} \ra -\cos{\delta_P} ~~{\rm and} ~~~\cos{\delta_P} \ra \sin{\delta_P}\}~~.$$

\section{$A_\pm$ Solutions~: The $APPP$ and  $VPPP$ Lagrangians}
\label{PPP}
\indentB Beside the Lagrangian pieces associated with the triangle anomalies
reminded in the Appendix just above, those associated with the so-called 
box anomalies
play an important role in the $\eta/\etp \ra \pi^+ \pi^- \gam$ decays and
in the $e^+ e^- \ra \pi^+ \pi^-\pi^0$ annihilation thoroughly considered 
in our \cite{ExtMod8}.
We find it helpful to provide their expressions while the Kroll conditions
are applied.  The $APPP$ and  $VPPP$ Lagrangian pieces introduce a new
HLS parameter ($c_1-c_2$) which is not fixed by the model and should be derived from fits.

As for the $VVP$ interactions reminded in Appendix 
\ref{AAP-VVP}, the neutral vector fields occurring in the $VPPP$ interaction
Lagrangian are their ideal combinations; they should be expressed in terms
of {\it physical} vector fields as developed in \cite{ExtMod7} in practical
applications.

 \subsection{The $APPP$  Lagrangian} 
\label{APPP}
\indent \indent The $APPP$ Lagrangian is given by~:
\be
\displaystyle 
{\cal L}_{APPP}= D
~\epsilon^{\mu \nu \alpha \beta} A_\mu
\mathrm{Tr} \left [Q \pa_\nu P \pa_\alpha   P \pa_\beta   P\right ]
~~, ~~ \displaystyle 
D = -i\frac{N_c e}{3 \pi^2 f_\pi^3} \left[1-\frac{3}{4}(c_1-c_2+c_4)\right]
\,,
\label{CC1}
\ee
Regarding the phenology we address, the relevant $APPP$ Lagrangian 
piece to be considered is~:
\be
\hspace{-1.cm}
\displaystyle 
{\cal L}_{APPP}^1= D
\epsilon^{\mu \nu \alpha \beta} A_\mu
\left \{
g_{\gamma \pi^0} \partial_\nu \pi^0 +
g_{\gamma \eta} \partial_\nu \eta +
g_{\gamma \etp} \partial_\nu \etp 
\right \} ~
\partial_\alpha \pi^- \partial_\beta \pi^+ 
\,,
\label{CC2}
\ee
in terms of fully renormalized PS fields. Requiring the $A_\pm$ Kroll 
conditions, these $g_{\gam P}$ couplings can be written~:
\be
\left \{
\begin{array}{ll}
\displaystyle g_{\gamma \pi^0} = -\frac{1}{4} \left [
1 - \frac{A_\pm}{2} -\frac{\lambda_0^2}{3}
-\epsilon\sin{\delta_P} +\epsilon^\prime \cos{\delta_P}
\right]\\[0.5cm]
 \displaystyle g_{\gamma \eta} =~~~~ 
 \left [ 1 -\frac{A_\pm}{2}  -\frac{3 \lambda_0^2}{4} \right ]
\frac{\sin{\delta_P}}{4} + \frac{\epsilon}{4}\\[0.5cm]

\displaystyle g_{\gamma \etp} =    
 - \left [  1 -\frac{A_\pm}{2}  -\frac{3 \lambda_0^2}{4} \right ]
 \frac{\cos{\delta_P}}{4}  +
\frac{\epsilon^\prime}{4}
\end{array}
\right.
\label{CC4}
\ee
keeping only the leading order terms in the breakings.

\subsection{The $VPPP$ Lagrangian}
\label{VPPP}
\indentB  The $VPPP$ anomalous  HLS Lagrangian is~:
 \be
 {\cal L}_{VPPP}= \displaystyle - i \frac{N_c g }{4 \pi^2 f_\pi^3} (c_1-c_2-c_3)
   \epsilon^{\mu \nu \alpha \beta}{\rm Tr}[V_\mu \partial_\nu P \partial_\alpha P \partial_\beta P]
\label{AA1new}
\ee
where the $c_i$ are the FKTUY parameters not fixed by the model. $N_c$ 
is the number of colors fixed to 3.The $V$ and $P$ field matrices are the 
bare ones.

The relevant part of ${\cal L}_{VPPP}$ within the present context is~:
\be
\left \{
\begin{array}{ll}
\displaystyle 
{\cal L}_{VP_0 \pi^+ \pi^-}=E\epsilon^{\mu \nu \alpha \beta}  \left\{
\left [ g_{\rho \pi}^0 \partial_\nu \pi^0+g_{\rho \eta}^0 \partial_\nu \eta
+g_{\rho \eta^\prime}^0 \partial_\nu \eta^\prime \right ] ~\rho^{I}_\mu\right.\\[0.5cm]
\displaystyle 
~~~+\left [ g_{\omg \pi}^0 \partial_\nu \pi^0+g_{\omg \eta}^0 \partial_\nu \eta+
+g_{\omg \eta^\prime}^0 \partial_\nu \eta^\prime \right ]~\omega_\mu^{I}
+ g_{\phi \pi}^0\partial_\nu \pi^0 ~\phi_\mu^{I} 
 \left\} \partial_\alpha\pi^-  \partial_\beta \pi^+ \right.\\[0.5cm]
\displaystyle 
{\rm ~~with~~} E=-i \frac{ 3 g (c_1-c_2-c_3)}{4 \pi^2 f_\pi^3}
\end{array}
\right.
\label{AA3new}
\ee
in terms of the {\it physical} pseudoscalar fields.
Keeping only the $A_\pm$ solutions and the leading-order breaking terms,
the couplings just defined  are~:
  \be
\left \{
\begin{array}{ll}
\displaystyle  
g_{\rho \pi^0}^0= \frac{1}{4} \left [\frac{A_\pm}{2} 
+ \epsilon\sin{\delta}_P
- \epsilon^\prime\cos{\delta}_P
\right ]\\[0.5cm]
\displaystyle 
g_{\rho \eta}^0=\frac{1}{4} 
\left [1 +\xi_3 - \frac{3}{4} \lambda_0^2 \right]\sin{\delta}_P
\\[0.5cm]
\displaystyle 
g_{\rho \eta^\prime}^0= -\frac{1}{4}
\left [1 +\xi_3 - \frac{3}{4} \lambda_0^2 \right] \cos{\delta}_P
\end{array}
\right.
\label{AA4new-1}
\ee
and~:
 \be
\left \{
\begin{array}{ll}
\displaystyle  
g_{\omg \pi^0}^0= -\frac{3}{4} 
\left [1 + \frac{2 \xi_0+\xi_8}{3} - \frac{1}{3}\lambda_0^2 \right]
\\[0.5cm]
\displaystyle 
 g_{\omg \eta}^0=   \frac{3}{4}
 \left\{ \epsilon - \frac{A_\pm}{2}  \sin{\delta_P} \right \}
\\[0.5cm]
\displaystyle 
g_{\omg\eta^\prime}^0= \frac{3 }{4}
 \left\{ \epsilon^\prime +\frac{A_\pm}{2} \cos{\delta_P} \right \}
\\[0.5cm]
\displaystyle 
g_{\phi \pi}^0= -\frac{\sqrt{2}}{4}\left [\xi_0-\xi_8 \right]
~~~,~~g_{\phi \eta}^0= 0,~~g_{\phi \eta^\prime}^0= 0 ~~.
\end{array}
\right.
\label{AA4new-2}
\ee
As pseudoscalar meson form factor values at origin imply \cite{ExtMod7}
$\xi_0=\xi_8$, one observes that no term involving $\phi^I$ survives
at leading order in breakings.

\newpage
                   \bibliographystyle{h-physrev}

                     \bibliography{vmd_2022_FINAL}

\end{document}